\documentclass[hyper]{JHEP3}
\usepackage{graphicx}
\usepackage{longtable}

\input{epsf}
\usepackage{epsfig}
\usepackage{amssymb}
\usepackage{amsfonts}
\usepackage{amsbsy}
\usepackage[all]{xy}
\usepackage{amsmath}
\usepackage{verbatim}

\usepackage{amssymb,amscd}
\usepackage{mathrsfs}
\usepackage{amsmath,amsthm}

\renewcommand{\Im}{{\rm Im}\,}
\renewcommand{\Re}{{\rm Re}\,}

\def\one{{\hbox{ 1\kern-.8mm l}}}
\def\tr{{\rm tr\,}}

\def\re{\mbox{Re }}
\def\im{\mbox{Im }}

\def\Tr{{\rm Tr}}

\def\CA {{\cal A}}
\def\CB {{\cal B}}

\def\CD {{\cal D}}

\def\CF {{\cal F}}

\def\CH {{\cal H}}
\def\CI {{\cal I}}
\def\CJ {{\cal J}}

\def\CL {{\cal L}}
\def\CM {{\cal M}}
\def\CN {{\cal N}}
\def\CO {{\cal O}}
\def\CP {{\cal P}}

\def\CW {{\cal W}}

\def\TT{{\Bbb{T}}}

\newcommand{\C}{{\mathbb{C}}}
\newcommand{\R}{{\mathbb{R}}}
\newcommand{\Z}{{\mathbb{Z}}}
\newcommand{\IH}{{\mathbb{H}}}

\newcommand{\M}{{{\cal M}}_{{\rm flat}}}

\newcommand\be{\begin{equation}}
\newcommand\ee{\end{equation}}
\newcommand\bes{\begin{equation*}}
\newcommand\ees{\end{equation*}}
\newcommand{\bea}{\begin{eqnarray}}
\newcommand{\eea}{\end{eqnarray}}
\newcommand{\beas}{\begin{eqnarray*}}
\newcommand{\eeas}{\end{eqnarray*}}

\newcommand{\Vol}{{\mbox{Vol\,}}}
\def\={\;=\;}
\def\Li{\text{Li}_2}
\def\a{\alpha}
\def\t{\tau}
\def\ve{\varepsilon}
\def\e{\bold e}
\def\Rp{\mathbb R^{(+)}}
\def\sm{\smallsetminus}

\title{Exact Results for Perturbative Chern-Simons Theory with Complex Gauge Group}

\author{Tudor Dimofte,$^1$ Sergei Gukov,$^{1,2}$ Jonatan Lenells,$^3$ and Don Zagier$^{4,5}$
\\ ~
\\
$^1$ California Institute of Technology, Pasadena, CA 91125, USA \\
$^2$  Department of Physics and Department of Mathematics,\\
~ University of California, Santa Barbara, CA 93106, USA\\
$^3$ Centre for Mathematical Sciences, Wilberforce Road,\\
~ Cambridge, CB3 0WA, United Kingdom\\
$^4$ Max-Planck-Institut f\"ur Mathematik, Vivatsgasse 7, D-53111 Bonn, Germany\\
$^5$ Coll\`ege de France, 3 rue d'Ulm, F-75005 Paris, France}

\abstract{We develop several methods that allow us to compute
all-loop partition functions in perturbative Chern-Simons theory
with complex gauge group $G_{\C}$, sometimes in multiple ways.
In the background of a non-abelian irreducible flat connection,
perturbative $G_{\C}$ invariants turn out to be interesting topological invariants,
which are very different from finite type (Vassiliev)
invariants obtained in a theory with compact gauge group $G$.
We explore various aspects of these invariants and present
an example where we compute them explicitly to high loop order.
We also introduce a notion of ``arithmetic TQFT'' and conjecture (with supporting
numerical evidence) that $SL(2,\C)$ Chern-Simons theory is an example of such a theory.
\\
\\
\\
\\
\\
\\
{\tt CALT-68-2716} }

\begin{document}


\section{Introduction and summary}
\label{sec:intro}

Three-dimensional Chern-Simons gauge theory is a prominent example of
a topological quantum field theory (TQFT).
By now, Chern-Simons theory with a compact gauge group $G$
is a mature subject with a history going back to the 1980's
(see {\it e.g.} \cite{freed-2008,labastida-1999} for excellent reviews)
and with a wide range of applications,
ranging from invariants of knots and 3-manifolds \cite{witten-1989}
on one hand, to condensed matter physics \cite{heinonen-1998,murthy-2003}
and string theory \cite{marino-2004} on the other.

In this paper, we will be interested in a version of Chern-Simons
gauge theory with complex gauge group $G_{\C}$.
Although at first it may appear merely as a variation on the subject,
the physics of this theory is qualitatively different from that
of Chern-Simons gauge theory with compact gauge group.
For example, one important difference is that
to a compact Riemann surface $\Sigma$ Chern-Simons theory with compact
gauge group associates a finite-dimensional Hilbert space $\CH_{\Sigma}$,
whereas in a theory with non-compact (and, in particular, complex)
gauge group the Hilbert space is infinite-dimensional.
Due to this and other important differences that will be explained in further detail below,
Chern-Simons gauge theory with complex gauge group
remains a rather mysterious subject. First steps toward understanding
this theory were made in \cite{witten-1991} and, more recently, in \cite{gukov-2003}.

As in a theory with a compact gauge group, the classical action
of Chern-Simons gauge theory with complex gauge group $G_{\C}$
is purely topological --- that is, independent of the metric on the
underlying 3-manifold $M$.
However, since in the latter case the gauge field $\CA$
(a $\frak g_{\C}$-valued 1-form on $M$) is complex,
in the action one can write two topological terms,
involving $\CA$ and $\bar \CA$:
\bea
S & \= & {t \over 8 \pi}
\int_M \Tr
  \Big( \CA \wedge d\CA + \frac23\,\CA \wedge \CA \wedge \CA \Big) \label{csaction} \\
  & & +\;{\bar t \over 8 \pi} \int_M \Tr \Big( \bar \CA \wedge d \bar \CA
   + \frac23\bar \CA \wedge \bar \CA \wedge \bar \CA \Big)\,. \nonumber
\eea
Although in general the complex coefficients (``coupling constants'')
$t$ and $\bar t$ need not be complex conjugate to each other,
they are not entirely arbitrary.
Thus, if we write $t = k + \sigma$ and $\bar t = k - \sigma$,
then consistency of the quantum theory requires the ``level'' $k$
to be an integer, $k \in \Z$, whereas unitarity requires $\sigma$
to be either real, $\sigma \in \R$, or purely imaginary, $\sigma \in i \R$;
see {\it e.g.} \cite{witten-1991}.

Given a 3-manifold $M$ (possibly with boundary), Chern-Simons theory
associates to $M$ a ``quantum $G_{\C}$ invariant'' that we denote as $Z(M)$.
Physically, $Z(M)$ is the partition function of the Chern-Simons
gauge theory on $M$, defined as a Feynman path integral
  \be\label{pathint} Z (M) \= \int \,e^{iS}\;\CD\CA  \ee
with the classical action \eqref{csaction}.
Since the action \eqref{csaction} is independent
of the choice of metric on $M$, one might expect that
the quantum $G_{\C}$ invariant $Z(M)$ is a topological invariant of $M$.
This is essentially correct even though independence of metric is less
obvious in the quantum theory, and $Z(M)$ turns out to be an interesting invariant.
How then does one compute $Z(M)$?

One approach is to use topological invariance of the theory.
In Chern-Simons theory with compact gauge group $G$,
the partition function $Z(M)$ can be efficiently computed
by cutting $M$ into simple ``pieces,'' on which the path
integral \eqref{pathint} is easy to evaluate.
Then, via ``gluing rules,'' the answers for individual
pieces are assembled together to produce $Z(M)$.
In practice, there may exist many different ways to decompose $M$
into basic building blocks, resulting in different ways of
computing $Z(M)$.

Although a similar set of gluing rules should exist in
a theory with complex gauge group $G_{\C}$, they are expected
to be more involved than in the compact case.
The underlying reason for this was
already mentioned: in Chern-Simons theory with complex
gauge group the Hilbert space is infinite dimensional (as opposed to
a finite-dimensional Hilbert space in the case of compact gauge group $G$).
One consequence of this fact is that finite sums which appear in
gluing rules for Chern-Simons theory with compact group $G$
turn into integrals over continuous parameters in a theory
with non-compact gauge group.
This is one of the difficulties one needs to face in computing
$Z(M)$ non-perturbatively, {\it i.e.} as a closed-form function
of complex parameters $t$ and $\bar t$.

A somewhat more modest goal is to compute $Z(M)$
perturbatively, by expanding the integral \eqref{pathint}
in inverse powers of $t$ and $\bar t$ around a saddle point
(a classical solution).
In Chern-Simons theory, classical solutions are flat gauge
connections, that is gauge connections $\CA$ which obey
\be
d \CA + \CA \wedge \CA \= 0\,,
\label{aflat}
\ee
and similarly for $\bar \CA$.
A flat connection on $M$ is determined
by its holonomies, that is by a homomorphism
\be
\rho : \pi_1 (M) \to G_{\C}\,.
\label{rhofirst}
\ee
Of course, this homomorphism is only defined modulo gauge transformations,
which act via conjugation by elements in $G_{\C}$.

Given a gauge equivalence class of the flat connection $\CA$, or, equivalently,
a conjugacy class of the homomorphism $\rho$,
one can define a ``perturbative partition function'' $Z^{(\rho)} (M)$
by expanding the integral \eqref{pathint} in inverse powers of $t$ and $\bar t$.
Since the classical action \eqref{csaction} is a sum of two terms,
the perturbation theory for the fields $\CA$ and $\bar \CA$ is independent.
As a result, to all orders in perturbation theory, the partition
function $Z^{(\rho)} (M)$ factorizes into a product of ``holomorphic''
and ``antiholomorphic'' terms:
\be
Z^{(\rho)} (M) \= Z^{(\rho)} (M;t) Z^{(\rho)} (M;\bar t)\,.
\label{zfactorization}
\ee
This holomorphic factorization is only a property of
the perturbative partition function.
The exact, non-perturbative partition function $Z(M)$
depends in a non-trivial way on both $t$ and $\bar t$,
and the best one can hope for is that it can be written in the form
\be
Z(M) \= \sum_{\rho} Z^{(\rho)} (M;t) Z^{(\rho)} (M;\bar t)\,,
\ee
where the sum is over classical solutions  \eqref{aflat} or,
equivalently, conjugacy classes of homomorphisms \eqref{rhofirst}.

In the present paper, we study the perturbative partition
function $Z^{(\rho)} (M)$. Due to the factorization \eqref{zfactorization},
it suffices to consider only the holomorphic part $Z^{(\rho)} (M;t)$.
Moreover, since the perturbative expansion is in the inverse
powers of $t$, it is convenient to introduce a new expansion parameter
\be
\hbar \= {2 \pi i \over t}\,,
\label{hbdef}
\ee
which plays the role of Planck's constant.
Indeed, the semiclassical limit corresponds to $\hbar \to 0$.
In general, the perturbative partition function $Z^{(\rho)} (M;\hbar)$
is an asymptotic power series in $\hbar$. To find its general form
one applies the stationary phase approximation to the integral \eqref{pathint}:
\be\label{zpert}
Z^{(\rho)} (M;\hbar)
 \= \exp\left( \frac{1}{\hbar} S_0^{(\rho)} - \frac{1}{2}\delta^{(\rho)} \log \hbar
 \,+\,\sum_{n=0}^\infty S_{n+1}^{(\rho)} \hbar^n \right)\,.
\ee
This is the general form of the perturbative partition function in
Chern-Simons gauge theory with any gauge group, compact or otherwise.
It follows simply by applying the stationary phase approximation
to the integral \eqref{pathint}, which basically gives the standard
rules of perturbative gauge theory \cite{witten-1989,as-1992,barnatan-1991,barnatan-1991w}
that will be discussed in more detail below.
For now, we note that the leading term $S_0^{(\rho)}$ is the value
of the classical Chern-Simons functional evaluated on a flat gauge
connection $\CA^{(\rho)}$ associated with a homomorphism $\rho$,
\be
S_0^{(\rho)} \= - {1 \over 4} \int_M \Tr
\Big( \CA^{(\rho)} \wedge d \CA^{(\rho)}
+ {2 \over 3} \CA^{(\rho)} \wedge \CA^{(\rho)} \wedge \CA^{(\rho)} \Big)\,.
\label{szero}
\ee
Moreover, the coefficient of the next-leading term, $\delta^{(\rho)}$,
is an integer (which, like all other terms, depends on the 3-manifold $M$,
the gauge group $G_{\C}$, and the classical solution $\rho$).
Each coefficient $S_{n}^{(\rho)}$ is obtained by summing over
Feynman diagrams with $n$ loops.
For example, the ``two-loop term'' $S_2^{(\rho)}$ is obtained by
adding contributions of the two kinds of Feynman diagrams shown in Figure \ref{twoloopfig}.
Although certain features of the Feynman rules in Chern-Simons theory
with complex gauge group $G_{\C}$ are similar to those in a theory with
compact gauge group $G$, we will see that there exist important differences.

\EPSFIGURE{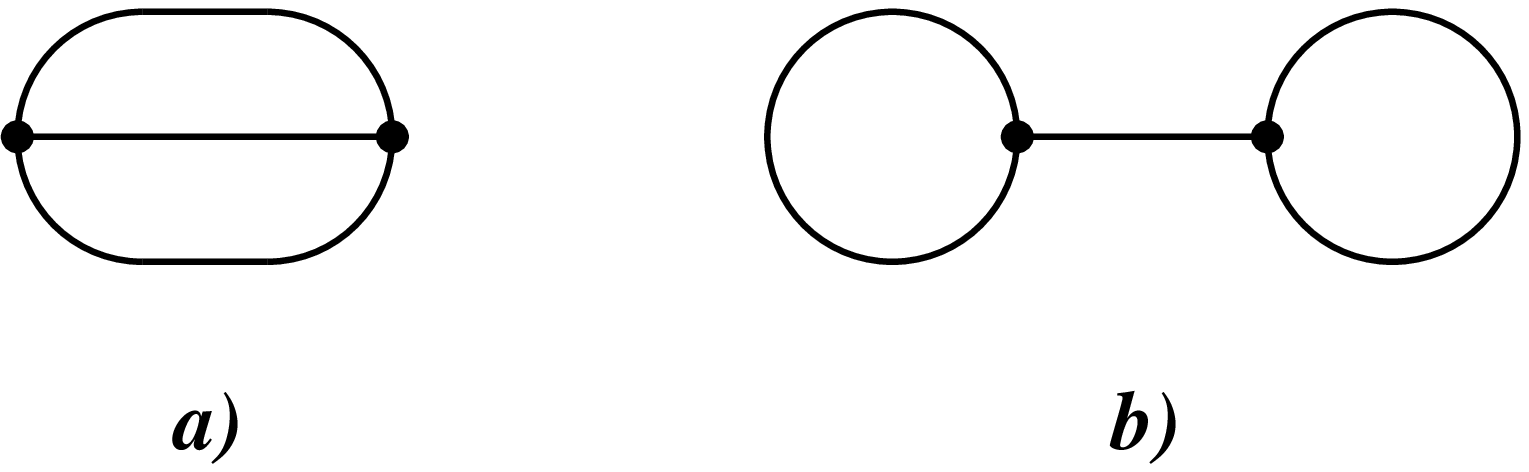,height=3cm,angle=0,trim=0 0 0 0}%
{Two kinds of 2-loop Feynman diagrams that contribute to $S_{2}^{(\rho)}$
in Chern-Simons gauge theory.
\label{twoloopfig} }

In Chern-Simons theory with compact gauge group, one often
develops perturbation theory in the background of a trivial (or reducible)
flat connection $\CA^{(\rho)}$. As a result, the perturbative
coefficients $S_{n}^{(\rho)}$ have a fairly simple structure;
they factorize into a product of topological invariants of $M$ --- the finite
type (Vassiliev) invariants and varations thereof --- and group theory factors \cite{labastida-1999}.
In particular, they are rational numbers.

The situation is very different in Chern-Simons theory
with complex gauge group; the $S_{n}^{(\rho)}$ are completely new,
unexplored invariants of $M$. From the physics point of view,
one novelty of the complex case is that it involves perturbation theory
in the background of a genuinely non-abelian flat connection,
a problem that has not been properly addressed in the existing literature.
One of our goals below will be to make a few modest steps in this direction.
As we shall see, in general, the information about the gauge group $G_{\C}$
and the 3-manifold $M$ is mixed within $S_n^{(\rho)} (M)$ in a non-trivial way.
In particular, in this case the $S_n^{(\rho)} (M)$'s are {\it not} finite type invariants
and are typically not valued in $\mathbb{Q}$.

If we take $\rho$ to be the trivial representation, denoted as $\rho=0$
(and corresponding to a trivial flat connection $\CA^{(0)} = 0$),
the perturbative invariants $S_n^{(0)} (M)$ are actually identical
to those in Chern-Simons theory with compact gauge group $G$.
Then $Z^{(0)} (M;\hbar)$ can be obtained from
the perturbative partition function of Chern-Simons theory
with gauge group $G$ simply by allowing $\hbar$ to be
complex,\footnote{In a theory with compact gauge group,
one frequently uses the ``level'' $k$ as the coupling constant.
It is related to $\hbar$ via $\hbar = i \pi / (k + h^{\vee})$,
where $h^{\vee}$ is the dual Coxeter number of $G$, {\it e.g.} $h^{\vee} = N$ for $G = SU(N)$.}
\be
Z_{G_{\C}}^{(0)} (M;\hbar) \= Z_G^{(0)} (M;\hbar)\,.
\ee
Both sides of this equation have the form of the asymptotic series \eqref{zpert}
with $S_0^{(0)} = 0$.

This example, however, is very special, as one can see {\it e.g.}
by considering hyperbolic 3-manifolds, which, in a sense,
constitute the richest and the most interesting class of 3-manifolds.
A hyperbolic structure on a 3-manifold $M$ corresponds to a discrete
faithful representation of the fundamental group $\pi_1 (M)$ into
${\rm Isom}^+ ( \IH^3 ) \cong PSL (2,\C)$, the group of orientation-preserving
isometries of 3-dimensional hyperbolic space $\IH^3$.
One can compose this representation with a morphism $\phi$ from $PSL (2)$
to a larger algebraic group to obtain a group homomorphism
\be
\rho : \pi_1 (M) \hookrightarrow  PSL(2,\C) \overset\phi\longrightarrow G_{\C}\,.
\label{rhogeom}
\ee
The flat connection $\CA^{(\rho)}$ associated with this
homomorphism is genuinely non-abelian and the corresponding
perturbative invariants $S_{n}^{(\rho)}$ are interesting new
invariants of the hyperbolic 3-manifold $M$.
See Table \ref{tab:figeightgeom} on page \pageref{tab:figeightgeom}
for the simplest example of this type.

Standard textbooks on quantum field theory tell us how
to systematically compute the perturbative invariants $S_{n}^{(\rho)}$
for any $n$, starting with the classical action \eqref{csaction}
and a classical solution \eqref{aflat}, and using the Feynman diagrams mentioned above.
This computation will be reviewed in section 2 for the problem at hand.
However, direct computation of Feynman diagrams becomes prohibitively complicated
when the number of loops $n$ becomes large.
Therefore, while Feynman diagrams provide the first-principles approach
to calculating $Z^{(\rho)} (M;\hbar)$, it is useful to look for alternative
ways of computing and defining the perturbative $G_{\C}$ invariants of $M$.

Such alternative methods often arise from equivalent physical descriptions of a given system.
In some cases, interesting equivalent descriptions are obtained merely via different gauge choices.
A relevant example is the much-studied Chern-Simons gauge theory
with compact gauge group $G$.
Perturbative expansion of this theory around the trivial flat connection $\CA^{(0)} = 0$
in the covariant Landau gauge leads to the configuration space integrals
for the Vassiliev invariants, while in the non-covariant light-cone gauge
it leads to the Kontsevich integral \cite{labastida-1999}.
This example illustrates how a simple equivalence of two physical descriptions
(in this case, based on different gauge choice) leads to an interesting mathematical statement.
Much deeper examples involve non-trivial duality symmetries in quantum field theories.
For example,
duality was the key element in the famous work of Seiberg and Witten \cite{sw-1994}
that led to an exact description of the moduli space of vacua in $\CN=2$ gauge theory
and to a new way of computing Donaldson invariants of 4-manifolds.

Similarly, in the present case of Chern-Simons gauge theory with complex gauge group,
there exist several ways to quantize the theory. When combined with topological
invariance, they lead to multiple methods for computing the perturbative invariant $Z^{(\rho)} (M;\hbar)$.
Mathematically, each of these methods can be taken as a definition of $Z^{(\rho)} (M;\hbar)$;
however, it is important to realize that they all have a common origin and
are completely equivalent.
In the mathematical literature, this is usually stated in the form of a conjecture.
To wit, the following are equivalent:

\begin{itemize}

\item {\bf Feynman diagrams:}
This is the most traditional approach, which was already mentioned earlier
and which will be discussed in more detail in Section~\ref{sec:loops}.
It gives $Z^{(\rho)} (M;\hbar)$ as an asymptotic power series of the form \eqref{zpert}
where every coefficient $S_{n}^{(\rho)}$ can be systematically computed by evaluating Feynman diagrams.

\item {\bf Quantization of $\M (G_{\C}, \Sigma)$:}
According to the general axioms of topological quantum field theory \cite{atiyah-1990},
Chern-Simons gauge theory associates a Hilbert space $\CH_{\Sigma}$
to a closed surface $\Sigma$ (not necessarily connected),
and, similarly, it associates a vector in $\CH_{\Sigma}$
to every closed 3-manifold $M$ with boundary $\Sigma$.
In Chern-Simons theory with complex gauge group $G_{\C}$,
the space $\CH_{\Sigma}$ is obtained by quantizing the classical phase space
$\M (G_{\C}, \Sigma)$, which is the moduli space of flat $G_{\C}$ connections on $\Sigma$.
Therefore, in this approach $Z^{(\rho)} (M;\hbar)$
is obtained as the wave function of a state associated with $M$.
(If $M$ is a closed 3-manifold without boundary,
one can consider {\it e.g.} a Heegaard decomposition $M = M_+ \cup_{\Sigma} M_-$ along $\Sigma$
and apply this approach to each handlebody, $M_+$ and $M_-$.)
In particular, as we explain in Section~\ref{sec:quantization},
$Z^{(\rho)} (M;\hbar)$ obeys a system of Schr\"odinger-like equations
\be
\widehat{A}_i ~Z^{(\rho)} (M;\hbar) \= 0\,,
\label{aionz}
\ee
which, together with appropriate boundary conditions, uniquely determine $Z^{(\rho)} (M;\hbar)$.
One advantage of Chern-Simons theory with complex gauge group
is that the classical phase space $\M (G_{\C}, \Sigma)$
is a hyper-K\"ahler manifold, a fact that considerably simplifies
the quantization problem in any of the existing frameworks
(such as geometric quantization~\cite{woodhouse-1992},
deformation quantization~\cite{bayen-1978,kontsevich-2006},
or ``brane quantization''~\cite{gukov-2008}).

\item {\bf ``Analytic continuation'':}
Since $G_{\C}$ is a complexification of $G$, one might expect that
Chern-Simons theories with gauge groups $G$ and $G_{\C}$ are closely related.
This is indeed the case \cite{gukov-2003}.
For example, if $M = {\bf S}^3 \sm K$ is the complement of a knot (or a link)
in the 3-sphere, one can compare the perturbative $G_{\C}$ invariant $Z^{(\rho)} (M;\hbar)$
with the expectation values of a Wilson loop supported on $K \subset {\bf S}^3$
in Chern-Simons theory with compact gauge group $G$.
The definition of the Wilson loop operator, $W_R (K) := \Tr_{R} ~P \exp \oint_{K} A$,
involves a choice of a representation $R$ of $G$,
and the expectation value
  \be \label{wilsonp} P_{G,R} (K;q) \;:= \;\langle W_R (K) \rangle  \ee
is a certain polynomial invariant of $K$.
Letting $R = R_{\lambda}$ be the irreducible representation of $G$
with highest weight $\lambda$, the invariant quadratic form $- \Tr$
on the Lie algebra $\frak g = {\rm Lie} (G)$ identifies $\lambda$
with an element $\lambda^*$ of the Cartan subalgebra $\frak t \subset \frak g$.
Then, one considers a double-scaling limit
such that
\be
\hbar \to 0\,, \qquad
\lambda^* \to \infty\,, \qquad
u := \hbar \lambda^* = {\rm fixed}\,.
\label{dslimit}
\ee
Taking this limit and analytically continuing to complex values of $u$,
one has \cite{gukov-2003}
\be
\frac{Z^{(\rho)}(M;\hbar)}{Z({\bf S}^3;\hbar)}
\= {\rm ~asymptotic~expansion~of~} P_{G,R_{\lambda}} (K;q)\,,
\label{zviap}
\ee
where $q = e^{2 \hbar}$ and $\rho$ is determined by $u$.
Further details will be explained in Section~\ref{sec:analytic}.

\item {\bf State sum model:}
Finally, using the topological nature of the theory, one can construct
$Z^{(\rho)} (M;\hbar)$ by decomposing $M$ into more elementary pieces.
In particular, one can consider a decomposition of $M$ into tetrahedra,
{\it i.e.} a triangulation,
\be M \= \bigcup_{i=1}^N \Delta_i\,.
\label{mtriang}
\ee
Since each tetrahedron $\Delta_i$ is a 3-manifold with boundary,
it defines a state (a vector) in the Hilbert space associated
to the boundary of $\Delta_i$.
Then, gluing the tetrahedra together as in \eqref{mtriang}
leads to an expression for $Z(M)$ as a sum over different states.
Of course, the result cannot depend on the choice of the triangulation.
A well-known example of such a state sum model
is the Turaev-Viro invariant \cite{turaev-1992}.

In general, states associated to the tetrahedra $\Delta_i$
are labeled by representations of the gauge group, which means
that the set of labels is discrete in a theory with compact gauge
group, but is continuous in a theory with non-compact gauge group.
Therefore, in Chern-Simons theory with complex gauge group $G_{\C}$
one might hope to define the perturbative $G_{\C}$ invariant $Z^{(\rho)} (M;\hbar)$
as an {\it integral} over states, starting from the initial data of
a triangulation \eqref{mtriang} and a homomorphism $\rho$.
We will therefore use the word ``state integral'' instead of
using the more familiar term ``state sum'' for invariants defined in this way.
Constructing such an invariant for the simplest theory with $G_{\C} = SL(2,\C)$
is one of the main goals of the present paper, and will be considered
in detail in section~\ref{sec:statesum}.

\end{itemize}
Of the different approaches just outlined, the first three have been used previously
to tackle Chern-Simons theory with complex gauge group, while the fourth is new.
In principle one can compute $Z^{(\rho)} (M;\hbar)$ following any
of these four approaches individually.
However, as we show in Section~\ref{sec:pert},
combining them together dramatically improves computational power.
For example, when $M$ is a knot complement, one can combine approaches 2 and 3
to learn that the operators $\widehat{A}_i$ in \eqref{aionz} should also
annihilate the polynomial invariants $P_{G,R} (K;q)$ computed by
Chern-Simons gauge theory with compact gauge group $G$,
\be
\widehat{A}_i ~P_{G,R} (K;q) \= 0\,.
\label{aionp}
\ee
(The normalization by $Z({\bf S}^3)$ is not important here.)
This provides a very simple method for finding the $\widehat{A}_i$'s
--- see sections \ref{sec:quantization} and \ref{sec:figeightquant}
for further details and a concrete example.

In turn, combining approaches 2 and 3 with the first approach implies
that, in the limit \eqref{dslimit}, the asymptotic expansion
of the polynomial invariant $P_{G,R} (K;q)$ has the form \eqref{zpert}.
For $G = SU(2)$ (whose complexification is $G_{\C} = SL(2,\C)$)
this statement is already known as a generalization
of the Volume Conjecture, first proposed in \cite{gukov-2003}
and further refined in \cite{murakami-2006,gukov-2006}.
In particular, if $G=SU(2)$ and $R$ is its $n$-dimensional representation,
the polynomial $P_{G,R} (K;q)$ is called the ``$n$-colored Jones polynomial''
of $K$ and is denoted $J_n (K;q)$.
The proposed relation \eqref{zviap} then says that
the asymptotic expansion of the colored Jones polynomial
reproduces the perturbative $SL(2,\C)$ invariant $Z^{(\rho)} (M;\hbar)$;
in particular,
the expansion has the structure \eqref{zpert} with a very interesting
dependence on both $\rho$ and $\hbar$.

In Section~\ref{sec:statesum}, we describe the fourth approach.
Building on the work of Hikami \cite{hikami-2001-16, hikami-2006},
we construct a state sum model for perturbative $SL(2,\C)$ Chern-Simons theory
on a hyperbolic 3-manifold $M$ (possibly with boundary).
Schematically, the state sum model has the form
  \be Z^{(\rho)} (M;\hbar) \= \int_{C_{\rho}} e^{\frac{1}{2 \hbar} f({\bf p})}\;
   \prod_{j=1}^{N} \Phi_{\hbar} \big( \Delta_j \big)^{\pm1}
   \; \prod_{i=1}^{N-b_0(\Sigma)} \frac{dp_i}{\sqrt{4\pi\hbar}}\,, \ee
where $\Phi_{\hbar} (\Delta_j)$ is the quantum dilogarithm function,
associated to every tetrahedron $\Delta_i$ in a triangulation of $M$,
$b_0 (\Sigma)$ is the number of connected components of $\Sigma$,
and $f({\bf p})$ is a simple quadratic polynomial of the integration variables~$p_i$.

One limitation of the construction of the state sum model, in the form stated here,
is that it applies only to hyperbolic 3-manifolds.
Although we expect that it can be extended to arbitrary 3-manifolds
with boundary, we will not presently attempt to seek such a generalization.
Hyperbolic 3-manifolds do provide the richest and, in a sense, the most
interesting laboratory for studying $SL(2,\C)$ Chern-Simons theory.
There are various other questions that we do not address here.
One obvious example is to define a state sum model
for perturbative Chern-Simons theory with complex gauge group $G_{\C}$
not equal to $SL(2,\C)$ (or $PSL(2,\C) = SL(2,\C)/ \{ \pm 1 \}$).
A much deeper issue would be to go beyond perturbation theory,
even in the simplest case of $G_{\C} = SL(2,\C)$.
Indeed, once we have a state sum model that allows the computation
of all perturbative invariants $S^{(\rho)}_n$ to arbitrary order $n$,
the next natural question is whether there exists a state sum
model that allows a non-perturbative computation of $Z(M)$, as a function of $q = e^{2 \hbar}$.
We leave these questions for future work.

By the end of this paper we will have developed several tools that allow us to solve
perturbative Chern-Simons gauge theory with complex gauge group {\it exactly},
in some cases in more than one way.
By an exact solution we mean
a straightforward method that explicitly determines
(at least in principle) all perturbative invariants $S_{n}^{(\rho)}$.
In Section~\ref{sec:example}, we illustrate how these methods work in the case
of the figure-8 knot complement (the simplest hyperbolic knot complement)
and compare the results of different calculations.
As expected, we find perfect agreement.

Even in this simplest example,
the perturbative invariants $S_{n}^{(\rho)}$ turn out to be highly non-trivial
(see {\it e.g.} Table~\ref{tab:figeightgeom}).
As we illustrate in this example, the new perturbative invariants $S_{n}^{(\rho)}$
computed by Chern-Simons gauge theory with complex gauge group $G_{\C}$
in the background of a non-abelian irreducible flat connection $\CA^{(\rho)}$
contain much interesting information about the geometry of $M$.
We believe that studying these invariants further, both physically
and mathematically, can lead to exciting developments in both subjects.


\section{Perturbation theory around a non-trivial flat connection}
\label{sec:pert}

In this section we pursue traditional approaches to Chern-Simons gauge theory
based on the evaluation of Feynman diagrams and the quantization of moduli spaces
of flat connections.
In particular, we explain how one can systematically compute the perturbative
$G_{\C}$ invariant $Z^{(\rho)} (M;\hbar)$ to arbitrary order in the $\hbar$-expansion
for non-trivial flat connections.
Through the analysis of Feynman diagrams, we will find that
the coefficients $S_n^{(\rho)}$ have a very special structure,
motivating us to define the notion of ``arithmetic QFT.''


\subsection{Feynman diagrams and arithmetic TQFT}
\label{sec:loops}

Let us examine more carefully each term in the perturbative expansion \eqref{zpert}:
\be
Z^{(\rho)} (M;\hbar)
\= \exp\left( \frac{1}{\hbar} S_0^{(\rho)} - \frac{1}{2}\delta^{(\rho)} \log \hbar
+ \sum_{n=0}^\infty S_{n+1}^{(\rho)} \hbar^n \right).
\label{zzpert}
\ee
As already mentioned in the introduction, the leading term $S_0^{(\rho)}$
is the value of the classical Chern-Simons functional \eqref{szero} evaluated
on a flat gauge connection $\CA^{(\rho)}$ that corresponds to a homomorphism
$\rho : \pi_1 (M) \to G_{\C}$.
It is also easy to understand the integer coefficient $\delta^{(\rho)}$ in \eqref{zzpert}.
The homomorphism $\rho$ defines a flat $G_{\C}$ bundle over $M$,
which we denote as $E_{\rho}$.
Letting $H^i (M;E_{\rho})$ be the $i$-th cohomology group of $M$
with coefficients in the flat bundle $E_{\rho}$, the coefficient $\delta^{(\rho)}$ is given by
\be
\delta^{(\rho)} \= h^1 - h^0\,,
\label{dhh}
\ee
where $h^i := \dim H^i (M;E_{\rho})$.
Both this term and $S_1^{(\rho)}$ come from the ``one-loop'' contribution
to the path integral \eqref{pathint}; $S_1^{(\rho)}$ can be expressed in terms
of the Ray-Singer torsion of $M$ with respect to the flat
bundle $E_{\rho}$ ({\it cf.} \cite{witten-1989,barnatan-1991w,gukov-2006}),
\be
S_1^{(\rho)} \= \frac{1}{2} \log \left( {T (M;E_{\rho}) \over 2} \right).
\label{sonetorsion}
\ee

The geometric interpretation of the higher-order terms $S_n^{(\rho)}$
with $n>1$ is more interesting, yet less obvious.
To understand it better, we note that the saddle point approximation
to the path integral \eqref{pathint} gives an expression for $S_n^{(\rho)}$
as a sum of Feynman diagrams with $n$ loops.
Since the Chern-Simons action \eqref{csaction} is cubic, all vacuum diagrams
(that is, Feynman diagrams with no external lines) are closed trivalent graphs;
lines in such graphs have no open end-points.
Therefore, a Feynman diagram with $n$ loops ($n>1$) has
$3(n-1)$ line segments with end-points meeting at $2(n-1)$ trivalent vertices.
Each such diagram contributes an integral of the form ({\it cf.} \cite{as-1992})
\be
\int_{M^{2n-2}} L^{3n-3}\,,
\label{snint}
\ee
where $M^{2n-2}$ denotes a product of $2n-2$ copies of $M$
and $L^{3n-3}$ denotes a wedge product of
a 2-form $L (x,y) \in \Omega^2 (M_x \times M_y; \frak g_{\C} \otimes \frak g_{\C})$.
The 2-form $L(x,y)$, called the ``propagator,''
is a solution to the first-order PDE,
\be
d_{\CA^{(\rho)}} L (x,y) \= \delta^3 (x,y)\,,
\label{propag}
\ee
where $\delta^3 (x,y)$ is a $\delta$-function 3-form supported
on the diagonal in $M_x \times M_y$ and $d_{\CA^{(\rho)}}$
is the exterior derivative twisted by a flat connection $\CA^{(\rho)}$
on the $G_{\C}$ bundle $E_{\rho}$.

For example, suppose that $M$ is a geodesically complete hyperbolic 3-manifold
of finite volume. As we pointed out in the introduction,
such 3-manifolds provide some of the most interesting examples for Chern-Simons
theory with complex gauge group.
Every such $M$ can be represented as a quotient
\be
M \= \IH^3 / \Gamma
\label{mquotient}
\ee
of the hyperbolic space $\IH^3$
by a discrete, torsion-free subgroup $\Gamma \subset PSL(2,\C)$,
which is a holonomy representation of the fundamental group $\pi_1 (M)$
into ${\rm Isom}^+ ( \IH^3 ) = PSL (2,\C)$.
In what follows, we call this representation ``geometric''
and denote the corresponding flat $PSL(2,\C)$ connection as $\CA^{({\rm geom})}$.

In order to explicitly describe the flat connection $\CA^{({\rm geom})}$,
recall that $\IH^3$ can be defined as the upper half-space with the standard
hyperbolic metric
\be
ds^2 \= \frac{1}{x_3^2} (dx_1^2 + dx_2^2 + dx_3^2)\,, \qquad x_3>0\,.
\label{hthreemetric}
\ee
The components of the vielbein and spin connection
corresponding to this metric can be written as
\begin{alignat*}{2}
e^1 & = \frac{d x_1}{x_3} \quad \quad \quad \quad w^1 & = & \frac{d x_2}{x_3} \\
e^2 & = \frac{d x_2}{x_3} \quad \quad \quad \quad w^2 & = & -\frac{d x_1}{x_3} \\
e^3 & = \frac{d x_3}{x_3} \quad \quad \quad \quad w^3 & = & 0\,.
\end{alignat*}
These satisfy $de^a + \epsilon^{abc} w_b \wedge e_c = 0$.
It is easy to check that the corresponding $PSL(2,\C)$ connection
\be
\CA^{({\rm geom})} \=  -(w + ie) \= \frac{1}{2x_3}
\begin{pmatrix}
 d x_3 & ~2 dx_1 - 2 id x_2 \\
0 & -d x_3
\end{pmatrix}
\label{ageom}
\ee
is flat, {\it i.e.} obeys \eqref{aflat}.
In Chern-Simons theory with gauge group $G_{\C} = PSL(2,\C)$,
this gives an explicit expression for the flat gauge connection
that corresponds to the hyperbolic structure on $M$.
In a theory with a larger gauge group, one can also define
a geometric connection $\CA^{({\rm geom})}$ by embedding \eqref{ageom}
into a larger matrix with extra rows and columns of zero entries.
We note, however, that $\CA^{({\rm geom})}$ constructed in this way
is not unique and depends on the choice of embedding, {\it cf.} \eqref{rhogeom}.
Nevertheless, we will continue to use the notation $\CA^{({\rm geom})}$
even in the higher-rank case whenever the choice of embedding
introduces no confusion.

The action of $\Gamma$ on $\IH^3$ can be conveniently expressed by
identifying a point $(x_1, x_2, x_3) \in \IH^3$ with a quaternion
$q = x_1 + x_2 i + x_3 j$ and defining
\be
\gamma : q \mapsto (aq + b)/(cq + d)\;, \qquad
\gamma = 
\begin{pmatrix}  ~a~ & ~b~ \\ c & d  \end{pmatrix}
\in PSL (2,\C)\,.
\ee
Explicitly, setting $z = x_1 + i x_2$, we find
\be
\gamma (z + x_3 j) \= z' + x_3' j~,
\ee
where
\be
z' \= \frac{(az + b)(\bar{c} \bar{z} + \bar{d})
+ a\bar{c} x_3^2}{\vert cz + d \vert^2 + \vert c \vert^2 x_3^2}
\;, \qquad
x_3' \= \frac{x_3}{\vert cz + d \vert^2 + \vert c \vert^2 x_3^2}~.
\ee

Let $L_0 (x,y)$ be the propagator for $\IH^3$,
{\it i.e.} a solution to equation \eqref{propag} on $\IH^3 \times \IH^3$
with the non-trivial flat connection $\CA^{({\rm geom})}$.
Then, for a hyperbolic quotient space \eqref{mquotient}, the propagator $L(x,y)$
can simply be obtained by summing over images:
\be
L (x,y) \= \sum_{\gamma \in \Gamma} L_0 (x , \gamma y)\,.
\label{lll}
\ee

Now, we would like to consider what kind of values the perturbative invariants
$S_n^{(\rho)}$ can take. For $n \ge 1$, the $S_n^{(\rho)}$'s are given by sums over
Feynman diagrams, each of which contributes an integral of the form \eqref{snint}.
{\it A priori} the value of every such integral can be an arbitrary complex number
(complex because we are studying Chern-Simons theory with complex gauge group)
that depends on the 3-manifold $M$, the gauge group $G_{\C}$, and the classical
solution $\rho$.
However, for a hyperbolic 3-manifold $M = \IH^3 / \Gamma$
and for the flat connection $\CA^{({\rm geom})}$ associated with the hyperbolic structure on $M$,
we find that the $S_n^{({\rm geom})}$'s are significantly more restricted.

Most basically, one might expect that the values of $S_n^{({\rm geom})}$'s are {\it periods} \cite{kontsevich-2001}
\be
S_n^{({\rm geom})} \in \CP.
\ee
Here $\CP$ is the set of all periods, satisfying
\be
\mathbb{Q} \subset \overline{\mathbb{Q}} \subset \CP \subset \C.
\label{qqpc}
\ee
By definition, a period is a complex number whose real and imaginary parts
are (absolutely convergent) integrals of rational functions with rational coefficients,
over domains in $\mathbb{R}^n$ defined by polynomial inequalities with rational
coefficients \cite{kontsevich-2001}.
Examples of periods are powers of $\pi$, special values of $L$-functions,
and logarithmic Mahler measures of polynomials with integer coefficients.
Thus, periods can be transcendental numbers, but they form a countable set, $\CP$.
Moreover, $\CP$ is an algebra; a sum or a product of two periods is also a period.

Although the formulation of the perturbative invariants $S_{n \ge 1}^{({\rm geom})}$
in terms of Feynman diagrams naturally leads to integrals of the form \eqref{snint},
which have the set of periods, $\CP$, as their natural home,
here we make a stronger claim and conjecture that for $n>1$ the $S_n^{({\rm geom})}$'s are algebraic numbers,
{\it i.e.} they take values in $\overline{\mathbb{Q}}$.
As we indicated in \eqref{qqpc}, the field $\overline{\mathbb{Q}}$ is contained in $\CP$,
but leads to a much stronger condition on the arithmetic nature
of the perturbative invariants $S_n^{({\rm geom})}$.
In order to formulate a more precise conjecture, we introduce the following definition:\\

\noindent \textbf{Definition}: A perturbative quantum field theory
is called {\it arithmetic} if, for all $n>1$,
the perturbative coefficients $S_n^{(\rho)}$ take values in
some algebraic number field $\mathbb{K}$,
\be
S_n^{(\rho)} \;\in\; \mathbb{K}\,,
\ee
and
$S_1^{(\rho)} \in \mathbb{Q} \cdot\log\mathbb{K}\,.$

\bigskip

Therefore, to a manifold $M$ and a classical solution $\rho$
an arithmetic topological quantum field theory (arithmetic TQFT for short)
associates an algebraic number field $\mathbb{K}$,
\be
(M,\rho) \leadsto \mathbb{K}\,.
\ee
This is very reminiscent of arithmetic topology,
a program proposed in the sixties by D.~Mumford, B.~Mazur, and Yu.~Manin,
based on striking analogies between number theory and low-dimensional topology.
For instance, in arithmetic topology, 3-manifolds correspond to
algebraic number fields and knots correspond to primes.

Usually, in perturbative quantum field theory the normalization
of the expansion parameter is a matter of choice.
Thus, in the notations of the present paper, a rescaling of the coupling
constant $\hbar \to \lambda \hbar$ by a numerical factor $\lambda$
is equivalent to a redefnition $S_n^{(\rho)} \to \lambda^{1-n} S_n^{(\rho)}$.
While this transformation does not affect the physics of the perturbative
expansion, it certainly is important for the arithmetic aspects discussed here.
In particular, the above definition of arithmetic QFT is preserved
by such a transformation only if $\lambda \in \mathbb{Q}$.
In a theory with no canonical scale of $\hbar$,
it is natural to choose it in such a way that makes the arithmetic nature
of the perturbative coefficients $S_n^{(\rho)}$ as simple as possible.
However, in some cases (which include Chern-Simons gauge theory),
the coupling constant must obey certain quantization conditions,
which, therefore, can lead to a ``preferred'' normalization of $\hbar$
up to irrelevant $\mathbb{Q}$-valued factors.

We emphasize that our definition of arithmetic QFT is perturbative.
In particular, it depends on the choice of the classical solution $\rho$.
In the present context of Chern-Simons gauge theory with complex
gauge group $G_{\C}$, there is a natural choice of $\rho$
when $M$ is a geodesically complete hyperbolic 3-manifold,
namely the geometric representation that corresponds to $\CA^{({\rm geom})}$.
In this case, we conjecture:\\

\noindent \textbf{Conjecture 1 (Arithmeticity conjecture)}: {\it As in \eqref{mquotient}, let $M$ be a geodesically
complete hyperbolic 3-manifold of finite volume, and let $\rho = {\rm geom}$
be the corresponding discrete faithful representation of $\pi_1 (M)$ into $PSL(2,\C)$.
Then the perturbative Chern-Simons theory with complex gauge group $G_{\C} = PSL(2,\C)$
(or its double cover, $SL(2,\C)$)
in the background of a non-trivial flat connection $\CA^{({\rm geom})}$ is arithmetic on $M$.} \\

In fact, we can be a little bit more specific.
In all the examples that we studied, we find that, for $M$ as in \eqref{mquotient}
and for all values of $n>1$, the perturbative invariants $S_n^{({\rm geom})}$
take values in the trace field of $\Gamma$,
\be
S_n^{({\rm geom})} \;\in\; \mathbb{Q} (\tr \Gamma)\,,
\label{snintrace}
\ee
where, by definition, $\mathbb{Q} (\tr \Gamma)$ is
the minimal extension of $\mathbb{Q}$ containing $\tr \gamma$ for all $\gamma \in \Gamma$.
We conjecture that this is the case in general, namely that
the $SL(2,\C)$ Chern-Simons theory on a hyperbolic 3-manifold $M = \IH^3 / \Gamma$
is arithmetic with $\mathbb{K} = \mathbb{Q} (\tr \Gamma)$.
This should be contrasted with the case of a compact gauge group,
where one usually develops perturbation theory in the background
of a trivial flat connection, and the perturbative invariants $S_n^{(\rho)}$
turn out to be rational numbers.

Of course, in this conjecture it is important that the representation $\rho$
is fixed, {\it e.g.} by the hyperbolic geometry of $M$ as in the case at hand.
As we shall see below, in many cases the representation $\rho$ admits continuous
deformations or, put differently, comes in a family.
For geometric representations, this does not contradict the famous rigidity of
hyperbolic structures because the deformations correspond
to incomplete hyperbolic structures on $M$.
In a sense, the second part of this section is devoted to studying
such deformations. As we shall see, the perturbative $G_{\C}$ invariant
$Z^{(\rho)} (M;\hbar)$ is a function of the deformation parameters,
which on the {\it geometric branch}\footnote{{\it i.e.} on the branch containing
the discrete faithful representation.} can be interpreted as shape parameters of
the associated hyperbolic structure.

In general, one might expect the perturbative coefficients $S_{n>1}^{(\rho)}$
to be rational functions of these shape parameters. Note that, if true, this statement
would imply the above conjecture, since at the point corresponding to the complete
hyperbolic structure the shape parameters take values in the trace field $\mathbb{Q} (\tr \Gamma)$.
This indeed appears to be the case, at least for several simple examples of
hyperbolic 3-manifolds that we have studied.

For the arithmeticity conjecture to hold, it is important that $\hbar$
is defined (up to $\mathbb{Q}$-valued factors) as in \eqref{hbdef},
so that the leading term $S_{0}^{(\rho)}$ is a rational multiple of
the classical Chern-Simons functional, {\it cf.} \eqref{szero}.
This normalization is natural for a number of reasons.
For example, it makes the arithmetic nature of the perturbative
coefficients $S_{n}^{(\rho)}$ as clear as possible.
Namely, according to the above arithmeticity conjecture,
in this normalization $S_{1}^{({\rm geom})}$ is a period,
whereas $S_{n>1}^{({\rm geom})}$ take values in $\overline{\mathbb{Q}}$.
However, although we are not going to use it here, we note that
another natural normalization of $\hbar$ could be obtained
by a redefinition $\hbar \to \lambda \hbar$ with $\lambda \in (2 \pi i)^{-1} \cdot \mathbb{Q}$.
As we shall see below, this normalization is especially natural from
the viewpoint of the analytic continuation approach.
In this normalization, the arithmeticity conjecture says that all
$S_{n>1}^{({\rm geom})}$ are expected to be periods.
More specifically, it says that $S_{n}^{(\rho)} \in (2 \pi i)^{n-1} \cdot \overline{\mathbb{Q}}$,
suggesting that the $n$-loop perturbative invariants $S_{n}^{({\rm geom})}$
are periods of (framed) mixed Tate motives $\mathbb{Q} (n-1)$.
In this form, the arithmeticity of perturbative Chern-Simons invariants
discussed here is very similar to the motivic interpretation
of Feynman integrals in \cite{bloch-2006}.

Finally, we note that, for some applications, it may be convenient to normalize
the path integral \eqref{pathint} by dividing the right-hand side by $Z({\bf S}^3;\hbar)$.
(Since $\pi_1 ({\bf S}^3)$ is trivial, we have $Z({\bf S}^3;\hbar) = Z^{(0)} ({\bf S^3};\hbar)$.)
This normalization does not affect the arithmetic nature of the perturbative
coefficients $S_n^{({\rm geom})}$ because, for $M = {\bf S}^3$, all the $S_n^{(0)}$'s
are rational numbers. Specifically,
\be
Z({\bf S}^3;\hbar) \= \left( {\hbar \over i \pi} \right)^{r/2}
\left( {\Vol \Lambda_{{\rm wt}} \over \Vol \Lambda_{{\rm rt}}} \right)^{1/2}
\prod_{\alpha \in \Lambda_{{\rm rt}}^+} 2 \sinh \left( \hbar (\alpha \cdot \varrho) \right),
\label{zsthree}
\ee
where the product is over positive roots $\alpha \in \Lambda_{{\rm rt}}^+$,
$r$ is the rank of the gauge group, and $\varrho$ is half the sum of the positive roots,
familiar from the Weyl character formula.
Therefore, in the above conjecture and in eq. \eqref{snintrace} we can use
the perturbative invariants of $M$ with either normalization.

The arithmeticity conjecture discussed here is a part of a richer
structure: the quantum $G_{\C}$ invariants are only the special case $x=0$
of a collection of functions indexed by rational numbers $x$
which each have asymptotic expansions in $\hbar$ satisfying
the arithmeticity conjecture and which have a certain kind
of modularity behavior under the action of $SL(2,\Z)$  on $\mathbb{Q}$ \cite{zagMod}.
A better understanding of this phenomenon and its interpretation
will appear elsewhere.


\subsection{Quantization of $\M (G_{\C}, \Sigma)$}
\label{sec:quantization}

Now, let us describe another approach to Chern-Simons gauge theory,
based on quantization of moduli space spaces of flat connections.
We focus on a theory with complex gauge group,
which is the main subject of the present paper.\footnote{We will
see, however, that quantization of Chern-Simons theory with complex gauge group
$G_{\C}$ is closely related to quantization for compact group $G$.
Indeed, as we shall describe in Section~\ref{sec:analytic}, this relation
may be used to justify the ``analytic continuation'' approach to computing
perturbative $G_{\C}$ invariants.}

As we already mentioned in the introduction, Chern-Simons gauge theory
(with any gauge group) associates a Hilbert space $\CH_{\Sigma}$
to a closed Riemann surface $\Sigma$ and a vector in $\CH_{\Sigma}$
to every 3-manifold $M$ with boundary $\Sigma$.
We denote this vector as $\vert M \rangle \in \CH_{\Sigma}$.
If there are two such manifolds, $M_+$ and $M_-$,
glued along a common boundary $\Sigma$ (with matching orientation),
then the quantum invariant $Z(M)$ that Chern-Simons theory associates
to the closed 3-manifold $M = M_+ \cup_{\Sigma} M_-$
is given by the inner product of two vectors
$\vert M_+ \rangle$ and $\vert M_- \rangle$ in $\CH_{\Sigma}$
\be
Z(M) \= \langle M_+ \vert M_- \rangle\,.
\label{mmgluing}
\ee
Therefore, in what follows, our goal will be to understand Chern-Simons
gauge theory on manifolds with boundary, from which invariants of closed manifolds
without boundary can be obtained via \eqref{mmgluing}.

Since the Chern-Simons action \eqref{csaction} is first order in derivatives,
the Hilbert space $\CH_{\Sigma}$ is obtained by quantizing the classical phase space,
which is the space of classical solutions on the 3-manifold $\R \times \Sigma$.
According to \eqref{aflat}, such classical solutions are given precisely
by the flat connections on the Riemann surface $\Sigma$.
Therefore, in a theory with complex gauge group $G_{\C}$, the classical phase
space is the moduli space of flat $G_{\C}$ connections on $\Sigma$,
modulo gauge equivalence,
\be
\M (G_{\C}, \Sigma) \= {\rm Hom} (\pi_1 (\Sigma), G_{\C}) / {\rm conj.}
\label{mpspace}
\ee
As a classical phase space, $\M (G_{\C}, \Sigma)$ comes equipped
with a symplectic structure $\omega$, which can also be deduced from
the classical Chern-Simons action \eqref{csaction}.
Since we are interested only in the ``holomorphic'' sector of the theory,
we shall look only at the kinetic term for the field $\CA$ (and not $\bar \CA$);
it leads to a holomorphic symplectic 2-form on $\M (G_{\C}, \Sigma)$:
\be
\omega \= {i \over 4 \hbar} \int_{\Sigma} \Tr~ \delta \CA \wedge \delta \CA\,.
\label{msympl}
\ee
We note that this symplectic structure does not depend on the complex structure
of $\Sigma$, in accord with the topological nature of the theory.
Then, in Chern-Simons theory with complex gauge group $G_{\C}$,
the Hilbert space $\CH_{\Sigma}$ is obtained by quantizing the moduli
space of flat $G_{\C}$ connections on $\Sigma$ with symplectic structure \eqref{msympl}:
\be
{\rm quantization~of~} (\M (G_{\C}, \Sigma),\omega) ~~\leadsto~~ \CH_{\Sigma}\,.
\ee

Now, let us consider a closed 3-manifold $M$ with boundary $\Sigma = \partial M$,
and its associated state $\vert M \rangle \in \CH_{\Sigma}$.
In a (semi-)classical theory, quantum states correspond to Lagrangian submanifolds
of the classical phase space.
Recall that, by definition, a Lagrangian submanifold $\CL$ is a middle-dimensional
submanifold such that the restriction of $\omega$ to $\CL$ vanishes,
\be
\omega \vert_{\CL} \= 0\;.
\ee
For the problem at hand, the phase space is $\M (G_{\C}, \Sigma)$
and the Lagrangian submanifold $\CL$ associated to a 3-manifold $M$
with boundary $\Sigma = \partial M$ consists of the classical solutions on $M$.
Since the space of classical solutions on $M$ is the moduli space of flat
$G_{\C}$ connections on $M$,
\be
\M (G_{\C}, M) = {\rm Hom} (\pi_1 (M), G_{\C}) / {\rm conj.}\,,
\label{mflatm}
\ee
it follows that
\be
\CL \= \iota \big( \M (G_{\C}, M) \big)
\label{lpspace}
\ee
is the image of $\M (G_{\C}, M)$ under the map
\be
\iota : \M (G_{\C}, M) \to \M (G_{\C}, \Sigma)
\ee
induced by the natural inclusion map $\pi_1 (\Sigma) \to \pi_1 (M)$.
One can show that $\CL \subset \M (G_{\C}, \Sigma)$
is indeed Lagrangian with respect to the symplectic structure \eqref{msympl}.

Much of what we described so far is very general and has an obvious analogue
in Chern-Simons theory with arbitrary gauge group.
However, quantization of Chern-Simons theory with complex gauge group has
a number of good properties. In this case
the classical phase space $\M (G_{\C}, \Sigma)$ is an algebraic variety;
it admits a complete hyper-K\"ahler metric~\cite{hitchin-1987},
and the Lagrangian submanifold $\CL$ is a holomorphic subvariety of $\M (G_{\C}, \Sigma)$.
The hyper-K\"ahler structure on $\M (G_{\C}, \Sigma)$ can be obtained by interpreting
it as the moduli space $\CM_H (G,\Sigma)$ of solutions to Hitchin's equations on $\Sigma$.
Note that this requires a choice of complex structure on $\Sigma$,
whereas $\M (G_{\C}, \Sigma) \cong \CM_H (G,\Sigma)$
as a complex symplectic manifold does not.
Existence of a hyper-K\"ahler structure on $\M (G_{\C}, \Sigma)$
considerably simplifies the quantization problem in any of the existing frameworks,
such as geometric quantization~\cite{woodhouse-1992},
deformation quantization~\cite{bayen-1978,kontsevich-2006},
or ``brane quantization''~\cite{gukov-2008}.

The hyper-K\"ahler moduli space $\CM_H (G,\Sigma)$
has three complex structures that we denote as $I$, $J$, and $K = IJ$,
and three corresponding K\"ahler forms, $\omega_I$, $\omega_J$, and $\omega_K$.
In the complex structure usually denoted by $J$,
$\CM_H (G,\Sigma)$ can be identified
with $\M (G_{\C}, \Sigma)$ as a complex symplectic manifold
with the holomorphic symplectic form \eqref{msympl},
\be
\omega \= {1 \over \hbar} (\omega_K + i \omega_I)\,.
\ee
Moreover, in this complex structure, $\CL$ is an algebraic subvariety
of $\M (G_{\C}, \Sigma)$. To be more precise, it is a (finite) union
of algebraic subvarieties, each of which is defined by polynomial equations $A_i = 0$.
In the quantum theory, these equations are replaced by corresponding
operators $\widehat{A}_i$ acting on $\CH_{\Sigma}$
that annihilate the state $\vert M \rangle$.

Now, let us consider in more detail the simple but important case
when $\Sigma$ is of genus 1, that is $\Sigma = T^2$.
In this case, $\pi_1 (\Sigma) \cong \Z \times \Z$ is abelian, and
\be
\M (G_{\C}, T^2) \= ( \TT_{\C} \times \TT_{\C} ) / \CW\;,
\label{mflatttwo}
\ee
where $\TT_{\C}$ is the maximal torus of $G_{\C}$ and $\CW$ is the Weyl group.
We parametrize each copy of $\TT_{\C}$ by complex variables
$l = (l_1, \ldots, l_r)$ and $m = (m_1, \ldots, m_r)$, respectively.
Here, $r$ is the rank of the gauge group $G_{\C}$.
The values of $l$ and $m$ are eigenvalues of the holonomies
of the flat $G_{\C}$ connection over the two basic 1-cycles of $\Sigma = T^2$.
They are defined up to Weyl transformations, which act
diagonally on $\TT_{\C} \times \TT_{\C}$.

The moduli space of flat $G_{\C}$ connections on a 3-manifold $M$ with a single
toral boundary, $\partial M = T^2$, defines a complex Lagrangian submanifold
\be
\CL \subset ( \TT_{\C} \times \TT_{\C} ) / \CW\;.
\ee
(More precisely, this Lagrangian submanifold comprises the top-dimensional
(stable) components of the moduli space of flat $G_{\C}$ connections on $\Sigma$.)
In particular, a generic irreducible component of $\CL$
is defined by $r$ polynomial equations
\be
A_i (l,m) \= 0\,, \qquad i = 1, \ldots, r\,,
\label{aizero}
\ee
which must be invariant under the action of the Weyl group $\CW$
(which simultaneously acts on the eigenvalues $l_1, \ldots, l_r$ and $m_1, \ldots, m_r$).
In the quantum theory, these equations are replaced by the operator equations \eqref{aionz},
\be
\widehat{A}_i (\hat l, \hat m )\;Z (M) \= 0\,, \qquad i = 1, \ldots, r\,.
\label{almonz}
\ee
For $\Sigma = T^2$ the complex symplectic structure \eqref{msympl}
takes a very simple form
\be
\omega \= \frac{i}{\hbar} \sum_{i=1}^r du_i \wedge dv_i\;,
\label{sympluv}
\ee
where we introduce new variables $u$ and $v$
(defined modulo elements of the cocharacter lattice
$\Lambda_{{\rm cochar}} \= {\rm Hom} (U(1),\TT)$),
such that $l = - e^v$ and $m = e^u$.
In the quantum theory, $u$ and $v$ are replaced by operators $\hat u$ and $\hat v$
that obey the canonical commutation relations
\be
[\hat u_i , \hat v_j ] \= - \hbar \delta_{ij}\;.
\label{uvcomm}
\ee
As one usually does in quantum mechanics, we can introduce a complete
set of states $\vert u \rangle$ on which $\hat u$ acts by multiplication,
$\hat u_i \vert u \rangle = u_i \vert u \rangle$.
Similarly, we let $\vert v \rangle$ be a complete basis,
such that $\hat v_i \vert v \rangle = v_i \vert v \rangle$.
Then, we can define the wave function associated to
a 3-manifold $M$ either in the $u$-space or $v$-space representation,
respectively, as $\langle u \vert M \rangle$ or $\langle v \vert M \rangle$.
We will mostly work with the former and let $Z(M;u) := \langle u \vert M \rangle$.

We note that a generic value of $u$ does not uniquely specify
a flat $G_{\C}$ connection on $M$ or, equivalently, a unique point on the representation
variety \eqref{lpspace}.
Indeed, for a generic value of $u$, equations \eqref{aizero}
may have several solutions that we label by a discrete parameter $\alpha$.
Therefore, in the $u$-space representation, flat $G_{\C}$ connections on $M$
(previously labeled by the homomorphism $\rho \in \CL$)
are now labeled by a set of continuous parameters $u = (u_1, \ldots, u_r)$
and a discrete parameter $\alpha$:
\be
\rho \quad \longleftrightarrow \quad (u,\alpha)\;.
\label{rhoviaua}
\ee
The perturbative $G_{\C}$ invariant $Z^{(\rho)} (M;\hbar)$
can then be written in this notation as
$Z^{(\alpha)} (M;\hbar,u) = \langle u, \alpha \vert M \rangle$.
Similarly, the coefficients $S^{(\rho)}_n$ in the $\hbar$-expansion
can be written as $S^{(\alpha)}_n (u)$.

To summarize, in the approach based on quantization of $\M (G_{\C}, \Sigma)$
the calculation of $Z^{(\rho)} (M;\hbar)$ reduces to two main steps:
$i)$ the construction of quantum operators $\widehat{A}_i (\hat l, \hat m )$,
and $ii)$ the solution of Schr\"odinger-like equations \eqref{almonz}.
Below we explain how to implement each of these steps.


\subsubsection{Analytic continuation}
\label{sec:analytic}

For a generic 3-manifold $M$ with boundary $\Sigma = T^2$,
constructing the quantum operators $\widehat{A}_i (\hat l, \hat m )$
may be a difficult task.
However, when $M$ is the complement of a knot $K$ in the 3-sphere,
  \be \label{mkcomplement}  M = {\bf S}^3 \sm K\,,   \ee
there is a simple way to find the $\widehat{A}_i$'s. Indeed, according to \eqref{aionp}
these operators annihilate the polynomial knot invariants $P_{G,R} (K;q)$, which are
defined in terms of Chern-Simons theory with compact gauge group $G$,
\be
\widehat{A}_i (\hat l, \hat m ) ~P_{G,R} (K;q) = 0\,.
\label{almionp}
\ee
The operator $\hat l_i$ acts on the set of polynomial invariants $\{ P_{G,R} (K;q) \}$
by shifting the highest weight $\lambda = (\lambda_1, \ldots, \lambda_r)$
of the representation $R$ by the $i$-th basis elements of the weight lattice $\Lambda_{{\rm wt}}$,
while the operator $\hat m_j$ acts simply as multiplication by $q^{\lambda_j / 2}$.
Let us briefly explain how this comes about.

In general, the moduli space $\M (G_{\C}, \Sigma)$ is a complexification of $\M (G,\Sigma)$.
The latter is the classical phase space in Chern-Simons theory with compact gauge group $G$
and can be obtained from $\M (G_{\C}, \Sigma)$ by requiring all the holonomies to be ``real,''
{\it i.e.} in $G$.
Similarly, restricting to real holonomies in the definition of $\CL$
produces a Lagrangian submanifold in $\M (G,\Sigma)$ that corresponds to
a quantum state $\vert M \rangle$ in Chern-Simons theory with compact gauge group $G$.
In the present example of knot complements,
restricting to such ``real'' holonomies means replacing $\TT_{\C}$ by $\TT$ in \eqref{mflatttwo}
and taking purely imaginary values of $u_i$ and $v_i$ in equations \eqref{aizero}.
Apart from this,
the quantization problem is essentially the same for gauge groups $G$ and $G_{\C}$.
In particular, the symplectic structure \eqref{sympluv} has the same form
(with imaginary $u_i$ and $v_i$ in the theory with gauge group $G$)
and the quantum operators $\widehat{A}_i (\hat l, \hat m )$
annihilate both $P_{G,R} (K;q)$ and $Z^{(\alpha)} (M;\hbar,u)$
computed, respectively, by Chern-Simons theories with gauge groups $G$ and $G_{\C}$.

In order to understand the precise relation between the parameters in these theories,
let us consider a Wilson loop operator, $W_R (K)$,
supported on $K \subset {\bf S}^3$ in Chern-Simons theory with compact gauge group $G$.
It is labeled by a representation $R = R_{\lambda}$ of the gauge group $G$,
which we assume to be an irreducible representation with highest weight $\lambda \in \frak t^{\vee}$.
As we already mentioned earlier ({\it cf.} eq. \eqref{wilsonp}), the path integral in
Chern-Simons theory on ${\bf S}^3$ with a Wilson loop operator $W_R (K)$
computes the polynomial knot invariant $P_{G,R} (K;q)$, with $q = e^{2 \hbar}$.
Using \eqref{mmgluing}, we can represent this path integral as
$\langle R \vert M \rangle$,
where $\vert R \rangle$ is the result of the path integral on a solid torus containing
a Wilson loop $W_R (K)$, and $\vert M \rangle$ is the path integral on its complement, $M$.

In the semi-classical limit, the state $\vert R \rangle$ corresponds to
a Lagrangian submanifold of $\M (G,T^2) = (\TT \times \TT) / \CW$
defined by the fixed value of the holonomy $m = e^u$
on a small loop around the knot, where $u$ is given by \eqref{dslimit}.
The relation between $m=e^u$, which is an element of the maximal torus $\TT$ of $G$,
and the representation $R_{\lambda}$ is given by the invariant quadratic form $- \Tr$
(restricted to $\frak t$). Specifically,
\be
m \= \exp (\hbar \lambda^*) \in \TT \,,
\ee
where $\lambda^*$ is the unique element of $\frak t$ such that
$\lambda^* (x) = - \Tr \lambda x$ for all $x \in \frak t$.
In \eqref{zviap}, we analytically continue this relation to $m \in \TT_{\C}$.

For a given value of $m=e^u$, equations \eqref{aizero}
have a finite set of solutions $l_{\alpha}$, labeled by $\alpha$.
Only for one particular value of $\alpha$ is
the perturbative $G_{\C}$ invariant $Z^{(\alpha)} (M;\hbar,u)$
related to the asymptotic behavior of $P_{G,R} (K;q)$.
This is the value of $\alpha$ which maximizes $\Re \big( S_0^{(\alpha)} (M; u) \big)$.
For this $\alpha$, we have \eqref{zviap}:
\be
\frac{Z^{(\alpha)} (M;\hbar,u)}{Z({\bf S}^3;\hbar)}
\= {\rm ~asymptotic~expansion~of~} P_{G,R} (K;q)\,.
\label{zuviap}
\ee
For hyperbolic knots and $u$ sufficiently close to 0,
this ``maximal'' value of $\alpha$ is always $\alpha = {\rm geom}$.

It should nevertheless be noted that the analytic continuation described here is not as ``analytic'' as it sounds. In particular, the limit \eqref{dslimit} is very subtle and requires much care.
As explained in \cite{gukov-2003}, in taking this limit it is important that values of $q=e^{2\hbar}$ avoid roots of unity. If one takes the limit with $\hbar^{-1}\in\frac{1}{i\pi}\Z$, which corresponds to the allowed values of the coupling constant in Chern-Simons theory with compact gauge group $G$, then one can never see the exponential asymptotics \eqref{zpert} with $\Im (S_0^{(\rho)} ) >0$. The exponential growth characteristic to Chern-Simons theory with complex gauge group emerges only in the limit with $\hbar = u/\lambda^*$ and $u$ generic.


\subsubsection{A hierarchy of differential equations}
\label{sec:hierarchy}

The system of Schr\"odinger-like equations \eqref{almonz} determines
the perturbative $G_{\C}$ invariant $Z^{(\alpha)} (M;\hbar,u)$
up to multiplication by an overall function of $\hbar$,
which can be fixed by suitable boundary conditions.

In order to see in detail how the perturbative coefficients $S_n^{(\alpha)}(u)$
may be calculated and to avoid cluttering, let us assume that $r=1$.
(A generalization to arbitrary values of $r$ is straightforward.)
In this case, $A(l,m)$ is the so-called A-polynomial of $M$,
originally introduced in \cite{cooper-1994},
and the system \eqref{almonz} consists of a single equation
\be
\widehat{A} (\hat l, \hat m)\,Z^{(\alpha)} (M;\hbar,u) \= 0\,.
\label{ahatcondition}
\ee
In the $u$-space representation the operator $\hat m = \exp (\hat u)$
acts on functions of $u$ simply via multiplication by $e^u$,
whereas $\hat l = \exp (\hat v +i\pi)= \exp (\hbar\;{d \over du})$
acts as a ``shift operator'':
\be
\hat m f(u) \= e^u f(u) \,,\qquad
\hat l f(u) \= f(u + \hbar )\,.
\label{lmactonfofu}
\ee
In particular, the operators $\hat l$ and $\hat m$ obey the relation
\be
\hat l \hat m  \= q^{{1 \over 2}} \hat m \hat l\,,
\ee
which follows directly from the commutation relation \eqref{uvcomm}
for $\hat u$ and $\hat v$, with
\be
q \= e^{2 \hbar}\;.
\ee

We would like to recast eq. \eqref{ahatcondition} as an infinite hierarchy of differential
equations that can be solved recursively for the perturbative coefficients $S_n^{(\alpha)}(u)$.
Just like its classical limit $A(l,m)$, the operator $\widehat{A} (\hat l, \hat m )$
is a polynomial in $\hat l$.
Therefore, pushing all operators $\hat{l}$ to the right, we can write it as
\be
\widehat{A} (\hat{l},\hat{m}) \= \sum_{j=0}^d a_j (\hat{m},\hbar)\,\hat{l}^j
\label{hatAcoeffs}
\ee
for some functions $a_j (m,\hbar)$ and some integer $d$.
Using \eqref{lmactonfofu}, we can write eq. \eqref{ahatcondition} as
\be
\sum_{j=0}^d a_j(m,\hbar) Z^{(\alpha)} (M;\hbar,u + j \hbar ) \= 0\,.
\label{reccompJN}
\ee
Then, substituting the general form \eqref{zzpert} of $Z^{(\alpha)} (M;\hbar,u)$,
we obtain the equation
\be
\sum_{j=0}^d a_j(m,\hbar)
\exp \left[\frac{1}{\hbar} S_0^{(\alpha)} (u + j\hbar)
-\frac{1}{2} \delta^{(\alpha)} \log \hbar
+ \sum_{n=0}^{\infty} \hbar^n S_{n+1}^{(\alpha)} (u + j\hbar) \right] \= 0\;.
\label{ASintermed}
\ee
Since $\delta^{(\alpha)}$ is independent of $u$, we can just factor out
the $-\frac{1}{2}\delta^{(\alpha)} \log \hbar$ term and remove it from
the exponent.
Now we expand everything in $\hbar$. Let
\be
a_j (m,\hbar) \= \sum_{p=0}^{\infty} a_{j,p} (m) \hbar^p \label{aj-hbar}
\ee
and
\be
\sum_{n=-1}^{\infty} \hbar^n S_{n+1} ( u + j \hbar)
\= \sum_{r=-1}^{\infty} \sum_{m=-1}^r \frac{j^{r-m}}{(r-m)!} \hbar^r S_{m+1}^{(r-m)} (u)\,,
\label{Sn-hbar}
\ee
suppressing the index $\alpha$ to simplify notation.
We can substitute \eqref{aj-hbar} and \eqref{Sn-hbar}
into \eqref{ASintermed} and divide the entire expression by
$\exp\hspace{-.05cm}\left(\,\sum_n \hbar^n S_{n+1}(u)\,\right)$.
The hierarchy of equations then follows by expanding the exponential
in the resulting expression as a series in $\hbar$ and requiring that
the coefficient of every term in this series vanishes. The first four equations 
are shown in Table \ref{tab:Seqns}.

\begin{table}[htb]
\small
\begin{align*}
\sum_{j=0}^d& e^{j S_0'} a_{j,0} = 0
        \\
\sum_{j=0}^d& e^{j S_0'}\biggl(a_{j,1} + a_{j,0}\bigl(\frac{1}{2}j^2S_0'' + j S_1'\bigr)\biggr) = 0
        \\
\sum_{j=0}^d& e^{j S_0'}\biggl(a_{j,2} + a_{j,1}\bigl(\frac{1}{2}j^2S_0'' + j S_1'\bigr) + a_{j,0}\biggl(\frac{1}{2}\bigl(\frac{1}{2}j^2S_0'' + j S_1'\bigr)^2
+ \frac{j^3}{6}S_0''' + \frac{j^2}{2} S_1'' + jS_2'\biggr)\biggr) = 0
        \\
\sum_{j=0}^d&  e^{j S_0'}\biggl(a_{j,3} + a_{j,2}\bigl(\frac{1}{2}j^2S_0'' + j S_1'\bigr) + a_{j,1}\biggl(\frac{1}{2}\bigl(\frac{1}{2}j^2S_0'' + j S_1'\bigr)^2
+ \frac{j^3}{6}S_0''' + \frac{j^2}{2} S_1'' + jS_2'\biggr)
        \\
&+ a_{j,0}\biggl(\frac{1}{6}\bigl(\frac{1}{2}j^2S_0'' + j S_1'\bigr)^3 + \bigl(\frac{1}{2}j^2S_0'' + j S_1'\bigr)\bigl(\frac{j^3}{6}S_0''' + \frac{j^2}{2} S_1'' + jS_2'\bigr)
        \\
&+ \frac{j^4}{24} S_0^{(4)} + \frac{j^3}{6}S_1''' + \frac{j^2}{2}S_2'' + jS_3'\biggr)\biggr) = 0 \\ \ldots
\end{align*}
\caption{Hierarchy of differential equations derived from $\widehat{A} (\hat l, \hat m ) ~Z^{(\alpha)} (M;\hbar,u) = 0$.}
\label{tab:Seqns}
\end{table}

The equations in Table \ref{tab:Seqns} can be solved recursively for the $S_n(u)$'s,
since each $S_n$ first appears in the $(n+1)^{{\rm st}}$ equation, differentiated only once.
Indeed, after $S_0$ is obtained, the remaining equations feature the $S_{n\geq1}$
linearly the first time they occur, and so determine these coefficients uniquely up to
an additive constant of integration.

The first equation, however, is somewhat special. Since $a_{j,0}(m)$ is precisely the coefficient
of $l^j$ in the classical A-polynomial $A(l,m)$, we can rewrite this equation as
\be
A(e^{S_0'(u)},e^u)=0\,.
\ee
This is exactly the classical constraint $A (e^{v + i\pi} , e^u) = 0$ that defines
the complex Lagrangian submanifold $\CL$, with $S_0'(u) = v + i\pi$.
Therefore, we can integrate
along a branch $(l_\alpha=e^{v_\alpha+i\pi},m=e^u)$ of $\CL$ to get the value of
the classical Chern-Simons action \eqref{szero},
\be
S_0^{(\alpha)}(u) = {\rm const} + \int^u \theta \vert_{\CL}\,,
\label{szerowkb}
\ee
where $\theta \vert_{\CL}$ denotes a restriction to $\CL$
of the Liouville 1-form on $\M (G_{\C}, \Sigma)$,
\be
\theta = v du + i \pi du\,.
\label{liouvilleform}
\ee
The expression \eqref{szerowkb} is precisely the semi-classical approximation to
the wave function $Z^{(\alpha)} (M;\hbar,u)$ supported on the Lagrangian submanifold $\CL$,
obtained in the WKB quantization of the classical phase space $\M (G_{\C}, \Sigma)$.
By definition, the Liouville form $\theta$ (associated with a symplectic structure $\omega$)
obeys $d \theta = i \hbar \omega$, and it is easy to check that this is indeed the case for
the forms $\omega$ and $\theta$ on $\M (G_{\C}, \Sigma)$
given by eqs. \eqref{sympluv} and \eqref{liouvilleform}, respectively.

The semi-classical expression \eqref{szerowkb} gives the value of the classical
Chern-Simons functional \eqref{szero} evaluated on a flat gauge connection $\CA^{(\rho)}$,
labeled by a homomorphism $\rho$.
As we explained in \eqref{rhoviaua}, the dependence on $\rho$ is encoded in
the dependence on a continuous holonomy parameter $u$, as well as a discrete parameter $\alpha$
that labels different solutions $v_{\alpha} (u)$ to $A (e^{v + i\pi} , e^u) = 0$,
at a fixed value of $u$.
In other words, $\alpha$ labels different branches of the Riemann surface $A(l,m) = 0$,
regarded as a cover of the complex plane $\C$ parametrized by $m = e^u$,
\be
A (l_{\alpha} , m) = 0\,.
\label{abranches}
\ee
Since $A(l,m)$ is a polynomial in both $l$ and $m$, the set of values of $\alpha$
is finite (in fact, its cardinality is equal to the degree of $A(l,m)$ in $l$).
Note, however, that for a given choice of $\alpha$ there are infinitely many ways to
lift a solution $l_{\alpha} (u)$ to $v_{\alpha} (u)$; namely, one can add to $v_{\alpha} (u)$
any integer multiple of $2\pi i$.
This ambiguity implies that the integral \eqref{szerowkb} is defined only
up to integer multiples of $2 \pi i u$,
\be
S_0^{(\alpha)}(u) - {\rm const}
=  \int^u\log l_\alpha(u')\;du' = \int^u v_\alpha(u')\;du'
\;+\; i \pi u \qquad (\mbox{mod}\;2\pi i u)\;.
\label{S0_integral}
\ee
In practice, this ambiguity can always be fixed by imposing
suitable boundary conditions on $S_0^{(\alpha)}(u)$, and it never affects
the higher-order terms $S_n^{(\alpha)}(u)$.
Therefore, since our main goal is to solve the {\it quantum} theory
(to all orders in perturbation theory) we shall not worry about this
ambiguity in the classical term.
As we illustrate later (see Section \ref{sec:figeightquant}),
it will always be easy to fix this ambiguity in concrete examples.

Before we proceed, let us remark that if $M$ is a hyperbolic 3-manifold
with a single torus boundary $\Sigma = \partial M$ and $\CA^{({\rm geom})}$
is the ``geometric'' flat $SL(2,\C)$ connection associated with a hyperbolic
metric on $M$ (not necessarily geodesically complete),
then the integral \eqref{S0_integral} is essentially
the complexified volume function,
$i(\Vol (M;u) +i {\rm CS} (M;u))$,
which combines the (real) hyperbolic volume and Chern-Simons
invariants%
\footnote{For example, imaginary $u$ parametrizes a conical singularity.
See, {\it e.g.} \cite{thurston-1980, nz-1985, yoshida-1985} and our discussion in Section \ref{sec:ideal}
for more detailed descriptions of $\Vol(M;u)$ and ${\rm CS} (M;u)$. In part of the literature ({\it e.g.} in \cite{nz-1985}), the parameters $(u,v)$ are related to those used here by $2u_{\rm here}=u_{\rm there}$ and $2(v_{\rm here}+i\pi)=v_{\rm there}$. We include a shift of $i\pi$ in our definition of $v$ so that the complete hyperbolic structure arises at $(u,v)=(0,0)$.} %
of $M$. Specifically, the relation is \cite{gukov-2003,nz-1985}:
\be
S_0^{({\rm geom})} (u)
= \frac{i}{2}\Big[ \Vol (M;u) +  i {\rm CS} (M;u) \Big]
+ \,v_{{\rm geom}}(u)\,\Re(u) \,+\, i \pi u\,,
\label{S0_volume}
\ee
modulo the integration constant and multiples of $2\pi i u$.


\subsection{Classical and quantum symmetries}
\label{sec:symmetries}

A large supply of 3-manifolds with a single toral boundary can be obtained
by considering knot complements \eqref{mkcomplement};
our main examples in Section \ref{sec:example} are of this type.
As discussed above, the Lagrangian subvariety $\CL\in\CM_{\rm flat}(G_\C,\Sigma)$
for any knot complement $M$ is defined by polynomial equations \eqref{aizero}.
Such an $\CL$ contains multiple branches, indexed by $\alpha$, corresponding
to the different solutions to $\{A_i(l,m)=0\}$ for fixed $m$.
In this section, we describe relationships among these branches and
the corresponding perturbative invariants $Z^{(\alpha)}(M;\hbar,u)$
by using the symmetries of Chern-Simons theory with complex gauge group $G_{\mathbb{C}}$.

Before we begin, it is useful to summarize what we already know about
the branches of ${\cal L}$. As mentioned in the previous discussion,
there always exists a geometric branch --- or in the case of rank $r>1$
several geometric branches --- when $M$ is a hyperbolic knot complement.
Like the geometric branch, most other branches of ${\cal L}$ correspond
to genuinely nonabelian representations $\rho:\pi_1(\Sigma)\rightarrow G_{\mathbb{C}}$.
However, for any knot complement $M$ there also exists
an ``abelian'' component of $\CL$, described by the equations
\be
l_1 = \ldots = l_r = 1\,.
\label{abell}
\ee
Indeed, since $H_1 (M)$ is the abelianization of $\pi_1 (M)$,
the representation variety \eqref{mflatm} always has a component
corresponding to abelian representations that factor through $H_1 (M) \cong \Z$,
\be
\pi_1 (M) \to H_1 (M) \to G_{\C}\,.
\ee
The  corresponding flat connection, $\CA^{({\rm abel})}$,
is characterized by the trivial holonomy
around a 1-cycle of $\Sigma = T^2$ which is trivial in homology $H_1 (M)$;
choosing it to be the 1-cycle whose holonomy was denoted by $l = (l_1, \ldots, l_r)$
we obtain \eqref{abell}.
Note that, under projection to the $u$-space, the abelian component of $\CL$
corresponds to a single branch that we denote by $\alpha = {\rm abel}$.

The first relevant symmetry of Chern-Simons theory with complex gauge group $G_\C$ is conjugation. We observe that for every flat connection $\CA^{(\rho)}$ on $M$, with $\rho=(u,\alpha)$, there is a conjugate flat connection $\CA^{(\bar{\rho})}:= \overline{\CA^{(\rho)}}$ corresponding to a homomorphism $\bar{\rho}=(\bar{u},\bar{\alpha})$. We use $\bar{\alpha}$ to denote the branch of $\CL$ ``conjugate'' to branch $\alpha$; the fact that branches of $\CL$ come in conjugate pairs is reflected in the fact that eqs. \eqref{aizero} have real (in fact, integer) coefficients. The perturbative expansions around $\CA^{\rho}$ and $\CA^{\bar{\rho}}$ are very simply related. Namely, by directly conjugating the perturbative path integral and noting that the Chern-Simons action has real coefficients, we find%
\footnote{More explicitly, letting $I_{CS}(\hbar,\CA) = -\frac{1}{4\hbar}\int_M \Tr\big(\CA\wedge\CA+\frac{2}{3}\CA\wedge\CA\wedge\CA\big)$, we have
\be \overline{Z^{\alpha}(M;\hbar,u)} 
= \left(\int_{(u,\alpha)}\CD\CA\,e^{I_{CS}(\hbar,\CA)}\right)^* 
= \int_{(\bar{u},\bar{\alpha})} \CD\bar{\CA}\,e^{I_{CS}(\bar{\hbar},\bar{\CA})}
= Z^{(\bar{\alpha})}(M;\bar{\hbar},\bar{u})\,.\notag\ee} %
$\overline{Z^{\alpha}(M;\hbar,u)}=Z^{(\bar{\alpha})}(M;\bar{\hbar},\bar{u})$. The latter partition function is actually in the antiholomorphic sector of the Chern-Simons theory, but we can just rename $(\bar{\hbar},\bar{u})\mapsto (\hbar,u)$ (and use analyticity) to obtain a perturbative partition function for the conjugate branch in the holomorphic sector,
\be Z^{(\bar{\alpha})}(M;\hbar,u) = \overline{Z^{(\alpha)}}(M;\hbar,u)\,. \label{C-sym} \ee
Here, for any function $f(z)$ we define $\bar{f}(z):=\overline{f(\bar{z})}$. In particular, if $f$ is analytic, $f(z)=\sum f_n z^n$, then $\bar{f}$ denotes a similar function with conjugate coefficients, $\bar{f}(z)=\sum \bar{f}_n z^n$.

In the case $r=1$, the symmetry \eqref{C-sym} implies that branches of the classical $SL(2,\C)$ A-polynomial come in conjugate pairs $v_\alpha$ and $v_{\bar{\alpha}}(u)=\overline{v_\alpha}(u)$. Again, these pairs arise algebraically because the A-polynomial has integer coefficients.
 (See {\it e.g.} \cite{cooper-1994, cooper-1996} for a detailed discussion
of properties of $A(l,m)$.)  Some branches, like the abelian branch, may be self-conjugate. For the abelian branch, this is consistent with $S_0^{(abel)}=0$. The geometric branch, on the other hand, has a distinct conjugate because $\Vol(M;0)>0$\,; from \eqref{S0_volume} we see that its leading perturbative coefficient obeys
\be
S_0^{({\rm conj})}(u)
= \frac{i}{2}\Big[ - \Vol (M;\bar{u}) + i {\rm CS} (M;\bar{u}) \Big]
+\overline{v_{{\rm geom}}} (u) \Re(u) - i\pi u\,.
\label{S0_conj}
\ee
In general, we have
\be
S_0^{(\bar{\alpha})}(u) = \overline{S_0^{(\alpha)}}(u)
\qquad (\mbox{mod}\;2\pi u)\,.
\ee

Now, let us consider symmetries that originate from geometry,
{\it i.e.} symmetries that involve involutions of $M$,
\be
\tau : \quad M \to M\,.
\ee
Every such involution restricts to a self-map of $\Sigma = \partial M$,
\be
\tau \vert_{\Sigma} : \quad \Sigma \to \Sigma\,,
\ee
which, in turn, induces an endomorphism on homology, $H_i (\Sigma)$.
Specifically, let us consider an orientation-preserving involution $\tau$
which induces an endomorphism $(-1,-1)$ on $H_1 (\Sigma) \cong \Z \times \Z$.
This involution is a homeomorphism of $M$;
it changes our definition of the holonomies,
\be
m_i \to \frac{1}{m_i}
\qquad {\rm and} \qquad
l_i \to \frac{1}{l_i}\,,
\label{lmrefl}
\ee
leaving the symplectic form \eqref{sympluv} invariant.
Therefore, it preserves both the symplectic phase space $\M (G_{\C}, \Sigma)$
and the Lagrangian submanifold $\CL$ (possibly permuting some of its branches).

In the basic case of rank $r=1$, the symmetry \eqref{lmrefl} corresponds to the simple, well-known relation $A(l^{-1},m^{-1})=A(l,m)$,
up to overall powers of $l$ and $m$. Similarly, at the quantum level, $\widehat{A}(\hat{l}^{-1},\hat{m}^{-1}) = \widehat{A}(\hat{l},\hat{m})$ when $\widehat{A}(\hat{l},\hat{m})$ is properly normalized. Branches of the A-polynomial are individually preserved, implying that the perturbative partition functions (and the coefficients $S_n^{(\alpha)}$) are all \emph{even}:
\be Z^{(\alpha)}(M;\hbar,-u) = Z^{(\alpha)}(M;\hbar,u)\,, \label{even} \ee
modulo factors of $e^{2\pi u/\hbar}$ that are related to the ambiguity in $S_0^{(\alpha)}(u)$. Note that in the $r=1$ case one can also think of the symmetry \eqref{lmrefl} as the Weyl reflection.
Since, by definition, holonomies that differ by an element of the Weyl group
define the same point in the moduli space \eqref{mflatttwo}, it is clear that
both $\M (G_{\C},\Sigma)$ and $\CL$ are manifestly invariant under this symmetry. (For $r>1$, Weyl transformations on the variables $l$ and $m$ lead to new, independent relations among the branches of $\cal L$.)

Finally, let us consider a more interesting ``parity'' symmetry, an orientation-reversing involution
\be
P : \quad M \to \overline{M}
\ee
\be
P \vert_{\Sigma} : \quad \Sigma \to \overline{\Sigma}
\ee
that induces a map $(1,-1)$ on $H_1 (\Sigma) \cong \Z \times \Z$.
This operation by itself cannot be a symmetry of the theory because
it does not preserve the symplectic form \eqref{sympluv}.
We can try, however, to combine it with the transformation $\hbar \to - \hbar$
to get a symmetry of the symplectic phase space $(\M (G_{\C}, \Sigma), \omega)$.
We are still not done because this combined operation
changes the orientation of both $\Sigma$ and $M$,
and unless $\overline{M} \cong M$
the state $\vert M \rangle$ assigned to $M$ will be mapped
to a different state $\vert \overline{M} \rangle$.
But if $M$ is an amphicheiral\footnote{A manifold is called chiral or amphicheiral
according to whether the orientation cannot or can be reversed by a self-map.}
manifold, then both $m_i \to \frac{1}{m_i}$ and $l_i \to \frac{1}{l_i}$
(independently) become symmetries of the theory,
once combined with $\hbar \to - \hbar$.
This now implies that solutions come in signed pairs,
$v_{\alpha} (u)$ and $v_{\hat{\alpha}} (u) = -v_{\alpha} (u)$,
such that the corresponding perturbative $G_{\C}$ invariants satisfy
\be
Z^{(\hat{\alpha})} (M;\hbar,u) = Z^{(\alpha)} (M;-\hbar,u)\,.
\ee
For the perturbative coefficients, this leads to the relations
\begin{align}
S_0^{(\hat{\alpha})}(u) & = -S_0^{(\alpha)}(u) \qquad\quad (\mbox{mod}\;2\pi u)\,, \label{sym_sign} \\
S_n^{(\hat{\alpha})}(u) & = (-1)^{n+1} S_n^{(\alpha)}(u) \qquad n\ge 1\,.
\label{sym_signsn}
\end{align}

Assuming that the $\sim \pm\frac{i}{2} \Vol (M;u)$ behavior of the geometric
and conjugate branches is unique, their signed and conjugate pairs must coincide
for amphicheiral 3-manifolds.
$S_0^{(\rm geom)}$, then, is an even analytic function of $u$ with strictly real
series coefficients; at $u \in i \R$, the Chern-Simons invariant ${\rm CS} (M;u)$ will vanish.


\subsection{Brane quantization}

Now, let us briefly describe how the problem of quantizing
the moduli space of flat connections, $\M (G_{\C},\Sigma)$, would look
in the new approach \cite{gukov-2008} based on the topological A-model and D-branes.
Although it can be useful for a better understanding of Chern-Simons theory
with complex gauge group, this discussion is not crucial for the rest of the paper and
the reader not interested in this approach may skip directly to Section~\ref{sec:statesum}.

In the approach of \cite{gukov-2008}, the problem of quantizing
a symplectic manifold $N$ with symplectic structure $\omega$
is solved by complexifying $(N,\omega)$ into $(Y,\Omega)$
and studying the A-model of $Y$ with symplectic structure $\omega_Y = \Im \Omega$.
Here, $Y$ is a complexification of $N$, {\it i.e.} a complex manifold
with complex structure $\CI$ and an antiholomorphic involution
\be
\tau: Y \to Y \,,
\ee
such that $N$ is contained in the fixed point set of $\tau$ and $\tau^* \CI = - \CI$.
The 2-form $\Omega$ on $Y$ is holomorphic in complex structure $\CI$ and obeys
\be
\tau^* \Omega = \overline{\Omega}
\label{omtau}
\ee
and
\be
\Omega \vert_N = \omega\,.
\label{omrestr}
\ee
In addition, one needs to pick a unitary line bundle $\frak L \to Y$
(extending the ``prequantum line bundle'' $\frak L \to N$)
with a connection of curvature $\Re \Omega$.
This choice needs to be consistent with the action of the involution $\tau$,
meaning that $\tau: Y \to Y$ lifts to an action on $\frak L$,
such that $\tau|_N = {\rm id}$.
To summarize, in brane quantization the starting point involves the choice
of $Y$, $\Omega$, $\frak L$, and $\tau$.

In our problem, the space $N = \M (G_{\C},\Sigma)$ that we wish to quantize
is already a complex manifold. Indeed, as we noted earlier, it comes equipped
with the complex structure $J$ (that does not depend on the complex structure on $\Sigma$).
Therefore, its complexification\footnote{Notice,
since in our problem we start with a hyper-K\"ahler
manifold $\M (G_{\C},\Sigma)$, its complexification $Y$
admits many complex structures.
In fact, $Y$ has holonomy group $Sp(n) \times Sp(n)$,
where $n$ is the quaternionic dimension of $\M (G_{\C},\Sigma)$.}
is $Y = N \times \overline{N}$
with the complex structure on $\overline{N}$ being prescribed by $-J$
and the complex structure on $Y$ being $\CI=(J, -J)$.
The tangent bundle $TY=TN\oplus T\overline{N}$
is identified with the complexified tangent bundle of $N$,
which has the usual decomposition $\C TN=T^{1,0}N\oplus T^{0,1} N$.
Then, the ``real slice'' $N$ is embedded in $Y$ as the diagonal
\be
N \ni x \mapsto (x, x) \in N \times \overline{N}\,.
\label{ndiag}
\ee
In particular, $N$ is the fixed point set of the antiholomorphic involution
$\tau : Y \to Y$ which acts on $(x,y) \in Y$ as $\tau: (x,y) \mapsto (y, x)$.

Our next goal is to describe the holomorphic 2-form $\Omega$
that obeys \eqref{omtau} and \eqref{omrestr} with\footnote{Notice,
while in the rest of the paper we consider only the ``holomorphic'' sector
of the theory (which is sufficient in the perturbative approach),
here we write the complete symplectic form on $\M (G_{\C},\Sigma)$
that follows from the classical Chern-Simons action \eqref{csaction},
including the contributions of both fields $\CA$ and $\bar \CA$.}
\be
\omega = \frac{t}{2\pi i} (\omega_K + i \omega_I) - \frac{\bar t}{2\pi i} (\omega_K - i \omega_I)
\ee
Note, that $(\omega_K + i \omega_I)$ is holomorphic on $N = \M (G_{\C},\Sigma)$
and $(\omega_K - i \omega_I)$ is holomorphic on $\overline{N}$.
Moreover, if we take $\bar t$ to be a complex conjugate of $t$, the antiholomorphic
involution $\tau$ maps $t (\omega_K + i \omega_I)$ to $\bar t (\omega_K - i \omega_I)$,
so that $\tau^* \omega = \omega$.
Therefore, we can simply take
\be
\Omega = \frac{t}{2\pi i} (\omega_K^{(1)} + i \omega_I^{(1)}) - \frac{\bar t}{2\pi i} (\omega_K^{(2)} - i \omega_I^{(2)})
\ee
where the superscript $i=1,2$ refers to the first (resp. second) factor in $Y = N \times \overline{N}$.
It is easy to verify that the 2-form $\Omega$ defined in this way indeed obeys $(\Im \Omega)^{-1} \Re \Omega = \CI$.
Moreover, one can also check that if $\bar t$ is a complex conjugate of $t$
then the restriction of $\omega_Y = \Im \Omega$
to the diagonal \eqref{ndiag} vanishes, so that the ``real slice'' $N \subset Y$,
as expected, is a Lagrangian submanifold in $(Y,\omega_Y)$.

Now, the quantization problem can be realized in the A-model of $Y$
with symplectic structure $\omega_Y = \Im \Omega$. In particular,
the Hilbert space $\CH_{\Sigma}$ is obtained as the space of $(\CB_{cc},\CB')$ strings,
\be
\CH_{\Sigma} = {\rm ~space~of~} (\CB_{cc},\CB') {\rm ~strings}
\ee
where $\CB_{cc}$ and $\CB'$ are A-branes on $Y$ (with respect to the symplectic structure $\omega_Y$).
The brane $\CB'$ is the ordinary Lagrangian brane supported on
the ``real slice'' $N \subset Y$.
The other A-brane, $\CB_{cc}$, is the so-called canonical coisotropic brane
supported on all of $Y$. It carries a Chan-Paton line bundle of curvature $F = \Re \Omega$.
Note that for $[F]$ to be an integral cohomology class we need $\Re (t) \in \Z$.
Since in the present case the involution $\tau$ fixes the ``real slice'' pointwise,
it defines a hermitian inner product on $\CH_{\Sigma}$ which is positive definite.


\section{A state integral model for perturbative $SL(2,\C)$ Chern-Simons theory}
\label{sec:statesum}

In this section, we introduce a ``state integral'' model for $Z^{(\rho)} (M;\hbar)$
in the simplest case of $G_{\C} = SL(2,\C)$.
Our construction will rely heavily on the work of Hikami \cite{hikami-2001-16, hikami-2006},
where he introduced an invariant of hyperbolic 3-manifolds using ideal triangulations.
The resulting invariant is very close to the state integral model we are looking for.
However, in order to make it into a useful tool for computing  $Z^{(\rho)} (M;\hbar)$
we will need to understand Hikami's construction better and make a number of important modifications.
In particular, as we explain below, Hikami's invariant is defined as a certain integral
along a path in the complex plane (or, more generally, over a hypersurface in complex space)
which was not specified\footnote{One
choice, briefly mentioned in \cite{hikami-2001-16, hikami-2006}, is to integrate over
the real axis (resp. real subspace) of the complex parameter space.
While this choice is in some sense natural, a closer look shows that it cannot be the right one.}
in the original work \cite{hikami-2001-16, hikami-2006}.
Another issue that we need to address is how to incorporate in Hikami's construction
a choice of the homomorphisms \eqref{rhofirst},
\be
\rho: \pi_1 (M) \to SL(2,\C)\,.
\ee
(The original construction assumes very special choices of $\rho$
that we called ``geometric'' in Section~\ref{sec:pert}.)
It turns out that these two questions are not unrelated and can be addressed
simultaneously, so that Hikami's invariant can be extended to a state sum model
for $Z^{(\rho)} (M;\hbar)$ with an arbitrary $\rho$.

Throughout this section, we work in the $u$-space representation.
In particular, we use the identification \eqref{rhoviaua}
and denote the perturbative $SL(2,\C)$ invariant as $Z^{(\alpha)} (M;\hbar,u)$.


\subsection{Ideal triangulations of hyperbolic 3-manifolds}
\label{sec:ideal}

The construction of a state integral model described in the rest of this section
applies to orientable hyperbolic 3-manifolds of finite volume (possibly with boundary)
and uses ideal triangulations in a crucial way.
Therefore, we begin this section by reviewing some relevant facts from hyperbolic geometry
(more details can be found in \cite{thurston-1980, thurston-1982,
thurston-1999}).

Recall that hyperbolic 3-space $\IH^3$ can be represented as the upper half-space
$\{(x_1,x_2,x_3) \vert$ $x_3>0 \}$ with metric \eqref{hthreemetric} of constant curvature $-1$.
The boundary $\partial \IH^3$, topologically an ${\bf S}^2$,
consists of the plane $x_3=0$ together with the point at infinity.
The group of isometries of $\IH^3$ is $PSL(2,\C)$,
which acts on the boundary via the usual M\"obius transformations.
In this picture, geodesic surfaces are spheres of any radius which intersect $\partial \IH^3$ orthogonally.

An ideal tetrahedron $\Delta$ in $\IH^3$ has by definition all its faces along geodesic surfaces,
and all its vertices in $\partial \IH^3$ --- such vertices are called {\it ideal points}.
After M\"obius transformations, one can fix three of the vertices at $(0,0,0)$,
$(1,0,0)$, and infinity.
The coordinate of the fourth vertex $(x_1,x_2,0)$, with $x_2\geq 0$,
defines a complex number $z=x_1 + i x_2$ called the {\it shape parameter}
(sometimes also called {\it edge parameter}).
At various edges, the faces of the tetrahedron $\Delta$ form dihedral angles
$\arg z_j$ ($j=1,2,3$) as indicated in Figure \ref{fig:ideal}, with
\be z_1 = z\,, \quad z_2 = 1-\frac{1}{z}\,, \quad z_3 = \frac{1}{1-z}\,.  \label{zj} \ee
The ideal tetrahedron is noncompact, but has finite volume given by
  \be \label{volviad} \Vol (\Delta_z) \= D(z)\,,  \ee
where $D(z)$ is the Bloch-Wigner dilogarithm function, related to the usual dilogarithm $\Li$ (see Section \ref{sec:qdilog}) by
  \be\label{BW} D(z) = \Im\bigl(\Li (z)\bigr) + \arg(1-z)\,\log |z|\,. \ee
Note that any of the $z_j$ can be taken to be the shape parameter of $\Delta$,
and that $D(z_j) = \Vol (\Delta_z)$ for each~$j$.
We will allow shape parameters to be any complex numbers in $\C-\{0,1\}$,
noting that for $z\in\R$ an ideal tetrahedron is degenerate and that for
$\Im\,z<0$ it technically has negative volume due to its orientation.

\EPSFIGURE{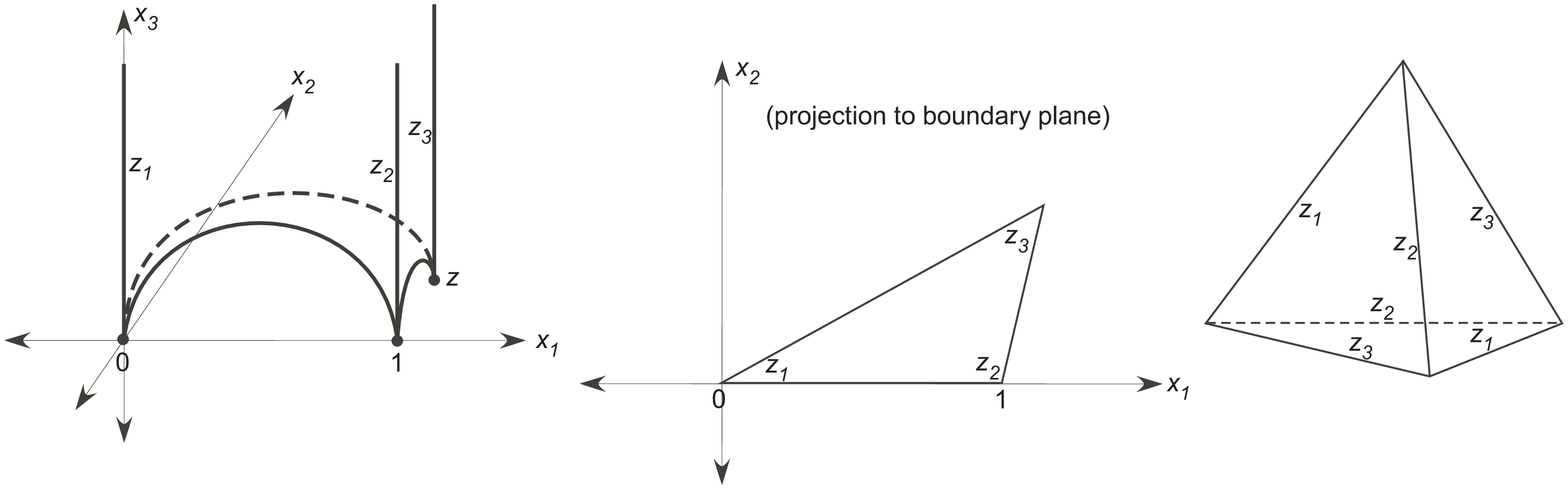,height=4.5cm,angle=0,trim=0 0 0 0}%
{An ideal tetrahedron in $\IH^3$.
\label{fig:ideal} }

A hyperbolic structure on a 3-manifold is a metric that is locally isometric to $\IH^3$.
A 3-manifold is called hyperbolic if it admits a hyperbolic structure that is geodesically complete and has finite volume.
Most 3-manifolds are hyperbolic, including the vast majority
of knot and link complements in ${\bf S}^3$.
Specifically, a knot complement is hyperbolic as long as the knot is not
a torus or satellite knot \cite{thurston-1986}.
Every closed 3-manifold can be obtained via Dehn surgery on
a knot in ${\bf S}^3$, and for hyperbolic knots
all but finitely many such surgeries yield hyperbolic manifolds \cite{thurston-1980}.

By the Mostow rigidity theorem \cite{mostow-1973, prasad-1973},
the complete hyperbolic structure on a hyperbolic manifold is unique. Therefore, geometric invariants like the hyperbolic volume are actually {topological} invariants.
For the large class of hyperbolic knot complements in ${\bf S}^3$,
the unique complete hyperbolic structure has a parabolic holonomy with unit eigenvalues around the knot.
In $SL(2,\C)$ Chern-Simons theory, this structure corresponds to the ``geometric'' flat connection $\CA^{({\rm geom})}$ with $u=0$. As discussed in Section \ref{sec:loops}, hyperbolic manifolds with complete hyperbolic structures can also be described as quotients $
\mathbb{H}^3/\Gamma$.

Given a hyperbolic knot complement, one can deform the hyperbolic metric in such a way that the holonomy $u$ is not zero.
Such deformed 
metrics are unique in a neighborhood of $u=0$, but they are not geodesically complete. 
For a discrete set of values of $u$, one can add in the ``missing'' geodesic, and the deformed metrics coincide
with the unique complete hyperbolic structures on closed 3-manifolds obtained
via appropriate Dehn surgeries on the knot in ${\bf S}^3$. For other values of $u$, the knot complement can be completed by adding either a circle $S^1$ or a single point, but the resulting hyperbolic metric will be singular. For example, if $u\in i\R$ one adds a circle and the resulting metric has a conical singularity. These descriptions can easily be extended to link complements (\emph{i.e.} multiple cusps),
using multiple parameters $u_k$, one for each link component.

Any orientable hyperbolic manifold $M$ is homeomorphic to the interior of a compact 3-manifold $\bar{M}$ with boundary consisting of finitely many tori. (The manifold $M$ itself can also be thought of as the union of $\bar{M}$ with neighborhoods of the cusps, each of the latter being homeomorphic to $T^2\times [0,\infty)$.)
All hyperbolic manifolds therefore arise as knot or link complements in closed 3-manifolds. Moreover, every hyperbolic manifold has an ideal triangulation, {\it i.e} a finite decomposition into (possibly degenerate\footnote{It is conjectured and widely believed that nondegenerate tetrahedra alone are always sufficient.}) ideal tetrahedra; see {\it e.g.} \cite{thurston-1980, petronio-2000}.

To reconstruct a hyperbolic 3-manifold $M$ from its ideal triangulation $\{\Delta_i\}_{i=1}^N$,
faces of tetrahedra are glued together in pairs.
One must remember, however, that vertices of tetrahedra are not part of $M$,
and that the combined boundaries of their neighborhoods in $M$ are not spheres, but tori.
(Thus, some intuition from simplicial triangulations no longer holds.)
There always exists a triangulation of $M$ whose edges can all be oriented in such a way that the boundary of every face (shared between two tetrahedra) has two edges oriented in the same direction (clockwise or counterclockwise) and one opposite. Then the vertices of each tetrahedron can be canonically labeled 0, 1, 2, 3 according to the number of edges entering the vertex, so that the tetrahedron can be identified in a unique way with one of the two numbered tetrahedra shown in Figure \ref{fig:tetr_momenta} of the next subsection. This at the same time orients the tetrahedron. The orientation of a given tetrahedron $\Delta_i$ may not agree with that of $M$; one defines $\epsilon_i=1$ if the orientations agree and $\epsilon_i=-1$ otherwise. The edges of each tetrahedron can then be given shape parameters $(z^{(i)}_1,z^{(i)}_2,z^{(i)}_3)$, running counterclockwise around each vertex (viewed from outside the tetrahedron) if $\epsilon_i=1$ and clockwise if $\epsilon_i=-1$.

For a given $M$ with cusps or conical singularities
specified by holonomy parameters $u_k$, the shape parameters $z_j^{(i)}$ of the tetrahedra $\Delta_i$
in its triangulation are fixed by two sets of conditions.
First, the product of the shape parameters $z_j^{(i)}$
at every edge in the triangulation must be equal to 1,
in order for the hyperbolic structures of adjacent tetrahedra to match.
More precisely, the sum of some chosen branches of $\log z_j^{(i)}$ (equal to the standard branch if one is near the complete structure) should equal $2\pi i$, so that the total dihedral angle at each edge is $2\pi$.
Second, one can compute holonomy eigenvalues around each torus boundary
in $M$ as a product of $z_j^{(i)}$'s by mapping out the neighborhood
of each vertex in the triangulation in a so-called developing map,
and following a procedure illustrated in, {\it e.g.}, \cite{nz-1985}.
There is one distinct vertex ``inside'' each boundary torus.
One then requires that the eigenvalues of the holonomy around the $k$th boundary are equal to $e^{\pm u_k}$.
These two conditions will be referred to, respectively,
as {\it consistency} and {\it cusp} relations.

Every hyperbolic 3-manifold has a well-defined class in the Bloch group \cite{neumann-1997}. This is a subgroup%
\footnote{Namely, the kernel of the map $[z]\mapsto 2z\wedge(1-z) \in \C^*\wedge_\Z \C^*$ acting on this quotient module.} %
of the quotient of the free $\Z$-module $\Z[\C-\{0,1\}]$ by the relations%
\be
[x]-[y]+[\frac{y}{x}]-[\frac{1-x^{-1}}{1-y^{-1}}]+[\frac{1-x}{1-y}]=0\,. \label{FT}
\ee
This five-term or pentagon relation accounts for the fact that a polyhedron with five ideal vertices can be decomposed into ideal tetrahedra in multiple ways. The five ideal tetrahedra in this polyhedron (each obtained by deleting an ideal vertex) can be given the five shape parameters $x,\,y,\,y/x,...$ appearing above. The signs of the different terms correspond to orientations. Geometrically, an instance of the five-term relation can be visualized as the 2-3 Pachner move, illustrated in Figure \ref{fig:Pachner}.

\EPSFIGURE{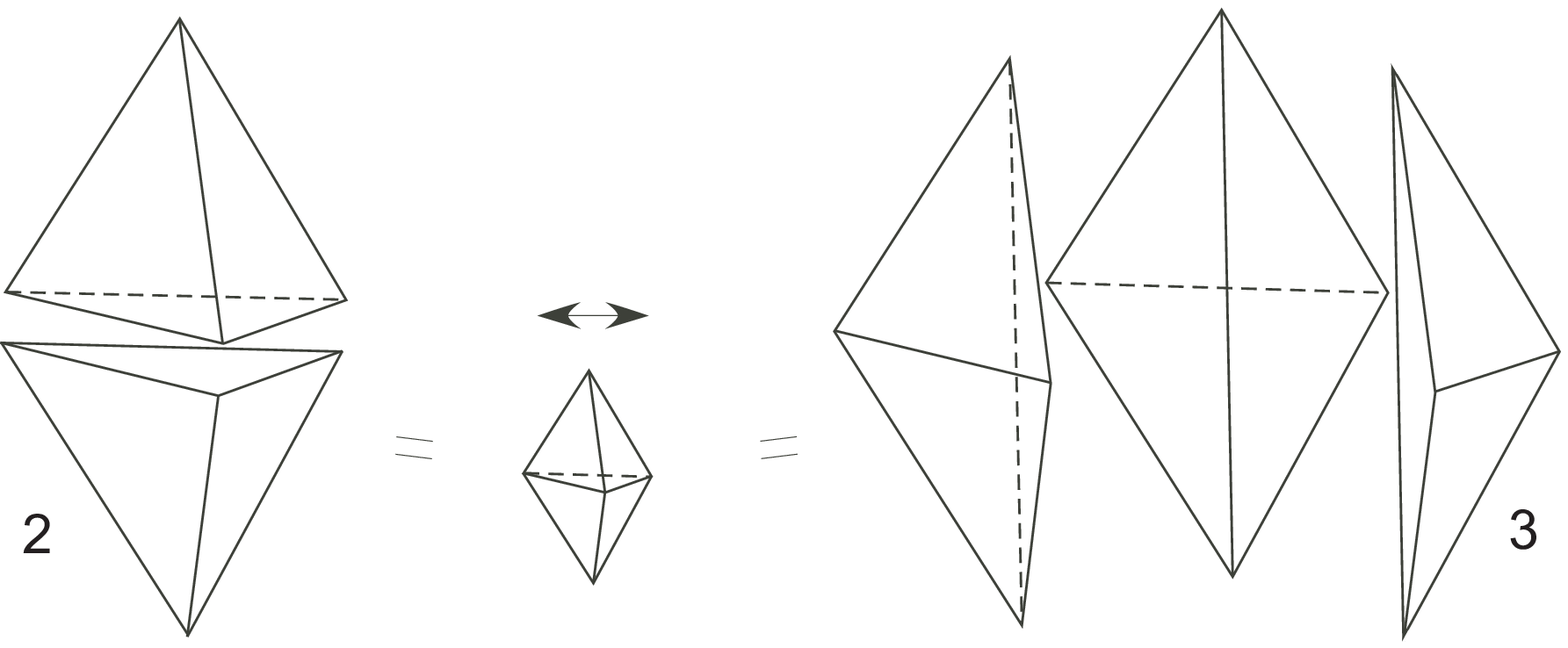,height=4cm,angle=0,trim=0 0 0 0}%
{The 2-3 Pachner move.
\label{fig:Pachner} }

The class $[M]$ of a hyperbolic 3-manifold $M$ in the Bloch group can be computed by summing (with orientation) the shape parameters $[z]$ of any ideal triangulation, but it is {\it independent} of the triangulation. Thus, hyperbolic invariants of 3-manifolds may be obtained by functions on the Bloch group --- {\it i.e.} functions compatible with \eqref{FT}. For example, the Bloch-Wigner function \eqref{BW} satisfies
\be D(x)-D(y)+D\big(\frac{y}{x}\big)-D\big(\frac{1-x^{-1}}{1-y^{-1}}\big)+D\big(\frac{1-x}{1-y}\big)=0\,, \ee
and the hyperbolic volume of a manifold $M$ triangulated by ideal tetrahedra $\{\Delta_i\}_{i=1}^{N}$ can be calculated as
\be \Vol(M) = \sum_{i=1}^{N}\epsilon_i D(z^{(i)})\,. \label{VolM} \ee
The symbols $\epsilon_i$ here could be removed if shape parameters were assigned to tetrahedra in a manner independent of orientation, noting that reversing the orientation of a tetrahedron corresponds to sending $z\mapsto 1/z$ and that $D(1/z)=-D(z)$. This is sometimes seen in the literature.

The complexified volume $i(\Vol(M)+i{\rm CS}(M))$ is trickier to evaluate. For a hyperbolic manifold with a spin structure, corresponding to full $SL(2,\C)$ holonomies, this invariant is defined modulo $4\pi^2$. Here, we will outline a computation of the complexified volume modulo $\pi^2$, following \cite{neumann-2004}; for the complete invariant modulo $4\pi^2$, see \cite{zickert-2008}. To proceed, one must first make sure that the three shape parameters $z^{(i)}=z_1^{(i)}$, $z_2^{(i)}$, and $z_3^{(i)}$ are specifically assigned to edges $[v_1^{(i)},v_2^{(i)}]$, $[v_1^{(i)},v_3^{(i)}]$, and $[v_1^{(i)},v_2^{(i)}]$ (respectively) in each tetrahedron $\Delta_i$ of an oriented triangulation of $M$, where $[v_a^{(i)},v_b^{(i)}]$ denotes the edge going from numbered vertex $v_a^{(i)}$ to numbered vertex $v_b^{(i)}$. One also chooses logarithms $(w_1^{(i)},w_2^{(i)},w_3^{(i)})$ of the shape parameters such that
\begin{subequations}
\begin{align}& e^{w^{(i)}_1}=\pm z^{(i)}_1\,,\quad e^{w^{(i)}_2}=\pm z^{(i)}_2\,, \quad e^{w^{(i)}_3}=\pm z^{(i)}_3\,, \\
& w^{(i)}_1+w^{(i)}_2+w^{(i)}_3 = 0\quad \forall\,i\,,
\end{align}
\end{subequations}
and defines integers $(q^{(i)},r^{(i)})$ by
\be w_1^{(i)}={\rm Log}(z^{(i)})+\pi i q^{(i)}\,,\qquad w_2^{(i)}=-{\rm Log}(1-z^{(i)})+\pi i r^{(i)}\,, \ee
where ${\rm Log}$ denotes the principal branch of the logarithm, with a cut from $0$ to $-\infty$. For a consistent labeling of the  triangulation, called a ``combinatorial flattening,'' the sum of log-parameters $w^{(i)}_j$ around every edge must vanish, and the (signed\footnote{Signs arise from tetrahedron orientations and~the~sense~in~which~a~path~winds~around~edges;~see~\cite{neumann-2004},~Def.~4.2.}) sum of log-parameters along the two paths generating $\pi_1(T^2)=\Z^2$ for any boundary (cusp) $T^2$ must equal twice the logarithm of the $SL(2,\C)$ holonomies around these paths.%
\footnote{\label{foot:v2pi}Explicitly, in the notation of Section \ref{sec:quantization} and above, the sum of log-parameters along the two paths in the neighborhood of the $k$th cusp must equal $2u_k$ and $2v_k+2\pi i$, respectively. 
} %
The complexified volume is then given, modulo $\pi^2$, as
\be i(\Vol(M)+i {\rm CS}(M)) \,=\, \sum_{i=1}^N \epsilon_i {\rm L}(z^{(i)};q^{(i)},r^{(i)}) - \sum_{{\rm cusps}\,k}(v_k\overline{u}_k+i\pi u_k)\,, \label{CSVolM}\ee
with
\be {\rm L}(z;q,r) = \Li(z)+\frac{1}{2}({\rm Log}(z)+\pi i q)({\rm Log}(1-z)+\pi i r) + \frac{\pi^2qr}{2} - \frac{\pi^2}{6}\,. \label{RogersL} \ee
The function ${\rm L} (z;q,r)$, a modified version of the Rogers dilogarithm, satisfies a five-term relation in an extended Bloch group that lifts \eqref{FT} in a natural way to the space of log-parameters.%
\footnote{The branch of $\Li$ in \eqref{RogersL} is taken to be the standard one, with a cut from $1$ to $+\infty$. Note, however, that we could also take care of ambiguities arising from the choice of dilogarithm branch by rewriting \eqref{RogersL} in terms of the function $\tilde{\rm L}(w)=\int_{-\infty}^w \frac{t\,dt}{1-e^{-t}}$. This is a well-defined holomorphic function :\;$\C\rightarrow \C/4\pi^2\Z$ because all the residues of $t/(1-e^t)$ are integer multiplies of $2\pi i$, and it coincides with a branch of the function $\Li(e^w)+w\log(1-e^w)$.}


\subsection{Hikami's invariant}
\label{sec:Hikami}

We can now describe Hikami's geometric construction.
Roughly speaking, to compute the invariant for a hyperbolic manifold $M$,
one chooses an ideal triangulation of $M$, assigns an infinite-dimensional
vector space $V$ or $V^*$ to each tetrahedron face, and assigns a matrix element in
$V\otimes V\otimes V^* \otimes V^*$ to each tetrahedron.
These matrix elements depend on a small parameter $\hbar$,
and in the classical $\hbar \to 0$ limit they capture the hyperbolic structure of the tetrahedra.
The invariant of $M$ is obtained by taking inner products
of matrix elements on every pair of identified faces (gluing the tetrahedra back together),
subject to the cusp conditions described above in the classical limit.

To describe the process in greater detail,
we begin with an orientable hyperbolic manifold 
that has an oriented ideal triangulation $\{\Delta_i\}_{i=1}^N$,
and initially forget about the hyperbolic structures of these tetrahedra. As discussed in Section \ref{sec:ideal} and indicated in Figure \ref{fig:tetr_momenta}, each tetrahedron comes with one of two possible orientations of its edges, which induces an ordering of its vertices $v_j^{(i)}$ (the subscript $j$ here is not to be confused with the shape parameter subscript in \eqref{zj}), an ordering of its faces, and orientations on each face.
The latter can be indicated by inward or outward-pointing normal vectors. The faces (or their normal vectors) are labelled by $p_j^{(i)}$, in correspondence with opposing vertices.
The normal face-vectors of adjacent tetrahedra match up head-to-tail
(and actually define an oriented dual decomposition) when tetrahedra are glued to form $M$.

\EPSFIGURE{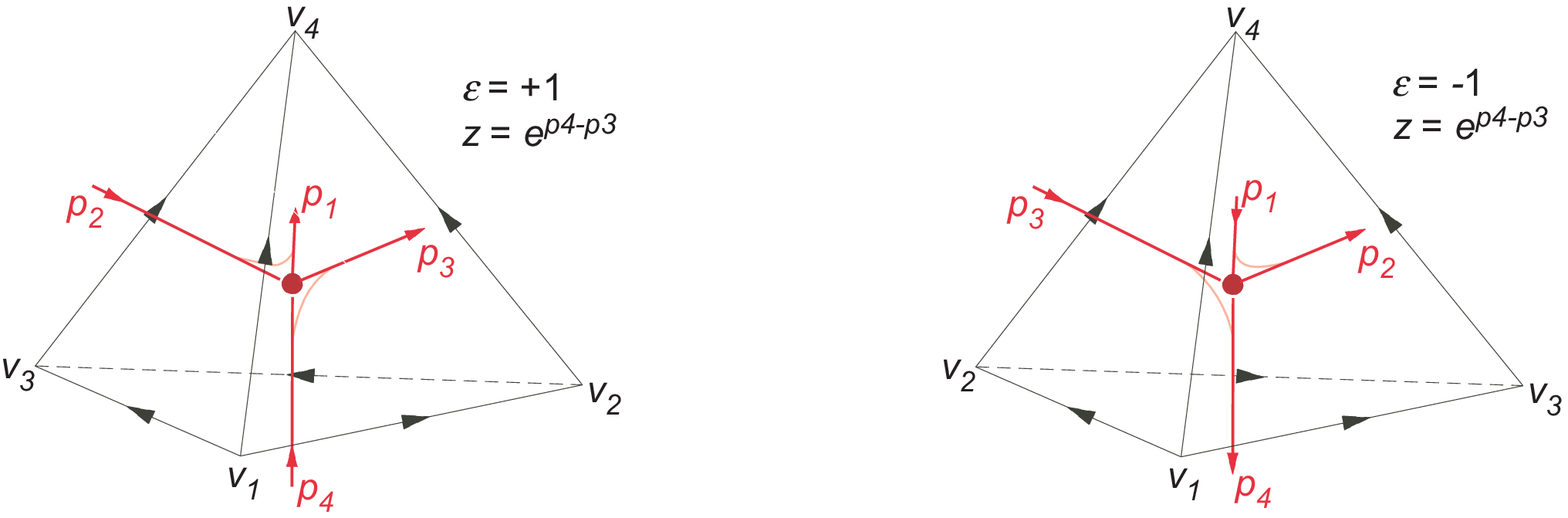,height=4.5cm,angle=0,trim=0 0 0 0}%
{\small Oriented tetrahedra, to which matrix elements
$\langle p_1,p_3|{\bf S}|p_2,p_4\rangle$ (left) and
$\langle p_2,p_4|{\bf S}^{-1}|p_1,p_3\rangle$ (right) are assigned.
\label{fig:tetr_momenta} }

Given such an oriented triangulation, one associates a vector space $V$
to each inward-oriented face, and the dual space $V^*$ to each outward-oriented face.
(Physicists should think of these spaces as ``Hilbert'' spaces obtained
by quantizing the theory on a manifold with boundary.)
The elements of $V$ are represented by complex-valued functions in one variable,
with adjoints given by conjugation and inner products given by integration.
Abusing the notation, but following the very natural set of conventions of \cite{hikami-2001-16, hikami-2006},
we denote these complex variables by $p_j^{(i)}$,
which we used earlier to label the corresponding faces of the tetrahedra.
As a result, to the boundary of every tetrahedron $\Delta_i$ one associates a vector space
$V\otimes V\otimes V^* \otimes V^*$, represented by functions of its four face labels,
$p_1^{(i)}$, $p_2^{(i)}$, $p_3^{(i)}$, and~$p_4^{(i)}$.

To each tetrahedron one assigns a matrix element $\langle p_1,p_3|{\bf S}|p_2,p_4\rangle$
or $\langle p_2,p_4|{\bf S}^{-1}|p_1,p_3\rangle$,
depending on orientation as indicated in Figure \ref{fig:tetr_momenta}.
Here, the matrix ${\bf S}$ acts on functions $f(p_1,p_2) \in V\otimes V$ as
\be
{\bf S} \= e^{\hat{q}_1\hat{p}_2/2\hbar}\; \Phi_\hbar ( \hat{p}_1 + \hat{q}_2 - \hat{p}_2)\,,
\label{S-matrix}
\ee
where $\hat{p}_i f =p_i f$ and $\hat{q}_i f = 2\hbar \frac{\partial}{\partial p_i} f$.
The function $\Phi_\hbar$ is the quantum dilogarithm, to be described in the next subsection.
Assuming that
$\int \frac{idq}{4\pi\hbar} |q \rangle \langle q| = \int dp\, |p\rangle \langle p| = 1$
and $\langle p | q \rangle = e^{\frac{pq}{2i\hbar}}$
(the exact normalizations are not important for the final invariant), one obtains via Fourier transform
\small
\begin{align}
\langle p_1,p_3|{\bf S}|p_2,p_4\rangle
& = \frac{\delta(p_1 + p_3 - p_2)}{\sqrt{-4\pi i\hbar}} \Phi_{\hbar} (p_4 - p_3 + i\pi + \hbar)
\, e^{\frac{1}{2\hbar} \left( p_1 (p_4 - p_3) + \frac{i\pi\hbar}{2} - \frac{\pi^2-\hbar^2}{6} \right) }\,,
\label{S_element1} \\
\langle p_2,p_4|{\bf S}^{-1}|p_1,p_3\rangle
& = \frac{\delta(p_1 + p_3 - p_2)}{\sqrt{-4\pi i\hbar}} \frac{1}{\Phi_{\hbar}
(p_4 - p_3 - i\pi-\hbar)} e^{\frac{1}{2\hbar} \left( p_1 (p_3 - p_4) - \frac{i\pi\hbar}{2}
+ \frac{\pi^2-\hbar^2}{6} \right)}\,.
\label{S_element2}
\end{align}
\normalsize
In the classical limit $\hbar \to 0$, the quantum dilogarithm has the asymptotic $\Phi_\hbar(p)\sim \frac{1}{2\hbar} \Li(-e^p)$. One therefore sees that the classical limits of the above matrix elements look very much like exponentials of $\frac{1}{2\hbar}$ times the complexified hyperbolic volumes of tetrahedra. For example, the asymptotic of \eqref{S_element1} coincides with $\exp(L(z;\cdot,\cdot)/(2\hbar))$ if we identify $e^{p_4-p_3}$ with $z$ and $e^{-2p_1}$ with $1/(1-z)$. For building a quantum invariant, however, only half of the variables $p_j$ really ``belong'' to a single tetrahedron. Hikami's claim \cite{hikami-2001-16, hikami-2006} is that if we only identify a shape parameter
\be
z^{(i)} = e^{p_4^{(i)}-p_3^{(i)}}\,,
\label{z_and_p}
\ee
for every tetrahedron $\Delta_i$, the classical limit of the resulting quantum invariant will completely reproduce the hyperbolic structure and complexified hyperbolic volume on $M$.%
\footnote{Eqn. \eqref{z_and_p} is a little different from the relation appearing in \cite{hikami-2001-16, hikami-2006}, because our convention for assigning shape parameters to edges based on orientation differs from that of \cite{hikami-2001-16, hikami-2006}.}

To finish calculating the invariant of $M$, one glues the tetrahedra back together and 
takes inner products in every pair $V$ and $V^*$ corresponding to identified faces.
This amounts to multiplying together all the matrix elements \eqref{S_element1} or \eqref{S_element2},
identifying the $p^{(i)}_j$ variables on identified faces (with matching head-to-tail normal vectors),
and integrating over the $2N$ remaining $p$'s.
To account for possible toral boundaries of $M$, however,
one must revert back to the hyperbolic structure. This allows one to write the holonomy eigenvalues $\{ e^{u_k} \}$ as products of shape parameters $(z^{(i)},\,1-1/z^{(i)},\,1/(1-z^{(i)}))$,
and, using \eqref{z_and_p}, to turn every cusp condition
into a linear relation of the form $\sum p$'s$\, = 2u_k$.
These relations are then inserted as delta functions in the inner product integral,
enforcing global boundary conditions.
In the end, noting that each matrix element \eqref{S_element1}-\eqref{S_element2}
also contains a delta function, one is left with $N-b_0(\Sigma)$ nontrivial integrals,
where $b_0 (\Sigma)$ is the number of connected components of $\Sigma = \partial M$.
For example, specializing to hyperbolic 3-manifolds with a single torus boundary $\Sigma = T^2$, the integration variables can be relabeled so that Hikami's invariant takes the form
\begin{align}
H (M; \hbar, u)
& = \frac{1}{(4\pi\hbar)^{N/2}} \int\,
\prod_{i=1}^{N} \Phi_{\hbar} \big( \, g_i ({\bf p}, 2u) +\epsilon_i (i \pi + \hbar) \,\big)^{\epsilon_i}
\,e^{f({\bf p}, 2u, \hbar)/2\hbar}\; dp_1 \ldots dp_{N-1}\,.
\label{gen_Hik_integral}
\end{align}
The $g_i$ are linear combinations of $(p_1, \ldots, p_{N-1},2u)$ with integer coefficients,
and $f$ is a quadratic polynomial, also with integer coefficients for all terms involving $p_k$'s or $u$. In the classical limit, this integral can naively be evaluated in a saddle point approximation, and Hikami's claim is that the saddle point relations coincide precisely with the consistency conditions for the triangulation on $M$. There is more to this story, however, as we will see in Section \ref{sec:statesumdef}.


\subsection{Quantum dilogarithm}
\label{sec:qdilog}

Since quantum dilogarithms play a key role here, we take a little time to discuss some of
their most important properties.

Somewhat confusingly, there are at least three distinct---though related---functions
which have occurred in the literature under the name ``quantum dilogarithm'':

$i)$ the function $\Li(x;q)$ defined for $x,\,q\in\C$ with $|x|,\,|q|<1$ by
  \be\label{Li2} \Li(x;q)\=\sum_{n=1}^\infty\frac{x^n}{n\,(1-q^n)}\,, \ee
whose relation to the classical dilogarithm function $\Li(x)=\sum\limits_{n=1}^\infty\dfrac{x^n}{n^2}$ is that
  \be\label{asymp1} \Li\bigl(x;\,e^{2\hbar}\bigr)\;\sim\;-\frac1{2\hbar}\,\Li(x)\qquad\text{as $\hbar\to0\,$;}\ee

$ii)$ the function $(x;q)_\infty$ defined for $|q|<1$ and all $x\in\mathbb C$ by
  \be\label{qPoch} (x;q)_\infty\=\prod_{r=0}^\infty\bigl(1-q^rx)\,,\ee
which is related to $\Li(x;q)$ for $|x|<1$ by
  \be\label{expLi2}(x;q)_\infty\;=\;\exp\bigl(-\Li(x;q)\bigr)\,;\ee
and finally

$iii)$ the function $\Phi(z;\t)$ defined for $\re(\t)>0$ and $2|\re(z)|<1+\re(\t)$ by
   \be\label{qdl_integral} \Phi(z;\t)\=\exp\biggl(\frac14\int_{\Rp}
   \frac{e^{2xz}}{\sinh x\,\sinh\t x}\,\frac{dx}x\biggr) \ee
(here $\Rp$ denotes a path from $-\infty$ to $\infty$ along the
real line but deformed to pass over the singularity at zero), which
is related to $(x;q)_\infty$ by
  \be\label{qdl_quot} \Phi(z;\t)\=\begin{cases}
  \phantom x\dfrac{\big(-\e(z+\t/2);\,\e(\t)\big)_\infty}{\big(-\e((z-1/2)/\t);\,\e(-1/\t)\big)_\infty\vphantom{\int_{R_R}}}
  &\text{if $\im(\t)>0$,} \\ \phantom{xx}
  \dfrac{\vphantom{\int^{\Rp}}\big(-\e((z+1/2)/\t);\,\e(1/\t)\big)_\infty}{\big(-\e(z-\t/2);\,\e(-\t)\big)_\infty}
  & \text{if $\im(\t)<0$.}    \end{cases}\ee
(Here and in future we use the abbreviation $\e(x)=e^{2\pi ix}$.)

It is the third of these functions, in the normalization
  \be \Phi_\hbar(z)\=\Phi\bigl(\frac z{2\pi i};\,\frac{\hbar}{i\pi}\bigr)\,, \label{qdl-hbar} \ee
which occurs in our ``state integral'' and which we will take as our basic ``quantum dilogarithm,''
but all three functions play a role in the analysis, so we will describe the main properties of all three here.
We give complete proofs, but only sketchily since none of this material is new.  For further discussion and proofs,
see, {\it e.g.}, \cite{faddeev-1994, faddeev-1999, faddeev-2001-219, volkov-2003, goncharov-2007},
and \cite{zagDilog} (subsection II.1.D).

\smallskip\noindent{\bf1.}
The asymptotic formula~\eqref{asymp1}  can be refined to the asymptotic expansion
  \be\label{asymp2} \Li\bigl(x;e^{2\hbar}\bigr)\,\sim\,-\frac1{2\hbar}\,\Li(x)-\frac12\log(1-x)
    -\frac x{1-x}\,\frac\hbar{6}+0\hbar^2+\frac{x+x^2}{(1-x)^3}\frac{\hbar^3}{90}+\cdots\ee
as $\hbar\to0$ with $x$ fixed, in which the coefficient of $\hbar^{n-1}$ for $n\ge2$ is the product
of $-2^{n-1}B_n/n!$ (here $B_n$ is the $n$th Bernoulli number) with the negative-index polylogarithm
$\text{Li}_{2-n}(x)\in\mathbb Q\bigl[\frac1{1-x}\bigr]\,$.  More generally, one has the asymptotic formula
  \be\label{asymp3} \Li\bigl(xe^{2\lambda\hbar};\,e^{2\hbar}\bigr)\;\,\sim\;\,
  -\sum_{n=0}^\infty\frac{2^{n-1}B_n(\lambda)}{n!}\,\text{Li}_{2-n}(x)\,\hbar^{n-1}\ee
as $\hbar\to0$ with $\lambda$ fixed, where $B_n(t)$ denotes the $n$th Bernoulli
polynomial.%
\footnote{This is the unique polynomial satisfying $\int_x^{x+1} B_n(t)dt = x^n$, and is a monic polynomial of degree $n$ with constant term $B_n$.} %
Both formulas are easy consequences of the Euler-Maclaurin summation formula. By combining \eqref{qdl_quot}, \eqref{expLi2}, and \eqref{asymp3}, one also obtains an asymptotic expansion
\begin{align} \Phi_\hbar(z+2\lambda\hbar) &= \exp\left(\sum_{n=0}^\infty \frac{2^{n-1}B_{n}(1/2+\lambda)}{n!}\hbar^{n-1}
\text{Li}_{2-n}(-e^z)\right)\,. \label{qdl-hbarexpansion} 
\end{align}
(To derive this, note that in \eqref{qdl_quot} $(-\e(z\pm1/2)/\tau;\e(\pm1/\tau))_\infty \sim 1$
to all orders in $\hbar$ as $\hbar\rightarrow 0$.)

\smallskip\noindent{\bf 2.} The function $(x;q)_\infty$ and its reciprocal have the Taylor
expansions
  \be\label{Taylor} (x;q)_\infty\=\sum_{n=0}^\infty\frac{(-1)^n}{(q)_n}\,q^{\frac{n(n-1)}2}\,x^n\,,
  \qquad\frac1{(x;q)_\infty}\=\sum_{n=0}^\infty\frac1{(q)_n}\,x^n \ee
around $x=0$, where
  \be\label{qPochn} (q)_n\=\frac{\bigl(q;q\bigr)_\infty}{\bigl(q^{n+1};q\bigr)_\infty}
  \=(1-q)(1-q^2)\cdots(1-q^n) \ee
is the $n$th $q$-Pochhammer symbol.  These, as well as formula \eqref{expLi2}, can be proved easily from
the recursion formula $(x;q)_\infty=(1-x)(qx;q)_\infty$, which together with the initial value
$(0;q)_\infty=1$ determines the power series $(x;q)_\infty$ uniquely.  (Of course, \eqref{expLi2} can
also be proved directly by expanding each term in $\sum_r\log(1-q^rx)$ as a power series in~$x$.)
Another famous result, easily deduced from~\eqref{Taylor} using the identity
$\sum\limits_{m-n=k}\dfrac{q^{mn}}{(q)_m(q)_n}=\dfrac1{(q)_\infty}$ for all $k\in\Z$, is the Jacobi triple product formula
  \be\label{triple} (q;q)_\infty\,(x;q)_\infty\,(qx^{-1};q)_\infty
  \=\sum_{k\in\mathbb Z}(-1)^kq^{\frac{k(k-1)}2}\,x^k\,,  \ee
relating the function $(x;q)_\infty$ to the classical Jacobi theta function.

\smallskip\noindent{\bf 3.} The function $\Phi(z;\t)$ defined (initially for $\Re(\t)>0$
and $|\Re(z)|<\frac12+\frac12\Re(\t)$) by~\eqref{qdl_integral} has several functional equations.
Denote by $I(z;\t)$ the integral appearing in this formula. Choosing for $\Rp$ the path
$(-\infty,-\ve]\cup\,\ve\exp([i\pi,0])\,\cup[\ve,\infty)$ and letting $\ve\to0$, we find
  \be\label{split} I(z;\t)\=\frac{2\pi i}\t\,\biggl(\frac{1+\t^2}{12}-z^2\biggr)
    \,+\,2\,\int_0^\infty\biggl(\frac{\sinh 2xz}{\sinh x\;\sinh\t x}\;-\;\frac{2z}{\t x}\biggr)\,\frac{dx}x\;. \ee
Since the second term is an even function of $z$, this gives
   \be\label{parity} \Phi(z;\t)\;\Phi(-z;\t)\=\e\bigl(\frac{\t^2-12z^2+1}{24\t}\bigr)\;. \ee
From \eqref{split} we also get
  \be I(z+1/2;\t)\,-\,I(z-1/2;\t)\=-\frac{4\pi iz}\t\,+\,4\int_0^\infty
    \biggl(\frac{\cosh 2xz}{\sinh\t x}\;-\;\frac1{\t x}\biggr)\,\frac{dx}x\;. \ee
The integral equals $-\log(2\cos(\pi z/\t))$ (proof left as an exercise).
Dividing by~4 and exponentiating we get the first of the two functional equations
 \be\label{lattice}\frac{\Phi(z-1/2;\t)}{\Phi(z+1/2;\t)}\=1\,+\,\e\bigl(z/\t\bigr)\,,
  \qquad\frac{\Phi(z-\t/2;\t)}{\Phi(z+\t/2;\t)}\=1\,+\,\e\bigl(z\bigr)\,,\ee
and the second can be proved in the same way or deduced from the first using the obvious symmetry property
\EPSFIGURE{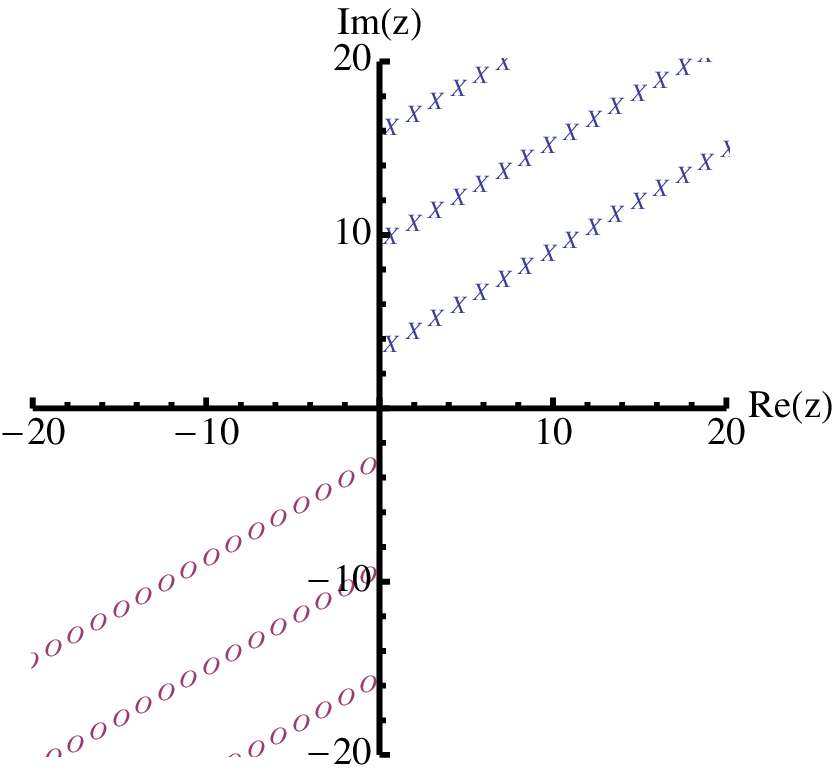,height=6.5cm,angle=0,trim=0 0 0 0}%
{The complex $z$-plane, showing poles ($X$'s) and zeroes ($O$'s) of $\Phi_\hbar(z)$ at $\hbar=\frac{3}{4}e^{i\pi/3}$.\\
\label{fig:minefield} }
  \be\label{scaling} \Phi(z;\t)\=\Phi(z/\t;1/\t)\,, \ee
of the function~$\Phi$. (Replace $x$ by $x/\t$ in~\eqref{qdl_integral}.)

\smallskip\noindent{\bf 4.}  The functional equations \eqref{lattice} show that $\Phi(z;\t)$, which in its
initial domain of definition clearly has no zeros or poles, extends (for fixed $\t$ with $\re(\t)>0)$ to a
meromorphic function of $z$ with simple poles at $z\in\Xi(\t)$ and simple zeros at $z\in-\Xi(\t)$, where
\be\label{poles} \Xi(\t)\=\bigl(\Z_{\ge0}+\tfrac12\bigr)\,\t\;+\;\bigl(\Z_{\ge0}+\tfrac12\bigr)\;\subset\;\C\,.\ee
In terms of the normalization \eqref{qdl-hbar}, this says that $\Phi_\hbar(z)$ has simple poles at $z\in\tilde{\Xi}(\hbar)$ and simple zeroes at $z\in-\tilde{\Xi}(\hbar)$, where
\be \label{polesh} \tilde{\Xi}(\hbar) = (2\Z_{\ge0}+1)i\pi+(2\Z_{\ge0}+1)\hbar\,. \ee
This is illustrated in Figure \ref{fig:minefield}. Equation~\eqref{qdl_quot} expressing~$\Phi$ in terms of the function $(x;q)_\infty$ also follows, because
the quotient of its left- and right-hand sides is a doubly periodic function of~$z$ with no zeros or poles
and hence constant, and the constant can only be $\pm1$ (and can then be checked to be $+1$ in several ways,
e.g., by evaluating numerically at one point) because the right-hand side of~\eqref{qdl_quot} satisfies
the same functional equation~\eqref{parity} as $\Phi(z;\t)$ by virtue of the Jacobi triple product
formula~\eqref{triple} and the well-known modular transformation properties of the Jacobi theta function.

\smallskip\noindent{\bf 5.} From \eqref{split} we also find the Taylor expansion of  $I(z;\t)$ at $z=0$,
  \bes\label{TaylorI}I(z;\t)\= 4\,\sum_{k=0}^\infty C_k(\t)\,z^k\,, \ees
 with coefficients $C_k(\t)=\t^{-k}C_k(1/\t)$ given by
\beas &&C_0(\t)\,=\,\frac{\pi i}{24}\bigl(\t+\t^{-1}\bigr)\,, \qquad
       C_1(\t)\,=\,\int_0^\infty\biggl(\frac1{\sinh x\;\sinh\t x}\,-\,\frac1{\t x^2}\biggr)\,dx\,, \\
  &&C_2(\t)\,=\,-\frac{\pi i}{2\t}\,,\qquad C_k(\t)\=0\quad\text{for $k\ge4$ even},\\
  &&C_k(\t)\=\frac{2^{k-1}}{k!}\,\int_0^\infty\frac{x^{k-1}\;dx}{\sinh x\;\sinh\t x} \quad\text{for $k\ge3$ odd.} \eeas
By expanding $1/\sinh(x)$ and $1/\sinh(\t x)$ as power series in $e^{-x}$ and $e^{-\t x}$ we can evaluate
the last of these expressions to get
  \bes C_k(\t)\=\frac{2^{k+1}}{k!}\sum_{m,n>0,\text{ odd}}\int_0^\infty e^{-mx-n\t x}\,x^{k-1}\,dx
     \=\frac2k\sum_{s\in\Xi(\t)}s^{-k}\quad(\text{$k\ge3$ odd})\ees
with $\Xi(\t)$ as in \eqref{poles}. Dividing by~4 and exponentiating gives the Weierstrass
product expansion
  \be\label{Weierstrass} \Phi(z;\t)\=\exp\biggl(\frac{\pi i}{24}\bigl(\t+\t^{-1}\bigr)\,+\,C_1(\t)\,z
    \,-\,\frac{\pi iz^2}{2\t}\biggr)\prod_{s\in\Xi(\t)}\biggl(\frac{s+z}{s-z}\,e^{-2z/s}\biggr)\ee
of $\Phi(z;\t)$. From this expansion, one finds that $\Phi(z;\t)$ extends meromorphically to
$\C\times\bigl(\C\sm(-\infty,0]\bigr)$ with simple
poles and simple zeros for $z\in\Xi(\t)$ and $z\in-\Xi(\t)$ and no other zeros or poles. (This analytic continuation can also be deduced
by rotating the path of integration in~\eqref{qdl_integral}, e.g.~by replacing $\int_{\Rp}$ by
$\int_{\Rp/\sqrt\t}$ for $z$ sufficiently small.)

\smallskip\noindent{\bf 6.}  The quantum dilogarithm is related via the Jacobi triple product
formula to the Jacobi theta function, which is a Jacobi form, i.e., it has transformation properties
not only with respect to the lattice translations $z\mapsto z+1$ and $z\mapsto z+\t$ but also with
respect to the modular transformations $\t\mapsto\t+1$ and $\t\mapsto-1/\t$.  The function $\Phi(z;\t)$
has the lattice transformation properties~\eqref{lattice} and maps to its inverse under
$(z,\t)\mapsto(z/\t,\,-1/\t)$, but it does not transform in a simple way with respect
to $\t\mapsto\t+1$.  Nevertheless, it has an interesting modularity property of a different
kind (cocycle property) which is worth mentioning here even though no use of it will be made
in the remainder of this paper.  Write~\eqref{qdl_quot} as
  \bes\label{StoPhi} \Phi(z;\t)\=\frac{S(z;\t)}{S(z/\t;-1/\t)}\,,\qquad S(z;\t)\=\begin{cases}
    \prod_{\text{$n>0$ odd}}\bigl(1+q^{n/2}\e(z)\bigr)&\text{ if $\im(\t)>0$}, \\
    \prod_{\text{$n<0$ odd}}\bigl(1+q^{n/2}\e(z)\bigr)^{-1}&\text{ if $\im(\t)<0$}, \end{cases} \ees
where $q=\e(\t)$.  The function $S(z;\t)$ has the transformation properties
  \bes S(z;\t)=S(z+1;\t)=(1+q^{1/2}\e(z))\,S(z+\t;\t)=S(z+\tfrac12;\t+1)=S(z;\t+2) \ees
amd from these we deduce by a short calculation the two three-term functional equations
 \be \Phi(z;\t)\=\Phi\bigl(z\pm\frac12,\t+1\bigr)\,\Phi\bigl(\frac{z\mp\t/2}{\t+1},\,\frac\t{\t+1}\bigr) \ee
of $\Phi$.  This is highly reminiscent of the fact (cf.~\cite{LewisZag}) that the holomorphic function
  \bes \psi(\t)\=f(\t)-\t^{-2s}f(-1/\t)\,,\qquad f(\t)\=\begin{cases}\phantom{-}
   \sum_{n>0}a_nq^n&\text{ if $\im(\t)>0$,} \\-\sum_{n<0}a_nq^n&\text{ if $\im(\t)<0$} \end{cases} \ees
associated to a Maass cusp form $u(\t)$ on $SL(2,\Z)$ with spectral parameter~$s$, where $a_n$ are the
normalized coefficients in the Fourier-Bessel expansion of $u$, satisfies the Lewis functional equation
$\psi(\t)=\psi(\t+1)+(\t+1)^{-2s}\psi\bigl(\frac\t{\t+1}\bigr)$ and extends holomorphically from its initial
domain of definition $\C\sm\R$ to $\C\sm(-\infty,0]\,$.

\smallskip\noindent{\bf 7.}  Finally, the quantum dilogarithm functions
satisfies various five-term relations, of which the classical five-term functional
equation of $\Li(x)$ is a limiting case, when the arguments are non-commuting variables.
The simplest and oldest is the identity
\be (Y;q)_\infty\,(X;q)_\infty\=(X;q)_\infty\,(-YX;q)_\infty\,(Y;q)_\infty \label{pre-pentagon}\ee
for operators $X$ and $Y$ satisfying $XY=qYX$. From this one deduces the ``quantum pentagon relation''
  \be \label{qdl_pentagon} \Phi_\hbar(\hat{p})\,\Phi_\hbar(\hat q)
  \= \Phi_\hbar(\hat q)\,\Phi_\hbar(\hat p+\hat q)\,\Phi_\hbar(\hat p) \ee
for operators $\hat{p}$ and $\hat{q}$ satisfying $[\hat{q},\hat{p}]=2\hbar$.
Letting ${\bf S}_{ij}$ be a copy, acting on the $i^{{\rm th}}$ and $j^{{\rm th}}$
factors of $V\otimes V\otimes V$, of the ${\bf S}$-matrix introduced in \eqref{S-matrix},
we deduce from \eqref{qdl_pentagon} the operator identity
  \be\label{S_pentagon}
  {\bf S}_{23} {\bf S}_{12} \= {\bf S}_{12} {\bf S}_{13} {\bf S}_{23}\,.\ee
It is this very special property which guarantees that the gluing procedure
used in the definition of \eqref{gen_Hik_integral}
is invariant under 2-3 Pachner moves on the underlying triangulations and produces a true hyperbolic invariant
\cite{hikami-2001-16,hikami-2006,kashaev-2000}.
This identity is also related to
the fact (not used in this paper) that the fifth power of the operator which
maps a nicely behaved function to the Fourier transform of its product
with $\Phi_\hbar(z)$ (suitably normalized) is a multiple of the identity~\cite{goncharov-2007}.


\subsection{A state integral model for $Z^{(\rho)} (M;\hbar)$}
\label{sec:statesumdef}

Now, let us return to the analysis of the integral \eqref{gen_Hik_integral}
and compare it with the perturbative $SL(2,\C)$ invariant $Z^{(\alpha)} (M;\hbar,u)$.
Both invariants compute quantum ({\it i.e.} $\hbar$\,-\,deformed)
topological invariants of hyperbolic 3-manifolds and, thus,
are expected to be closely related.
However, in order to establish a precise relation,
we need to face two problems mentioned in the beginning of this section:

$i)$ the integration contour is not specified in \eqref{gen_Hik_integral}, and

$ii)$ the integral \eqref{gen_Hik_integral} does not depend on the choice
of the classical solution $\alpha$.

\noindent
These two problems are related, and can be addressed by studying
the integral \eqref{gen_Hik_integral} in various saddle point approximations.
Using the leading term in \eqref{qdl-hbarexpansion}, we can approximate it to leading order as
\be
H (M;\hbar, u) \; \underset{\hbar \to 0}{\sim} \;
 \int\,e^{\frac{1}{\hbar} V(p_1, \ldots, p_{N-1}, u)}\; dp_1 \ldots dp_{N-1}\,,
\label{Hik_V_limit}
\ee
with the ``potential''
\be
V({\bf p},u) \= \frac{1}{2} \sum_{i=1}^N \epsilon_i \Li(-\exp(g_i({\bf p},2u) + i\pi \epsilon_i))+\frac{1}{2} f({\bf p},2u,\hbar=0)\,.
\label{Hik_gen_V}
\ee
As explained below \eqref{qdl-hbarexpansion}, the branches of $\Li$ must be chosen appropriately to coincide with the half-lines of poles and zeroes of the quantum dilogarithms in \eqref{gen_Hik_integral}. 
The leading contribution to $H (M;\hbar, u)$ will then come from
the highest-lying critical point through which a given contour can be deformed.

It was the observation of \cite{hikami-2006} that the potential $V$
always has one critical point that reproduces the classical Chern-Simons action
on the ``geometric'' branch,
that is a critical point ${\bf p}^{({\rm geom})} (u)$ such that
\be
l_{{\rm geom}} (u) = \exp \left[ \textstyle \frac{d}{du} V({\bf p}^{({\rm geom})}(u), u) \right]
\label{lVgeom}
\ee
and
\be
S_0^{({\rm geom})} (u) = V({\bf p}^{({\rm geom})} (u), u)\,.
\label{SVgeom}
\ee
This identification follows from the fact that both $S_0^{({\rm geom})} (u)$
and matrix elements \eqref{S_element1}-\eqref{S_element2} in the limit $\hbar \to 0$
are related to the complexified volume function
$i(\Vol (M;u) + i {\rm CS} (M;u))$.
We want to argue presently that in fact {\it every} critical point of $V$
corresponds to a classical solution in Chern-Simons theory
(that is, to a branch of $A(l,m)=0$) in this manner,
with similar relations
\be
l_{\alpha} (u) =  e^{\frac{d}{du} V({\bf p}^{(\alpha)} (u), u)}
\label{lValpha}
\ee
and
\be
S_0^{(\alpha)} (u) = V({\bf p}^{(\alpha)} (u), u)
\label{SValpha}
\ee
for some $\alpha$.
In particular, $l_{\alpha} (u)$ as given by \eqref{lValpha} and $m=e^u$ obey \eqref{abranches}.

To analyze generic critical points of $V$, observe that the critical point equations
take the form
\be
2 \frac{\partial}{\partial p_j} V({\bf p},u) =
-\sum_{i=1}^{N} \epsilon_i G_{ji} \log(1+\exp(g_i({\bf p},2u)+ i\pi \epsilon_i))
+ \frac{\partial}{\partial p_j} {f} ({\bf p},2u,0) = 0 \quad\;\; \forall\; j\,,
\label{p_saddles}
\ee
where $G_{ji} = \frac{\partial}{\partial p_j} g_i({\bf p},2u)$ are some constants and
the functions $\frac{\partial}{\partial p_j} {f} ({\bf p},2u,0)$ are linear. Again, the cuts of the logarithm must match the singularities of the quantum dilogarithm.
Exponentiating \eqref{p_saddles}, we obtain another set of conditions
\be
r_j ({\bf x},m) = 1 \quad \forall\; j\,,
\label{x_saddles}
\ee
where
\be
r_j({\bf x},m) = \exp \Big( {2 \frac{\partial}{\partial p_j} V({\bf p},u)}\Big)
= \prod_i (1-e^{g_i})^{-\epsilon_i G_{ji}}\exp\Big({\frac{\partial}{\partial p_j}f}\Big)
\ee
are all {\it rational} functions of the variables $x_j = e^{p_j}$ and $m = e^u$. Note that an entire family of points $\{{\bf p} + 2\pi i {\bf n} \,|\, {\bf n} \in \Z^{N-1} \}$ maps to a single $\bf x$, and that all branch cut ambiguities disappear in the simpler equations \eqref{x_saddles}. Depending on $\arg(\hbar)$ and the precise form of $f$, solutions to \eqref{x_saddles} either ``lift'' uniquely to critical points of the potential $V$, or they lift to a family of critical points at which $V$ differs only by integer multiples of $2\pi i u$.

Now, the system \eqref{x_saddles} is algebraic,
so its set of solutions defines a complex affine variety
\be
{\cal R} = \{ (x_1, \ldots, x_{N-1},m) \in \C^N \,|\, r_j(x_1, \ldots, x_{N-1},m) = 0 \quad \forall\, j\}\,,
\ee
which is closely related to the representation variety $\CL$ given by $A(l,m)=0$.
Both generically have complex dimension one.
Noting that $s(x_1, \ldots, x_{N-1}, m) = \exp \big( \frac{\partial}{\partial u} V \big)$
is also a rational function, we can define a rational map $\phi:\C^N \to \C^2$ by
\be
\phi (x_1, \ldots, x_{N-1}, m) = (s(x_1, \ldots, x_{N-1},m),m)\,.
\ee
The claim in \cite{hikami-2006} that one critical point of $V$
always corresponds to the geometric branch of $\CL$
means that $\overline{\phi(\cal R)}$ (taking an algebraic closure)
always intersects $\CL$ nontrivially, along a subvariety of dimension 1.
Thus, some irreducible component of $\overline{\phi(\cal R)}$,
coming from an irreducible component of $\cal R$,
must coincide with the entire irreducible component of $\CL$ containing the geometric branch.
Every solution ${\bf x}= {\bf x}(m)$ in this component of $\cal R$
corresponds to a branch of the A-polynomial.
Moreover, if such a solution ${\bf x}^{(\alpha)}$ (corresponding to branch $\alpha$)
can be lifted to a real critical point ${\bf p}^{(\alpha)} (u)$ of $V$,
then one must have relations \eqref{lValpha} and \eqref{SValpha}.

This simple algebraic analysis shows that {\it some} solutions of \eqref{x_saddles}
will cover an entire irreducible component of the curve $\CL$ defined by $A(l,m)=0$.
We cannot push the general argument further without knowing more about the reducibility of~$\cal R$.
However, we can look at some actual examples.
Computing $V ({\bf p},u)$ for thirteen hyperbolic manifolds with a single torus boundary,%
\footnote{Namely, the complements of hyperbolic knots
${\bf 4_1}$(${\bf k2_{1}}$), ${\bf 5_2}$(${\bf k3_{2}}$),
${\bf 12n_{242}}$ (${\bf 3_{1}}$, (-2,3,7)-pretzel knot),
${\bf 6_1}$(${\bf k4_{1}}$), ${\bf 6_3}$(${\bf k6_{43}}$),
${\bf 7_2}$(${\bf k4_{2}}$), ${\bf 7_3}$(${\bf k5_{20}}$),
${\bf 7_4}$(${\bf k6_{28}}$), ${\bf 10_{132}}$(${\bf K5_{9}}$),
${\bf 10_{139}}$(${\bf K5_{22}}$), and ${\bf 11n_{38}}$(${\bf K5_{13}}$),
as well as the one-punctured torus bundles $L^2R$ and $LR^3$ over ${\bf S}^1$
(also knot complements, but in a manifold other than ${\bf S}^3$).} %
we found in every case that solutions
of \eqref{x_saddles} { completely} covered all non-abelian branches $\alpha \neq {\rm abel}$; in other words,
$\overline{\phi(\cal R)} = \CL'$, with $\CL' = \{ A(l,m) / (l-1)=0 \}$.
For six of these manifolds, we found unique critical points
${\bf p} (u)$ corresponding to every non-abelian branch of $\CL$ at $\arg(\hbar)=i\pi$. Motivated by these examples, it is natural to state the following conjecture:\\

\noindent \textbf{Conjecture 2}: {\it
Every critical point of $V$ corresponds to some branch $\alpha$,
and all $\alpha \neq {\rm abel}$ are obtained in this way.
Moreover, for every critical point ${\bf p}^{(\alpha)}(u)$
(corresponding to some branch $\alpha$) we have
\eqref{lValpha}-\eqref{SValpha} and to all orders in perturbation theory:
\be
Z^{(\alpha)} (M; \hbar, u)
= \sqrt{2} \int_{C_{\alpha}}
\prod_{i=1}^{N} \Phi_{\hbar} \big( \, g_i ({\bf p}, 2u) +\epsilon_i (i \pi + \hbar) \,\big)^{\epsilon_i}\,
e^{\frac{1}{2\hbar} f({\bf p}, 2u, \hbar) - u} \; \prod_{j=1}^{N-1} \frac{dp_j}{\sqrt{4\pi\hbar}}\,,
\label{statesumdef}
\ee
where $C_{\alpha}$ is an arbitrary contour with fixed endpoints
which passes through ${\bf p}^{(\alpha)}(u)$ and no other critical point.}\\

A slightly more conservative version of this conjecture might state
that only those $\alpha$ that belong to the same irreducible component of $\CL$
as the geometric branch, $\alpha = {\rm geom}$, are covered by critical points of $V$.
Indeed, the ``abelian'' branch with $l_{{\rm abel}} = 1$ is not covered
by the critical points of $V$ and it belongs
to the separate component $(l-1)$ of the curve $A(l,m)=0$.
It would be interesting to study the relation between critical points of $V$
and irreducible components of $\CL$ further, in particular by looking at
examples with {reducible} A-polynomials aside from the universal $(l-1)$ factor.

The right-hand side of \eqref{statesumdef} is the proposed state integral model
for the {\it exact} perturbative partition function of $SL(2,\C)$ Chern-Simons
theory on a hyperbolic 3-manifold $M$ with a single torus boundary $\Sigma = T^2$.
(A generalization to 3-manifolds with an arbitrary number of boundary components
is straightforward.)
This state integral model is a modified version of Hikami's invariant \eqref{gen_Hik_integral}.
Just like its predecessor, eq. \eqref{statesumdef} is based
on an ideal triangulation $\{\Delta_i\}_{i=1}^N$
of a hyperbolic 3-manifold $M$ and inherits topological invariance from
the pentagon identity \eqref{qdl_pentagon} of the quantum dilogarithm.

However, in writing \eqref{statesumdef} we made two important
modifications to Hikami's invariant \eqref{gen_Hik_integral}.
First, we introduced contours $C_{\alpha}$ running across the sadle points ${\bf p}_\alpha$, which now encode the choice
of a classical solution in Chern-Simons theory.
Second,  in \eqref{statesumdef}
we introduced an extra factor of $\sqrt{8 \pi \hbar} e^{-u}$, which is needed
to reproduce the correct asymptotic behavior of $Z^{(\alpha)} (M; \hbar, u)$.
To understand this correction factor, we must look at the higher-order terms in the expansion of $Z^{(\alpha)}(M;\hbar,u)$.
By using \eqref{qdl-hbarexpansion}, one can continue the saddle point
approximations described above to arbitrary order in $\hbar$.
The result has the expected form \eqref{zpert},
\be
Z^{(\alpha)} (M;\hbar,u) =
\exp\left( \frac{1}{\hbar} S_0^{(\alpha)} (u) - \frac{1}{2}\delta^{(\alpha)} \log \hbar
+ \sum_{n=0}^\infty S_{n+1}^{(\alpha)} (u) \hbar^n  \right),
\label{zssspert}
\ee
with the correct leading term $S_0^{(\alpha)} (u)$
that we already analyzed above, {\it cf.} eq. \eqref{SValpha}.

Let us examine the next-leading logarithmic term.
Its coefficient $\delta^{(\alpha)}$ receives contributions from
two places: from the prefactor $(4\pi \hbar)^{-(N-1)/2}$ in \eqref{statesumdef},
and from the standard Gaussian determinant.
The former depends on the total number of tetrahedra, $N$,
in the triangulation of $M$ and therefore must be cancelled (at least partially)
since the total integral \eqref{statesumdef} is a topological invariant
and cannot depend on $N$. This is indeed what happens.
For example, in a saddle point approximation around a nondegenerate
critical point ${\bf p}^{(\alpha)}(u)$, the contribution of
the Gaussian determinant goes like $\sim \hbar^{(N-1)/2}$
and exactly cancels the contribution of the prefactor $\sim \hbar^{- (N-1)/2}$.
An example of such critical point is the critical point ${\bf p}^{({\rm geom})} (u)$
corresponding to the geometric branch.
Therefore, the asymptotic expansion of
the integral \eqref{statesumdef} around the critical point ${\bf p}^{({\rm geom})} (u)$
has the form \eqref{zssspert} with
\be
\delta^{({\rm geom})} = 0\,,
\label{dgeomzero}
\ee
which is the expected result.\footnote{Recall that throughout this paper
$Z^{(\rho)} (M;\hbar)$ (resp. $Z^{(\alpha)} (M;\hbar,u)$)
stands for the {\it unnormalized} perturbative $G_{\C}$ invariant.
A normalized version, obtained by dividing by $Z ({\bf S}^3)$,
has an asymptotic expansion of the same form \eqref{zpert} (resp. \eqref{zssspert})
with the value of $\delta^{(\alpha)}$ shifted by $\dim (G)$ for every $\alpha$.
This is easy to see from \eqref{zsthree}.}
Indeed, as explained {\it e.g.} in \cite{gukov-2003,gukov-2006},
the rigidity of the flat connection $\CA^{({\rm geom})}$ associated
with a hyperbolic structure on $M$ implies $h^0 = h^1 = 0$,
so that \eqref{dhh} gives $\delta^{({\rm geom})} = 0$.

Using \eqref{qdl-hbarexpansion}\,, one can also calculate the higher-order perturbative
coefficients $S_n^{(\alpha)}(u)$, with $n \ge 1$.
In the following section, we carry out this analysis to
high order for the figure-8 knot complement and find perfect agreement
with the results obtained by methods of Section~\ref{sec:pert}.
(Other interesting examples and further checks will appear elsewhere \cite{dimofte-2009}.)

Note that the $S_n^{(\alpha)}(u)$'s do not depend on the details of the contours $C_{\alpha}$.
The only part of \eqref{statesumdef} which actually depends on the details of $C_\alpha$
is 
exponentially suppressed and is not
part of the perturbative series \eqref{zssspert}.
Finally, we also note that
we use only those critical points of the full integrand \eqref{statesumdef}
which correspond to critical points of $V$ in the limit $\hbar \to 0$.
For any fixed $\hbar > 0$, the actual integrand has many other critical
points which become trapped in the half-line singularities of quantum dilogarithms
as $\hbar \to 0$,
so the integrals over them do not have a well-behaved limit.

We conclude this section by observing that Conjecture 2 implies Conjecture 1. From \eqref{qdl-hbarexpansion}, we can write an asymptotic double series expansion (for very small $p$):
\begin{align}
\Phi_{\hbar} (p_0 + p) &= \exp\left( \sum_{n=0}^\infty  B_n\bigl(\frac12+\frac
p{2\hbar}\bigr)\,\text{Li}_{2-n}\bigl(-e^{p_0}\bigr)
\,\frac{(2\hbar)^{n-1}}{n!}\right) \notag \\
 &= \exp \left( \sum_{k=-1}^\infty \sum_{j=0}^\infty
\frac{B_{k+1}(1/2)\,2^k}{(k+1)! j!} \text{Li}_{1-j-k} (-e^{p_0})\,\hbar^k p^j \right).
\label{qdl_exp}
\end{align}
Using this formula and taking into account the shifts by $\pm(i\pi+\hbar)$,
we expand every quantum dilogarithm appearing in
the integrand of \eqref{statesumdef} around a critical point ${\bf p}^{(\alpha)}$.
At each order in $\hbar$, the state integral model then reduces to an
integral of a polynomial in~${\bf p}$ with a Gaussian weight.
Due to the fact that $\text{Li}_k$ is a rational function for $k\le 0$,
the coefficients of these polynomials are all rational functions of
the variables ${\bf x}^{(\alpha)} = \exp ({\bf p}^{(\alpha)})$ and $m$.
Therefore, the resulting coefficients $S_n^{(\alpha)}(m)$, for $n>1$,
will also be rational functions of ${\bf x}^{(\alpha)}$ and $m$.
At $m=1$, the solutions ${\bf x}^{(\alpha)} (m)$ to the rational equations \eqref{x_saddles}
all belong to some algebraic number field $\mathbb{K} \subset \overline{\mathbb{Q}}$,
leading immediately to Conjecture 1.
In particular, for the geometric branch $\alpha = {\rm geom}$,
the field $\mathbb{K}$ is nothing but the trace field~$\mathbb{Q} (\tr \Gamma)$.


\section{Examples and computations}
\label{sec:example}

In the previous sections, we described several ways to approach perturbative
Chern-Simons theory with complex gauge group.
Our discussion was guided by the aim of computing the all-loop perturbative partition
function $Z^{(\rho)} (M;\hbar)$ in the background of a nontrivial
flat connection $\CA^{(\rho)}$ labeled by a homomorphism $\rho: \pi_1 (M) \to G_{\C}$.

In this section, our goal is to illustrate these methods in concrete examples
by calculating perturbative invariants $S_n^{(\rho)}$ (also denoted $S_n^{(\alpha)}(u)$)
to high loop order.
Which 3-manifolds shall we choose for our examples?
On one hand, the approach based on quantization and analytic continuation
most directly applies to 3-manifolds with boundary, such as knot complements.
(The invariants of closed 3-manifolds without boundary can be obtained
by gluing two manifolds, $M_+$ and $M_-$, along a common boundary $\Sigma$,
as in~\eqref{mmgluing}.)
On the other hand, the state integral model described in Section~\ref{sec:statesum}
does not require $M$ to have a boundary,
but in the present form applies only to hyperbolic 3-manifolds.
(We believe this state integral model can be extended to arbitrary 3-manifolds,
but we do not pursue it here.)
Therefore, if we wish to compare the results of different methods,
it is convenient to choose examples of 3-manifolds
that are knot complements {\it and} hyperbolic at the same time.
In other words, we take $M$ to be a complement of a hyperbolic knot $K$ in the 3-sphere:
\be
M = {\bf S}^3 \sm K\,.
\ee
Among hyperbolic knots with a small number of crossings,
the simplest ones are the figure-eight knot (also denoted ${\bf 4_1}$)
and the three-twist knot (also denoted ${\bf 5_2}$) shown in Figure~\ref{fig:knots}.
Among all hyperbolic knots, the complement of the figure-eight knot
has the least possible volume $\Vol ({\bf 4_1}) = 2.0298832128...$,
whereas the volume of the ${\bf 5_2}$ knot complement is $\Vol ({\bf 5_2}) = 2.8281220883...$.
In both cases, the knot group $\pi_1 (M)$ is generated by two elements $a$ and $b$,
such that $a^{-1} b a b^{-1} a = ba^{-1} ba b^{-1}$ for the figure-eight knot
and $waw^{-1}=b$, with $w = a^{-1} b a^{-1} b^{-1} a b^{-1}$, for the ${\bf 5_2}$ knot.
For the figure-eight knot, the trace field $\mathbb{Q}(\tr\Gamma)$
is the imaginary quadratic field $\mathbb{Q} (\sqrt{-3})$, while for the ${\bf 5_2}$ knot
it is a cubic field with one complex place and discriminant $-23$.

\EPSFIGURE{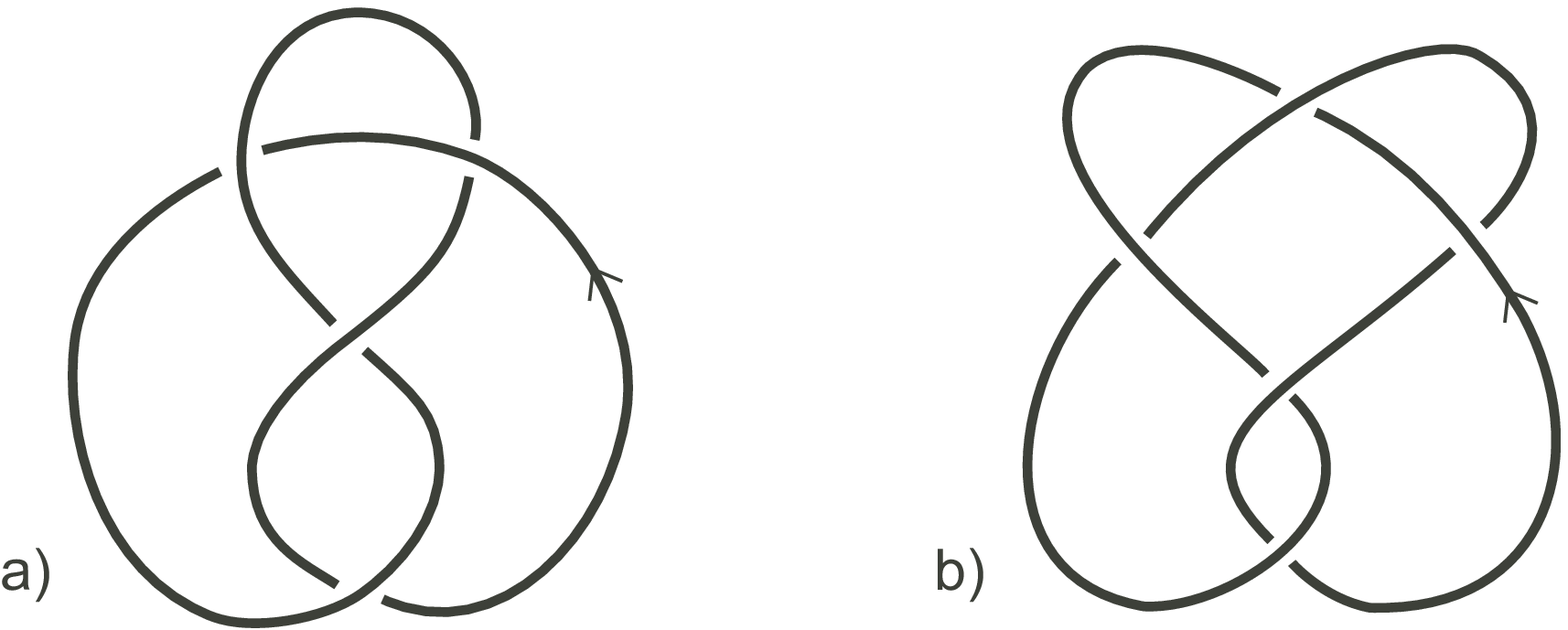,height=4cm,angle=0,trim=0 0 0 0}%
{$a)$ figure-8 knot ${\bf 4_1}$, and $b)$ three-twist knot ${\bf 5_2}$.
\label{fig:knots} }

This section consists of two parts.
In the first part (Section~\ref{sec:arith_example}), we fix the homomorphism $\rho = {\rm geom}$
and compute the perturbative $SL(2,\C)$ invariants $S_n^{({\rm geom})}$
for the simplest hyperbolic knot complements, $M = {\bf S}^3 \sm {\bf 4_1}$ and $M = {\bf S}^3 \sm {\bf 5_2}$.
In particular, we shall verify the arithmeticity conjecture formulated in Section~\ref{sec:loops}.
In the second part (Section~\ref{sec:figeight}),
we look more closely at the perturbative invariants of the figure-eight knot complement
and study how the $S_n^{(\rho)}$'s depend on the homomorphism $\rho$
(equivalently, we study each $S_n^{(\alpha)}(u)$ as a function of $u$ and $\alpha$).


\subsection{Numerical verification of the arithmeticity conjecture}
\label{sec:arith_example}

In this section, we explicitly calculate the perturbative $SL(2,\C)$ invariants
$S_n^{(\alpha)} (u)$ for the simplest hyperbolic knot complements and
for the special value of $u=0$ (corresponding to the discrete faithful representation $\rho = {\rm geom}$).
It was conjectured in~\cite{gukov-2003} that such invariants (in fact, for arbitrary value of $u$)
can be extracted from the asymptotic expansion of knot invariants computed by Chern-Simons theory
with compact gauge group $G$ in the double-scaling limit~\eqref{dslimit}.
In the present case, $G=SU(2)$ and the relevant knot invariant is
\be \CJ(K;N) =  \frac{J_N\bigl(K;\,e^{2\pi i/N}\bigr)}{J_N\bigl({\rm unknot};\,e^{2\pi i/N}\bigr)}\,, \ee
where $J_N(K;q)$ is the ``$N$-colored Jones polynomial'' of $K$,
normalized such that
\be
J_N ({\rm unknot};q) = \frac{q^{N/2}-q^{-N/2}}{q^{1/2}-q^{-1/2}}
\ee
is the ``quantum dimension'' of the $N$-dimensional representation of $SU(2)$.
As explained in the Introduction and then in more detail in Section \ref{sec:analytic},
the $N$-colored Jones polynomial $J_N(K;q)$ is computed by the Feynman path integral
in Chern-Simons gauge theory on~${\bf S}^3$ with a Wilson loop operator, $W_R (K)$,
supported on $K$ and labeled by the $N$-dimensional representation $R$ of $SU(2)$, {\it cf.}~\eqref{wilsonp}.
According to the Generalized Volume Conjecture~\cite{gukov-2003},
the invariant $\CJ(K;N)$ should have the asymptotics\footnote{Note that in our notations $e^{2\hbar}=q=e^{2\pi i/N}$.}
  \be\label{JKn} \CJ(K;N) \;\sim\; \frac{ Z^{\text{(geom)}}(M;\,i\pi/N)}
  {Z({\bf S}^1 \times {\bf D}^2;\,i\pi/N)}\;\sim\;N^{3/2}\, \exp\left( \sum_{n=0}^\infty s_n \biggl(\frac{2\pi i}N\biggr)^{n-1} \right)
  \ee
as $N\to\infty$, where
\begin{align}
S_n^{(\rm geom)}(0) & = s_n \cdot 2^{n-1} \qquad(n \ne 1) \label{Ss} \\
S_1^{(\rm geom)}(0) & = s_1 + \frac{1}{2} \log 2 \qquad(n=1) \nonumber
\end{align}
are the perturbative $SL(2,\C)$ invariants of $M = {\bf S}^3 \sm K$.
(Here, we view the solid torus, ${\bf S}^1 \times {\bf D}^2$, as the complement
of the unknot in the 3-sphere.) Specifically, we have
\be\label{VolConj} s_0\,=\,i\,\bigl(\text{Vol}(M)\,+\,i\,{\rm CS}(M)\bigr)\ee
(Volume Conjecture) and $s_1$ is the Ray-Singer torsion of $M$ twisted by a flat connection, {\it cf.} eq.~\eqref{sonetorsion}.
The arithmetic TQFT conjecture described in Section \ref{sec:loops} predicts that
\be\label{ArithConj} s_1\;\in\;\mathbb Q\cdot\log\mathbb K\,,\quad\qquad s_n\;\in\;\mathbb K\qquad(n\ge2),\ee
where $\mathbb K$ is the trace field of the knot. In this subsection, we will give numerical
computations supporting this conjecture for the two simplest hyperbolic knots ${\bf 4_1}$ and ${\bf 5_2}$.

The formulas for $\CJ(K;N)$ in both cases are known explicitly, see {\it e.g.} \cite{kashaev-1997}. We have
\be\label{Jones41} \CJ({\bf 4_1};N)=\sum_{m=0}^{N-1}(q)_m(q^{-1})_m\,, \ee
\be\label{Jones52} \CJ({\bf 5_2};N) = \sum_{m=0}^{N-1}\sum_{k=0}^m q^{-(m+1)k}(q)_m^2/(q^{-1})_k\,,\ee
where $q=e^{2\pi i/N}$ and
$(q)_m=(1-q)\cdots(1-q^m)$ is the $q$-Pochhammer symbol as in Section~\ref{sec:qdilog}.
The first few values of these invariants are
\begin{equation*}\label{tab:jntwoknots}
\renewcommand{\arraystretch}{1.3}
\begin{tabular}{|c||c|c|c| }
\hline  $N$& $\CJ({\bf 4_1};N)$& $\CJ({\bf 5_2};N)$
\\
\hline 1  & $ 1$ & $1$
\\
\hline 2  & $5$ & $7$
\\
\hline 3  & $13$ & $18-5q$
\\
\hline 4  & $27$ & $40-23q$
\\
\hline 5  & $44-4q^2-4q^3$ & $46-55q-31q^2+q^3$
\\
\hline 6  & $89$ & $120-187q$
\\
\hline 7  & $100-14q^2-25q^3-25q^4-14q^5$ & $-154q-88q^2+47q^3+58q^4+77q^5$
\\
\hline 8  & $187-45q-45q^3$ & $-84-407q-150q^2+96q^3$
\\
\hline
\end{tabular}
\end{equation*}

\medskip
Using formula~\eqref{Jones41} for $N$ of the order of 5000 and the numerical interpolation method
explained in \cite{zagVass} and \cite{GruenMor}, we computed the values of
$s_n$ for $0\le n\le 27$ to very high precision, finding that \eqref{JKn} holds with
\be\label{FirstTwo} s_0 =\frac1{2\pi^2i}\,D\bigl(e^{\pi i/3}\bigr),\qquad s_1=-\frac14\,\log3\,, \ee
the first in accordance with the volume conjecture and the second in accordance with
the first statement of~\eqref{ArithConj}, since $\mathbb K=\mathbb Q\bigl(\sqrt{-3}\bigr)$
in this case, and that the numbers
\be\label{rational} s'_n=s_n\cdot\bigl(6\sqrt{-3}\bigr)^{n-1}\qquad(n\ge2) \ee
were very close to rational numbers with relatively small and highly factored denominators:
\bes\label{tab:sjprime}
\renewcommand{\arraystretch}{1.3}
\begin{tabular}{|c||c|c|c|c|c|c|c|c| }
\hline  $n$ & $2$ & $3$ & $4$ & $5$ & $6$ & $7$ & $8$ & $9$
\\
\hline $s'_n$ &$\frac{11}{12}$ & 2 & $\frac{1081}{90}$ &
 98 & $\frac{110011}{105}$ & $\frac{207892}{15}$ & $\frac{32729683}{150}$ & $\frac{139418294}{35}$
\\
\hline
\end{tabular}
\ees
\bes\label{tab:sjprimecontd}
\renewcommand{\arraystretch}{1.3}
\begin{tabular}{|c||c|c|c| }
\hline  $n$ & $10$ & $\quad\dots\quad$ & $27$
\\
\hline $s'_n$ & $\frac{860118209659}{10395}$ & $\quad\dots\quad$ &
$\frac{240605060980290369529478710291172763261781986098552}{814172781296875}$
\\
\hline
\end{tabular}
\ees
confirming the second prediction in~\eqref{ArithConj}.
Here $D(z)$ is the Bloch-Wigner dilogarithm \eqref{BW}.
(That $s_n$ is a rational multiple of $(\sqrt{-3})^{n-1}$,
and not merely an element of $\mathbb Q(\sqrt{-3})$
is a consequence of~\eqref{JKn} and the fact that $\CJ({\bf 4_1};N)$ is real.)

Actually, in this case one can prove the correctness of the expansion rigorously: the two
formulas in~\eqref{FirstTwo} were proved in~\cite{andersen-hansen} and the rationality
of the numbers $s'_n$ defined by~\eqref{rational} in \cite{garoufalidis-costin, garoufalidis-geronimo}
and \cite{zagHans}, and can therefore check that the numerically determined
values are the true ones (see also \cite{zagMod} for a generalization of this analysis).
In the case of ${\bf 5_2}$, such an analysis has not been done,
and the numerical interpolation method is therefore needed.  If we try to do this directly
using eq.~\eqref{Jones52}, the process is very time-consuming because, unlike the figure-8
case, there are now O($N^2$) terms.  To get around this difficulty, we use the formula
\be\label{idA}
\sum_{k=0}^m\frac{q^{(m+1)k}}{(q)_k}=(q)_m\,\sum_{k=0}^m\frac{q^{k^2}}{(q)_k^2}\,,\ee
which is proved by observing that both sides vanish for $m=-1$ and satisfy the recursion
$t_m=(1-q^m)t_{m-1}\,+\,q^{m^2}/(q)_m\,$. This proof gives a way to successively compute
each $t_m$ in $\,\text O(1)\,$ steps (compute $y=q^m$ as $q$ times the previous $y$, $(q)_m$
as $1-y$ times its previous value, and then $t_m$ by the recursion) and hence to
compute the whole sum in~\eqref{Jones52} in only $\,\text O(N)\,$ steps.
A short PARI program for this is

\bigskip
{\obeylines\font\small=cmr10 scaled \magstephalf\parindent=0pt\hangafter=0\raggedright\tt
\hskip0pt \qquad $\{$f(x) = q=exp(2*Pi*I*x);y=1;p=1;s=1;t=1;\;
\hskip0pt  $\qquad$ for(n=1,denom(x)-1,y*=q;p*=1-y;t+=y\^{\kern.2pt}n/p\^{\kern.2pt}2;s+=p\^{\kern.2pt}2*conj(p*t)); s$\}$  \par\ \leftskip=0pt }

\noindent
and is very much faster than the computation based on the original formula.  For instance, to compute
$\CJ({\bf 5_2};N)$ to 200 digits for $N=100$ and $N=1000$ using~\eqref{Jones52} takes  1.6 seconds and
262 seconds, respectively, whereas with the improved algorithm these times are reduced to 0.03 seconds
and 0.4 seconds. The interpolation method can therefore be carried out
to just as high precision as in the figure-8 case.

The results are as follows.  The first coefficient is given to high precision by
\be s_0=-\frac3{2\pi}\,\bigl(\text{Li}_2(\a)+\frac12\log(\a)\log(1-\a)\bigr)\,+\,\frac{\pi}3\,,\ee
in accordance with the prediction~\eqref{VolConj}, where $\,\a=0.87743\cdots - 0.74486\dots i\;$
is the root of
\be \a^3-\a^2+1=0 \ee
with negative imaginary part.  The next four values are (again numerically to very high precision)
\bes s_1=\frac14\,\log\frac{1+3\a}{23}\,,\qquad s_2=\frac{198\a^2+1452\a-1999}{24\cdot23^2}, \ees
\bes s_3=\frac{465\a^2-465\a+54}{2\cdot23^3}\,,\qquad s_4=\frac{-2103302\a^2+55115\a+5481271}{240\cdot23^5}\,, \ees
in accordance with the arithmeticity conjecture since $\mathbb K=\mathbb Q(\a)$ in this case.

These coefficients are already quite complicated, and the next values even more so.
We can simplify them by making the rescaling
\be s_n'=s_n\,\lambda^{n-1} \ee
({\it i.e.}, by expanding in powers of $2\pi i/\lambda N$ instead of $2\pi i/N$), where
\be\lambda=\a^5(3\a-2)^3=\a^{-1}(\a^2-3)^3\,.\ee
(This number is a generator of $\mathfrak p^3$, where $\mathfrak p =(3\a-2)=(\a^2-3)$ is the unique
ramified prime ideal of $\mathbb K$, of norm~23.) We then find
\beas s_2' &=& -\tfrac1{24}\,\bigl(12\,\a^2-19\,\a+86\bigr),   \\
 s_3' &=&  - \tfrac32\,\bigl(2\,\a^2+5\,\a-4\bigr),     \\
 s_4' &=&   \tfrac1{240}\,\bigl(494\,\a^2+12431\,\a+1926\bigr),     \\
 s_5' &=&   -\tfrac18\,\bigl(577\,\a^2-842\,\a+1497\bigr),     \\
 s_6' &=&  \tfrac1{10080}\,\bigl(176530333\,\a^2-80229954\,\a-18058879\bigr),      \\
 s_7' &=&  -  \tfrac1{240}\,\bigl(99281740\,\a^2+40494555\,\a+63284429\bigr), \\
 s_8' &=&  -\tfrac1{403200}\,\bigl(3270153377244\,\a^2-4926985303821\,\a-8792961648103\bigr), \\
 s_9' &=&   \tfrac1{13440}\,\bigl(9875382391800\,\a^2 - 939631794912\,\a - 7973863388897\bigr), \\
 s_{10}' &=&  - \tfrac1{15966720}\,\bigl(188477928956464660\,\a^2+213430022592301436\,\a+61086306651454303\bigr), \\
 s_{11}' &=& -\tfrac1{1209600}\,\bigl(517421716298434577\,\a^2-286061854126193276\,\a-701171308042539352\bigr),  \eeas
with much simpler coefficients than before, and with each denominator dividing $(n+2)!\,$.
These highly nontrivial numbers give a strong experimental confirmation of the conjecture.

We observe that in both examples treated the first statement of the conjecture~\eqref{ArithConj} can
be strengthened to
\be \exp(4s_1)\;\in\;\mathbb K\,. \ee
It would be interesting to know if the same statement holds for all hyperbolic knot complements (or
even all hyperbolic 3-manifolds). Another comment in this vein is that \eqref{VolConj} also has an
arithmetic content: one knows that the right-hand side of this equation is in the image under the
extended regulator map of an element in the Bloch group (or, equivalently, the third algebraic $K$-group)
of the number field~$\mathbb K$.


\subsection{Perturbative $SL(2,\C)$ invariants for the figure-8 knot complement}
\label{sec:figeight}

In this subsection, we explicitly calculate the perturbative $SL(2,\C)$ invariants
$S_n^{(\rho)}$ for the figure-8 knot complement, first via quantization
of the moduli space of flat $SL(2,\C)$ connections and then, independently,
using the state integral model formulated in Section~\ref{sec:statesum}.
We find perfect agreement between the results, thus, supporting our proposal \eqref{statesumdef}.

Let $K$ be the figure-eight knot (shown in Figure~\ref{fig:knots}.$a$)
and let $M$ be its complement in the 3-sphere.
In order to compute the perturbative invariants~$S_n^{(\rho)}$
for an arbitrary $\rho:~\pi_1 (M)~\to~SL(2,\C)$ we first need to review the classical geometry of~$M$
in more detail and, in particular, to describe the moduli space of flat $SL(2,\C)$ connections on~$M$.
As we already mentioned earlier, 
the knot group $\pi_1 (M)$ is generated by two elements, $a$ and $b$,
such that $a^{-1} b a b^{-1} a = ba^{-1} ba b^{-1}$.
The corresponding representation into $SL(2,\C)$ is given by
\be
\rho (a) = \begin{pmatrix}  ~1~ & ~1~ \\ 0 & 1  \end{pmatrix}\,, \qquad
\rho (b) = \begin{pmatrix}  ~1~ & ~0~ \\ \zeta & 1  \end{pmatrix}\,,
\ee
where $\zeta = (-1 + \sqrt{-3})/2$ is the cube root of unity, $\zeta^3 = 1$.

The complement of the figure-eight knot can be also represented
as a quotient space $\IH^3/\Gamma$ \eqref{mquotient}, where the holonomy group $\Gamma$
is generated by the above two matrices. Specifically, we have
\be
\Gamma \cong PSL (2, \CO_{\mathbb{K}})\,,
\ee
where $\CO_{\mathbb{K}}$ is the ring of integers in the imaginary
quadratic field $\mathbb{K} = \mathbb{Q} (\sqrt{-3})$.
The fundamental domain, $\CF$, for $\Gamma$ is described by
a geodesic pyramid in $\IH^3$ with one vertex at infinity
and the other four vertices at the points:
\begin{align}
v_1 & = j \nonumber \\
v_2 & = \frac{1}{2} - \frac{\sqrt{3}}{6} i + \sqrt{\frac{2}{3}} j \\
v_3 & = \frac{1}{2} + \frac{\sqrt{3}}{6} i + \sqrt{\frac{2}{3}} j \nonumber \\
v_4 & = \frac{1}{\sqrt{3}} i + \sqrt{\frac{2}{3}} j\,. \nonumber
\end{align}
Explicitly, we have
\be
\CF = \{ z + x_3 j \in \IH^3 ~\vert~ z \in \CF_z, ~x_3^2 + |z|^2 \ge 1 \}\,,
\ee
where $z = x_1 + i x_2$ and
\begin{align}
\CF_z & = \{ z \in \C ~\vert~ 0 \le \Re (z),~
\frac{1}{\sqrt{3}} \Re (z) \le \Im (z),~
\Im (z) \le \frac{1}{\sqrt{3}} (1 - \Re (z)) \} \nonumber \\
& \cup \{ z \in \C ~\vert~ 0 \le \Re (z) \le \frac{1}{2},~
- \frac{1}{\sqrt{3}} \Re (z) \le \Im (z) \le \frac{1}{\sqrt{3}} \Re (z) \}
\end{align}
is the fundamental domain of a 2-torus with modular parameter $\tau = \zeta$.
The region of large values of $x_3$ in $\CF$ corresponds to the region near
the cusp of the figure-eight knot complement $M$.

The standard triangulation of the figure eight knot complement
comprises two ideal tetrahedra of opposite simplicial orientations, as in Figure \ref{fig:tetr_momenta},
glued together in the only nontrivial consistent manner possible,
\be
M = \Delta_z \cup \Delta_w\,.
\label{figeighttriang}
\ee
Here, $z$ and $w$ are complex numbers, representing the shapes of the ideal tetrahedra; we take $\Delta_z$ to be positively oriented and $\Delta_w$ to be negatively oriented.
As explained in Section~\ref{sec:ideal},
the shape parameters $z$ and $w$ must obey consistency relations,
which in the case of the figure-eight knot reduce
to a single algebraic relation (see {\it e.g.} Chapter 4 of \cite{thurston-1980} and Section 15 of \cite{neumann-2004}):
\be
(z-1)(w-1)=z^2w^2 \,.
\label{figeightgluing}
\ee
The shape parameters $z$ and $w$ are related to the $SL(2,\C)$ holonomy eigenvalue, $l$,
along the longitude\footnote{A 1-cycle on $\Sigma = T^2$ which is contractible
in the knot complement $M$.} of the knot in the following way
\be
\frac{z^2}{z-1} = -l\,, \qquad \frac{w^2}{w-1} = - l^{-1}\,,
\label{lholviashape}
\ee
which automatically solves the consistency condition \eqref{figeightgluing}.
Similarly, the holonomy eigenvalue $m=e^u$ around the noncontractible meridian of the torus is given by
\be z w =m^2. \label{mholviashape} \ee

Given the decomposition \eqref{figeighttriang} of the 3-manifold $M$ into
two tetrahedra, we can find its volume by adding the (signed) volumes of $\Delta_z$ and $\Delta_w$,
\be
\Vol (M;u) = \Vol (\Delta_z) - \Vol (\Delta_w) = D(z)-D(w)\,,\label{figeightvol}
\ee
where the volume of an ideal tetrahedron is given by \eqref{volviad}. Similarly, following the prescription in Section \ref{sec:ideal}, the complexified volume can be given by\footnote{This expression differs slightly from the one given in \cite{neumann-2004}, because we use $2v+2\pi i$ rather than $2v$ as the logarithm of the ``longitudinal'' holonomy, as mentioned in Footnote \ref{foot:v2pi} of Section \ref{sec:ideal}.}
\be
i(\Vol(M;u)+i{\rm CS}(M;u)) = L(z;0,0)-L(w;0,0)-v\bar{u}-i\pi u\,. \label{figeightcvol}
\ee
Notice that due to the consistency relation \eqref{figeightgluing}
the total volumes \eqref{figeightvol}, \eqref{figeightcvol} are functions of one complex parameter, or,
equivalently, a point on the zero locus of the A-polynomial, $A(l,m)=A(-e^v,e^u)=0$.

The classical A-polynomial of the figure-eight knot is
\be
A(l,m) = (l-1)(m^4 - (1-m^2-2m^4-m^6+m^8)l+m^4l^2)\,.
\ee
Observe that $A(l^{-1},m^{-1})\sim A(l,m)$ and that $A(l^{-1},m) \sim A(l,m)$,
the latter reflecting the fact that the figure-eight knot is amphicheiral.
The zero locus of this A-polynomial has three branches: the abelian branch
$l_{{\rm abel}} = 1$ (or $v_{{\rm abel}} = -i\pi$) from the first factor,
and two other branches from the second, given explicitly by
\be
l_{{\rm geom,conj}} (m) = \frac{1-m^{2}-2m^4-m^6+m^8}{2m^4}\pm \frac{1-m^4}{2m^2}\Delta(m)
\ee
or
\be
v_{{\rm geom,conj}} (u) = \log(-l_{{\rm geom,conj}}(e^u))\,,
\label{v41}
\ee
where we have defined
\be
\Delta(m) = i\sqrt{-m^{-4}+2m^{-2}+1+2m^2-m^4}\,.
\label{def_Delta}
\ee
Note that $\Delta (m)$ is an analytic function of $m=e^u$ around $u=0$,
as are $l(m)$ and $v(u)$. In particular, $\Delta(1)=\sqrt{-3}$ is the generator of the trace field $\mathbb{K}=\mathbb{Q}(\tr\Gamma)$ for the figure-eight knot. 

Since there are only two non-abelian branches,
they must (as indicated in Section~\ref{sec:symmetries})
be the geometric and conjugate ones.
Alternatively, eliminating shape parameters from the (geometric) eqs.\eqref{mholviashape} and \eqref{lholviashape} implies precisely that the non-abelian factor of the A-polynomial vanishes.
As expected for an amphicheiral knot, due to the symmetries discussed in Section \eqref{sec:symmetries}, $v_{{\rm geom}}(u) = -v_{{\rm conj}}(u)$. The exact labelling of branches (geometric vs. conjugate)
can be verified by computing the hyperbolic volume $\Vol (M;u)$
of the figure-eight knot complement via an ideal triangulation, as in \eqref{figeightcvol},
and comparing the result to the integral \eqref{S0_integral}
on the geometric branch. One obtains precisely
\begin{align} S_0^{(\rm geom)}(u) &= \frac{i}{2}(\Vol(M;0)+i{\rm CS}(M;0)) + \int_0^u (v_{\rm geom}(u')+i\pi)du' \notag \\
 &= \frac{1}{2}\big(\Li(z)-\Li(w)+\frac{1}{2}\log{z}\log(1-z)-\frac{1}{2}\log(w)\log(1-w)+uv+i\pi u\big) \notag \\
 &= \frac{i}{2}(\Vol(M;u)+i{\rm CS}(M;u))+v_{\rm geom}(u)\Re(u)+i\pi u\,, 
\end{align}
confirming \eqref{S0_volume}. Note that $CS(M;u)=0$ for any imaginary $u$.

{}From the relation \eqref{lholviashape} we find that the point $(l,m)=(-1,1)$
corresponding to the complete hyperbolic structure on $M$
is characterized by the values of $z$ and $w$ which solve the equation
\be
z^2 - z + 1 = 0\,.
\ee
In order to obtain tetrahedra of positive (signed) volume, we must choose $z$
to be the root of this equation with a positive imaginary part, and $w$ its inverse:
\be
z = \frac{1 + i \sqrt{3}}{2}\,,\qquad  w= \frac{1 - i \sqrt{3}}{2}\,.
\ee
These value correspond to regular ideal tetrahedra, and maximize (respectively, minimize)
the Bloch-Wigner dilogarithm function $D(z)$.


\subsubsection{Quantization}
\label{sec:figeightquant}

In order to test the proposed state integral model \eqref{statesumdef},
we first compute the perturbative
$SL(2,\C)$ invariants of $M$ via quantization of $\M (G_{\C},\Sigma)$,
as outlinted in Section~\ref{sec:quantization}.
Recall the expression \eqref{zpert} for the perturbative Chern-Simons partition function,
\be
Z^{(\alpha)} (M;\hbar,u) =
\exp\left( \frac{1}{\hbar} S_0^{(\alpha)} (u) - \frac{1}{2}\delta^{(\alpha)} \log \hbar
+ \sum_{n=0}^\infty S_{n+1}^{(\alpha)} (u) \hbar^n \right),
\ee
where for the figure-eight knot complement $\alpha \in \{{\rm geom},\,{\rm conj},\,{\rm abel}\}$,
and, as explained {\it e.g.} in \cite{gukov-2003,gukov-2006}, $\delta^{(\alpha \neq {\rm abel})} = 0$.

In Section~\ref{sec:quantization} we saw how quantization of the relation $A(l,m) = 0$
leads to the equation
\be
\widehat{A} (\hat l, \hat m ) ~Z^{(\alpha)} (M;\hbar,u) = 0\,,
\label{AZquantum}
\ee
which then determines the perturbative $G_{\C}$ invariant $Z^{(\alpha)} (M;\hbar,u)$.
As explained earlier, {\it cf.} eqs. \eqref{aionp} and \eqref{almionp},
a simple way to find the operator $\widehat{A} (\hat l, \hat m )$
is to note that it also annihilates the polynomial invariants (of the knot $K$)
computed by Chern-Simons theory with compact gauge group $G$.
In the present case, $G = SU(2)$ and the equation \eqref{almionp}
takes the form of a recursion relation on the set of colored Jones
polynomials, $\{ J_n (K;q) \}$.
Specifically, using the fact that $\hat l$ acts by shifting the value of
the highest weight of the representation
and writing $\widehat{A} (\hat{l},\hat{m}) = \sum_{j=0}^d a_j(\hat{m},q) \hat{l}^j$
as in \eqref{hatAcoeffs}, we obtain
\be
\sum_{j=0}^d a_j (q^{n/2}, q) J_{n+j} (K;q) = 0\,.
\ee
It is easy to verify that the colored Jones polynomials of the figure-eight
knot indeed satisfy such a recursion relation, with the coefficients
(see also\footnote{Note that \cite{garoufalidis-2004, garoufalidis-2006} look at asymptotics
of the colored Jones polynomial normalized by its value at the unknot,
while the $SL(2,\C)$ Chern-Simons partition function should agree with
the {unnormalized} colored Jones polynomial.
Hence, we must divide the expressions for $a_j$ there by $(m^2 q^{j/2} - q^{-j/2})$
to account for the difference, introducing a few factors
of $q^{1/2}$ in our formulas.}
\cite{garoufalidis-2004, garoufalidis-2006}):
\begin{subequations} \label{Aj}
\begin{align}
a_0(\hat{m},q) & = \frac{q\hat{m}^2}{(1+q\hat{m}^2)(-1+q\hat{m}^4)}\,, \\
a_1(\hat{m},q) & = \frac{1+(q^2-2q)\hat{m}^2-(q^3-q^2+q)\hat{m}^4-(2q^3-q^2)\hat{m}^6
+q^4\hat{m}^8}{q^{1/2}\hat{m}^2(1+q^2\hat{m}^2-q\hat{m}^4-q^3\hat{m}^6)}\,, \\
a_2(\hat{m},q) & = -\frac{1-(2q^2-q)\hat{m}^2-(q^5-q^4+q^3)\hat{m}^4+(q^7-2q^6)\hat{m}^6
+q^8\hat{m}^8}{q\hat{m}^2(1+q\hat{m}^2-q^5\hat{m}^4-q^6\hat{m}^6)}\,, \\
a_3(\hat{m},q) & = -\frac{q^4\hat{m}^2}{q^{1/2}(1+q^2\hat{m}^2)(-1+q^5\hat{m}^4)}\,.
\end{align}
\end{subequations}
Note that in the classical limit $q = e^{2\hbar} \to 1$, we have
\be
\widehat{A} (\hat{l},\hat{m}) ~~\underset{\hbar \to 0}{\longrightarrow}~~ \frac{A(l,m)}{m^2(m^2-1)(m^2+1)^2}\,.
\label{limit_hatA}
\ee
We could multiply all the $a_j$'s by the denominator of \eqref{limit_hatA}
(or a $q$-deformation thereof) to obtain a more direct correspondence
between $\widehat{A} (\hat{l},\hat{m})$ and $A(l,m)$,
but this does not affect any of the following calculations.

To find the exact perturbative invariant $Z^{(\alpha)} (M;\hbar,u)$ for each $\alpha$,
we reduce $\hat{m}$ to a classical variable $m$ (since in the $u$-space representation
the operator $\hat{m}$ just acts via ordinary multiplication),
expand each of the above $a_j$'s as $a_j(m,q) = \sum_{p=0}^\infty a_{j,p}(m)\hbar^p$,
and substitute the $a_{j,p}$ into the hierarchy of differential equations derived
in Section~\ref{sec:quantization} and displayed in Table \ref{tab:Seqns}.
The equations are then solved recursively on each branch $\alpha$
of the A-polynomial to determine the coefficients $S_n^{(\alpha)}(u)$.


\subsubsection*{Geometric branch}

For the geometric branch,
the solution to the first equation in Table \ref{tab:Seqns} is chosen to be
\be
S_0^{({\rm geom})} (u) = \frac{i}{2}\Vol({\bf 4_1};0) + \int_0^u du\, v_{{\rm geom}} (u) + i\pi u\,,
\label{GS0}
\ee
with $v_{{\rm geom}} (u)$ as in \eqref{v41}.
The integration constant $i(\Vol(0)+i\rm{CS}(0))$ is not important for determining the remaining coefficients
(since only derivatives of $S_0$ appear in the equations), but we have fixed it by requiring that
$S_0^{({\rm geom})}(0) = \frac{i}{2}(\Vol({\bf 4_1};0)+i {\rm CS} ({\bf 4_1};0))
= \frac{i}{2}\Vol({\bf 4_1};0) = (1.01494\ldots)i$,
as expected for the classical action of Chern-Simons theory.

Substituting the above $S_0^{({\rm geom})} (u)$
into the hierarchy of equations, the rest are readily solved%
\footnote{It is computationally advantageous to express everything in terms of $m=e^u$
and $m \frac{d}{dm}=\frac{d}{du}$, {\it etc.}, when implementing this on a computer.} %
for the subleading coefficients. The first eight functions $S_n^{({\rm geom})}(u)$ appear below:

\underline{\hspace{2in}}

\begin{longtable}{r@{\;}c@{\;}l}
$S_1(u)$ &=& $\displaystyle -\frac{1}{2}\log\left(\frac{-i\Delta(m)}{2}\right)$\,,
\vspace{.2cm}\\
$S_2(u)$ &=& $\displaystyle \frac{-1}{12\Delta(m)^3m^6}\big( 1 - m^2 - 2 m^4 + 15 m^6 - 2 m^8 - m^{10} + m^{12} \big) \,, $ \vspace{.2cm}\\
$S_3(u)$ &=& $\displaystyle\frac{2}{\Delta(m)^6m^6}\big(1 - m^2 - 2 m^4 + 5 m^6 - 2 m^8 - m^{10} + m^{12}\big) \,, $ \vspace{.2cm}\\
$S_4(u)$ &=& $\displaystyle\frac{1}{90\Delta(m)^9m^{16}}\big(1 - 4 m^2 - 128 m^4 + 36 m^6 $ \vspace{.2cm}\\ &&
$+ 1074 m^{8} - 5630 m^{10} + 5782 m^{12} +
   7484 m^{14} - 18311 m^{16} + 7484 m^{18} $ \vspace{.2cm}\\ &&
$+ 5782 m^{20} - 5630 m^{22} +
   1074 m^{24} + 36 m^{26} - 128 m^{28} - 4 m^{30} +
   m^{32}\big) \,,$ \vspace{.2cm} \\
$S_5(u)$ &=& $\displaystyle\frac{2}{3\Delta(m)^{12}m^{18}}\big(1 + 5 m^2 - 35 m^4 + 240 m^6 - 282 m^8 - 978 m^{10}
        $\vspace{.2cm} \\ && $
 + 3914 m^{12} -
     3496 m^{14} - 4205 m^{16} + 9819 m^{18} - 4205 m^{20} - 3496 m^{22}
        $\vspace{.2cm} \\ && $
  +
     3914 m^{24} - 978 m^{26} - 282 m^{28} + 240 m^{30} - 35 m^{32} + 5 m^{34} +
     m^{36}\big) \,,
        $\vspace{.2cm} \\
$S_6(u)$ &=& $\displaystyle\frac{-1}{945\Delta(m)^{15}m^{28}}\big(1 + 2 m^2 + 169 m^4 + 4834 m^6
        $\vspace{.2cm} \\ && $
 - 24460 m^{8} + 241472 m^{10} -
     65355 m^{12} - 3040056 m^{14} + 13729993 m^{16}
        $\vspace{.2cm} \\ && $
  - 15693080 m^{18} -
     36091774 m^{20} + 129092600 m^{22} - 103336363 m^{24}  $\vspace{.2cm} \\ && $
- 119715716 m^{26} + 270785565 m^{28} - 119715716 m^{30}
 -    103336363 m^{32}
        $\vspace{.2cm} \\ && $
+ 129092600 m^{34} - 36091774 m^{36} -
     15693080 m^{38}  + 13729993 m^{40} - 3040056 m^{42}
        $\vspace{.2cm} \\ && $
         - 65355 m^{44} +
     241472 m^{46} - 24460 m^{48} + 4834 m^{50} + 169 m^{52} + 2 m^{54} +
     m^{56}\big)\,,
        $\vspace{.2cm} \\
$S_7(u)$ &=& $\displaystyle\frac{4}{45\Delta(m)^{18}m^{30}}\big(1 + 47 m^2 - 176 m^4 + 3373 m^6 + 9683 m^8
        $\vspace{.2cm} \\ && $
- 116636 m^{10} +
     562249 m^{12} - 515145 m^{14} - 3761442 m^{16} + 14939871 m^{18}
        $\vspace{.2cm} \\ && $
   -  15523117 m^{20}- 29061458 m^{22} + 96455335 m^{24} - 71522261 m^{26}
        $\vspace{.2cm} \\ && $
    - 80929522 m^{28} + 179074315 m^{30}  - 80929522 m^{32} - 71522261 m^{34}
        $\vspace{.2cm} \\ && $
 + 96455335 m^{36} - 29061458 m^{38} - 15523117 m^{40} + 14939871 m^{42}
        $\vspace{.2cm} \\ && $
  -   3761442 m^{44} - 515145 m^{46} + 562249 m^{48} - 116636 m^{50}
        $\vspace{.2cm} \\ && $
     +
     9683 m^{52} + 3373 m^{54} - 176 m^{56} + 47 m^{58} + m^{60}\big) \,,$ \\
$S_8(u)$ &=& $\displaystyle\frac{1}{9450\Delta(m)^{21}m^{40}}\big(1 + 44 m^2 - 686 m^4
        $\vspace{.2cm} \\ && $
- 25756 m^6  + 25339 m^{8} - 2848194 m^{10} -
   28212360 m^{12}
        $\vspace{.2cm} \\ && $
   + 216407820 m^{14} - 1122018175 m^{16}  - 266877530 m^{18}
        $\vspace{.2cm} \\ && $
+ 19134044852 m^{20}  - 76571532502 m^{22} +
   75899475728 m^{24}
        $\vspace{.2cm} \\ && $
+ 324454438828 m^{26} - 1206206901182 m^{28}   + 1153211096310 m^{30}
        $\vspace{.2cm} \\ && $
+ 1903970421177 m^{32}  - 5957756639958 m^{34} +
   4180507070492 m^{36}
        $\vspace{.2cm} \\ && $
 + 4649717451712 m^{38} - 10132372721949 m^{40}  +  4649717451712 m^{42}
        $\vspace{.2cm} \\ && $
+ 4180507070492 m^{44} - 5957756639958 m^{46} +
   1903970421177 m^{48}
        $\vspace{.2cm} \\ && $
 + 1153211096310 m^{50} - 1206206901182 m^{52} + 324454438828 m^{54}
        $\vspace{.2cm} \\ && $
  + 75899475728 m^{56} - 76571532502 m^{58} +
   19134044852 m^{60}
        $\vspace{.2cm} \\ && $
- 266877530 m^{62} - 1122018175 m^{64}   + 216407820 m^{66} - 28212360 m^{68}
        $\vspace{.2cm} \\ && $
- 2848194 m^{70} + 25339 m^{72} -
   25756 m^{74} - 686 m^{76} + 44 m^{78} +
   m^{80}\big) \,. $ \vspace{0.6cm}\\
\caption{Perturbative invariants $S_n^{({\rm geom})}(u)$ up to eight loops.\label{tab:figeightgeom}}
\end{longtable}
\vspace{-0.8cm}
$\overline{\hspace{2in}}$ \\

According to \eqref{sonetorsion},
the coefficient $S_1^{({\rm geom})} (u)$ in the perturbative Chern-Simons partition function
should be related to the Reidemeister-Ray-Singer torsion of $M$ twisted by $\CA^{({\rm geom})}$,
which has been independently computed. Our function matches%
\footnote{To compare with \cite{gukov-2006}, note that
$k_{{\rm there}} = k_{{\rm here}} = i \pi/\hbar$, and $u_{{\rm there}} = 2u_{{\rm here}}$.
The shift by $-\log(\pi)$ is directly related to a jump in the asymptotics of the colored Jones polynomial at $u=0$.} %
that appearing in {\it e.g.} \cite{gukov-2006}, up to a shift by  $-\log\pi$.
The constants of integration for the remaining coefficients have been fixed
by comparison to the asymptotics of the colored Jones polynomial,
using \eqref{Ss} and the results of Section \ref{sec:arith_example}.


\subsubsection*{Conjugate branch}

For the ``conjugate'' branch, the solution for $S_0(u)$ is now chosen to be
\be
S_0^{({\rm conj})} (u) = -\frac{i}{2}\Vol({\bf 4_1};0) + \int_0^u du\, v_{{\rm conj}} (u) + i\pi u
\qquad (\mbox{mod $2\pi u$}),
\label{CS0}
\ee
so that $S_0^{({\rm conj})}(u) = -S_0^{({\rm geom})}(u)$.
As for the geometric branch, this is then substituted into remainder
of the hierarchy of equations. Calculating the subleading coefficients, the constants of integration can all be fixed so that
\be
S_n^{({\rm conj})} (u) = (-1)^{n+1} S_{n+1}^{({\rm geom})} (u),
\label{CGrel}
\ee
This is precisely what one expects for an amphicheiral knot when a conjugate pair of branches coincides with a ``signed'' pair, as discussed in Section \ref{sec:symmetries}.


\subsubsection*{Abelian branch}

For completeness, we can also mention perturbation theory around
an abelian flat connection $\CA^{({\rm abel})}$ on $M$,
although it has no obvious counterpart in the state integral model.

For an abelian flat connection $\CA^{({\rm abel})}$, the classical
Chern-Simons action \eqref{szero} vanishes.
This is exactly what one finds from \eqref{S0_integral}:
\be
S_0^{({\rm abel})} (u) =\int_0^u du\, v_{{\rm abel}} (u)
+ i\pi u = 0 \qquad (\mbox{mod $2\pi i u$})\,,
\ee
fixing the constant of integration so that $S_0^{(\rm abel)}(0)=0$.
From the hierarchy of differential equations, the first few subleading coefficients are
\begin{longtable}{r@{\;}c@{\;}l}
$S_1^{({\rm abel})}(u)$ &=& $\displaystyle\log\frac{m(m^2-1)}{1 - 3 m^2 + m^4}, $\vspace{.2cm} \\  $
S_2^{({\rm abel})}(u)$ &=& $0$\vspace{.2cm} \\  $
S_3^{({\rm abel})}(u)$ &=& $\displaystyle \frac{4(m^2-1)^2}{(1 - 3 m^2 + m^4)^3}\big(1-7m^2+16m^4-7m^6+m^8\big) , $\vspace{.2cm} \\  $
S_4^{({\rm abel})}(u)$ &=& $0$\vspace{.2cm} \\  $
S_5^{({\rm abel})}(u)$ &=& $\displaystyle \frac{4(m^2-1)^2}{3(1 - 3 m^2 + m^4)^6}\big(41 - 656 m^2 + 4427 m^4 - 16334 m^6 + 35417 m^8 - 46266 m^{10}$ \vspace{.2cm} \\ && $ + 
 35417 m^{12} - 16334 m^{14} + 4427 m^{16} - 656 m^{18} + 41 m^{20}\big) $\vspace{.2cm} \\  $
S_6^{({\rm abel})}(u)$ &=& $0$ \;\ldots. \vspace{0.5cm}\\
\caption{Perturbative invariants $S_n^{({\rm abel})}(u)$ up to six loops.\label{tab:figeightabel}}
\end{longtable}

\noindent
In the language of Section \ref{sec:symmetries},
the abelian branch must be its own ``signed pair,''
guaranteeing that all even $S_{2k}^{(\rm abel)}(u)$ vanish.

Having determined the perturbative coefficients
for the three branches of classical solutions (in principle to any order desired),
we can now compare them with the computation of the state integral model.


\subsubsection{State integral model}
\label{sec:figeightsum}

The state integral model \eqref{statesumdef} for the figure-eight knot complement gives:
\be
Z^{(\alpha)} (M; \hbar, u) = \frac{1}{\sqrt{2 \pi \hbar}} \int_{C_{\alpha}} dp
\frac{\Phi_{\hbar} (p + i\pi +\hbar)}{\Phi_{\hbar}(-p-2u-i\pi-\hbar) }
e^{-\frac{2}{\hbar}u(u+p) -u}\;.
\label{base_Hik_41}
\ee
There are two tetrahedra ($N=2$) in the standard triangulation of $M$,
and so two quantum dilogarithms in the integral.
There is a single integration variable $p$, and we can identify
$g_1(p,u) = p$,\, $g_2(p,u) = -p-2u$, and $f(p,2u,\hbar) = -4u(u+p)$.

\EPSFIGURE{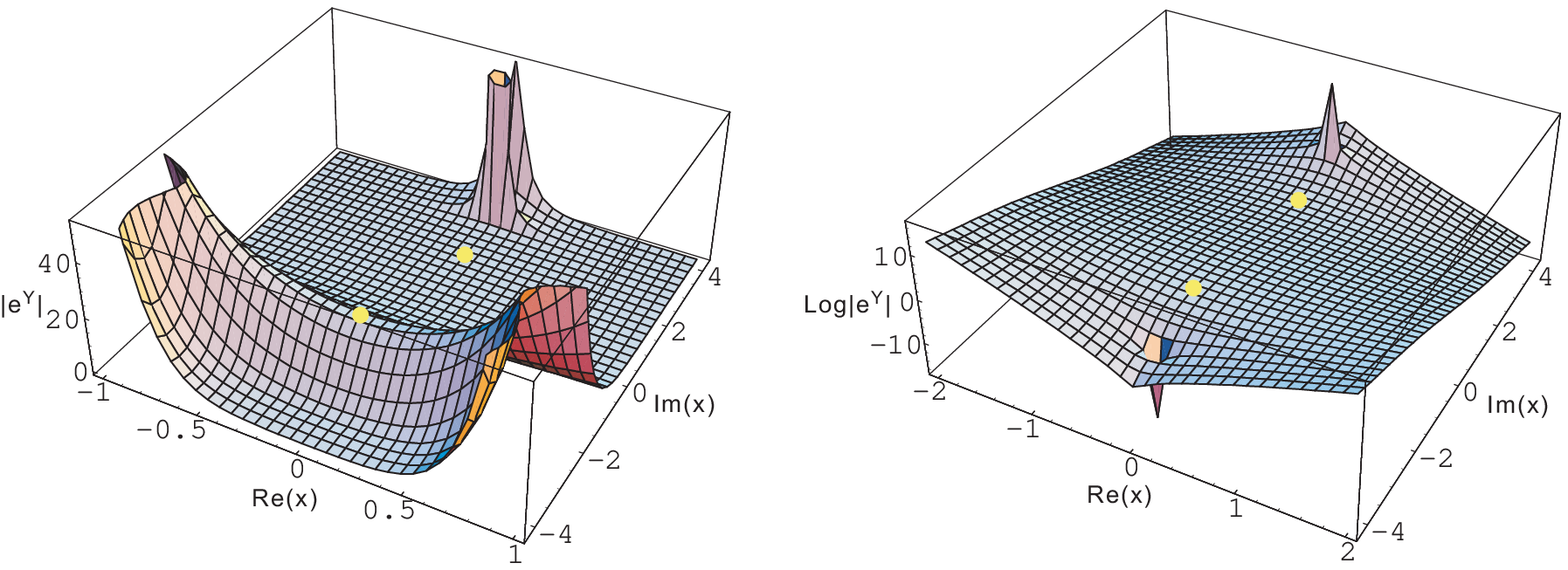,height=4.5cm,angle=0,trim=0 0 0 0}%
{Plots of $|e^{\Upsilon(\hbar,p,u)}|$ and its logarithm at $u=0$ and $\hbar = \frac{i}{3}$.
\label{fig:Lambda0} }

It will be convenient here to actually change variables $p\mapsto p-u-i\pi-\hbar$,
removing the $(i\pi+\hbar)$ terms in the quantum dilogarithms,
and obtaining the somewhat more symmetric expression
\begin{align}
Z^{(\alpha)} (M; \hbar, u)
&= \frac{1}{\sqrt{2 \pi \hbar}}
e^{\frac{2\pi iu}{\hbar}+u}\int_{C_{\alpha}} dp \frac{\Phi_{\hbar}(p-u)}{\Phi_{\hbar}(-p-u)} e^{-\frac{2pu}{\hbar}}
\label{Hik41} \\
&= \frac{1}{\sqrt{2 \pi \hbar}}e^{\frac{2\pi iu}{\hbar}+u}\int_{C_{\alpha}} dp\, e^{\Upsilon(\hbar,p,u)}\;.
\end{align}
We define $e^{\Upsilon(\hbar,p,u)} = \frac{\Phi_{\hbar}(p-u)}{\Phi_{\hbar}(-p-u)}e^{-\frac{2pu}{\hbar}}$\,.
Figures \ref{fig:Lambda0} and \ref{fig:Lambda12} show plots of
$|e^{\Upsilon(\hbar,p,u)}|$ and $\log|e^{\Upsilon(\hbar,p,u)}| = \Re\,\Upsilon(\hbar,p,u)$
at $\hbar = i/3$ and two values of $u$.
The half-lines of poles and zeroes of the two quantum dilogarithms combine
into similar singularities for $\Upsilon(\hbar,p,u)$, as is depicted in Figure \ref{fig:MinefieldY};
note the splitting of these poles and zeroes by an amount $2u$.

\EPSFIGURE{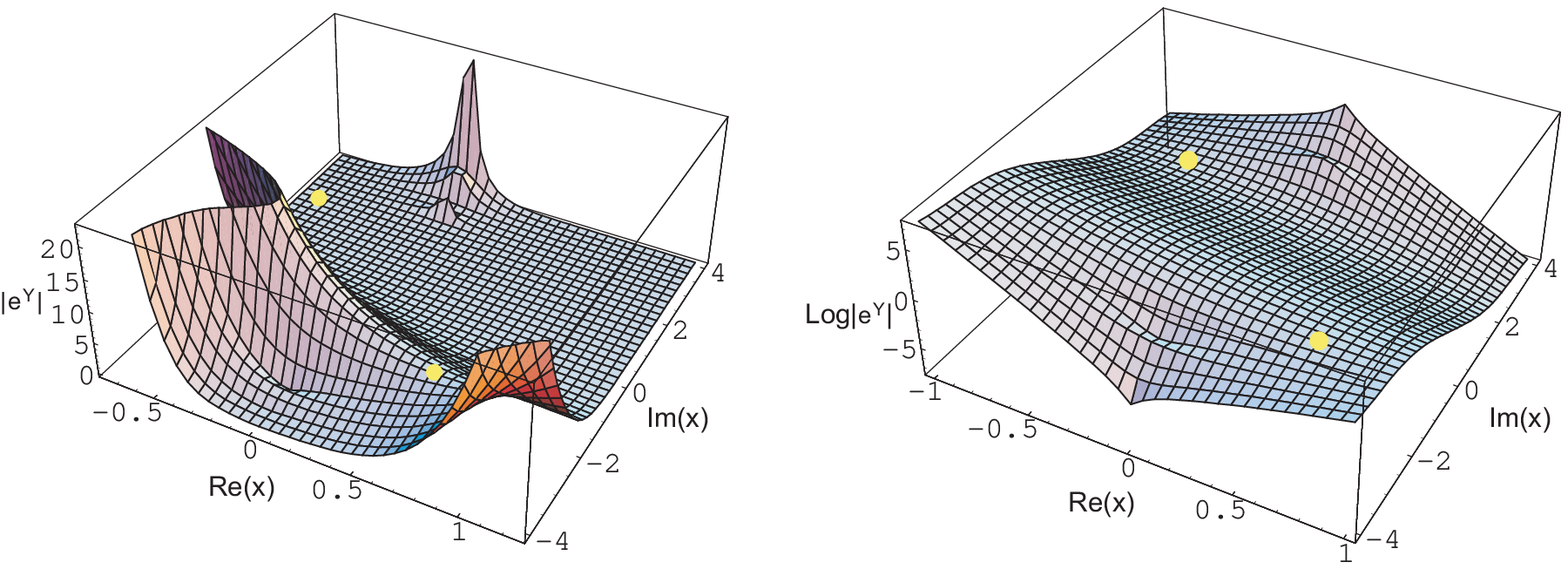,height=4.5cm,angle=0,trim=0 0 0 0}%
{Plots of $|e^{\Upsilon(\hbar,p,u)}|$ and its logarithm at $u=\frac{1}{2}i$ and $\hbar = \frac{i}{3}$.
\label{fig:Lambda12} }

After our change of variables, the ``potential'' function $V(p,u)$ as in \eqref{Hik_gen_V} is now seen to be
\be
V(p,u) = \frac{1}{2}\left[\Li(-e^{p-u})-\Li(-e^{-p-u})-4p u + 4\pi i u\right]\,.
\label{Hik41V}
\ee
Instead of looking directly at $\frac{\partial}{\partial p} V=0$
to find its critical points, we consider the simpler equation
\be
r(x,m) = e^{2\frac{\partial}{\partial p}V} = \frac{x}{m^2(m+x)(1+mx)} = 1\,,
\ee
in terms of $x=e^p$ and $m=e^u$.
This clearly has two branches of solutions,
which both lift to true critical points of $V$, given by
\be
p^{({\rm geom},{\rm conj})}(u) = \log\left[\frac{1-m^2-m^4\mp m^2\Delta(m)}{2m^3}\right]\,,
\label{HikpGC}
\ee
with $\Delta(m)$ defined in \eqref{def_Delta}. The lift is unique if (say) $\hbar \in i\,\R_{>0}$.
We claim that these %
\EPSFIGURE{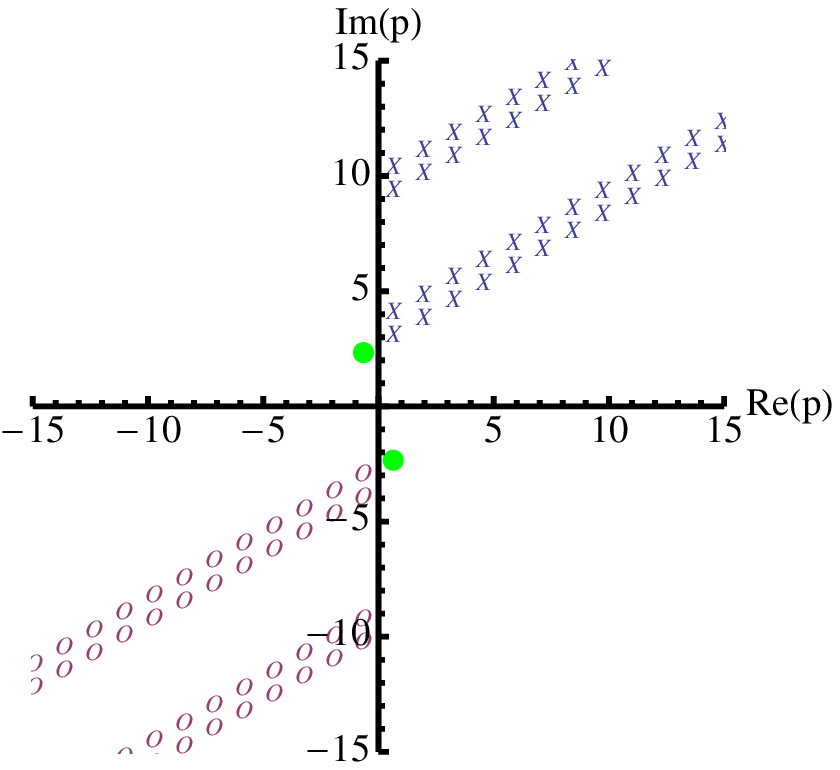,height=7.5cm,angle=0,trim=0 0 0 0}%
{Poles, zeroes, and critical points of $e^{\Upsilon(\hbar,p,u)}$ for $u=\frac{1}{2}i$ and $\hbar = \frac{3}{4}e^{i\pi/6}$.\\
\label{fig:MinefieldY} } %
\noindent correspond to the geometric and conjugate branches of the A-polynomial,
which can be verified by calculating%
\footnote{Here $s(x,m)$ is the square root of a rational function rather than
a rational function itself due to our redefined variables.
Using \eqref{base_Hik_41} directly, we would have gotten a pure rational expression.} %
$s(x,m) = \exp\big( \frac{\partial}{\partial u} V(p,u) \big) = \big[\frac{m+x}{x^3+mx^4}\big]^{1/2}$,
and checking indeed that $s(x^{({\rm geom}, {\rm conj})}(m),m) = l_{{\rm geom}, {\rm conj}}(m)$.

The two critical points of $\Lambda_{\hbar}$ which correspond to the critical points of $V$
(as $\hbar \to 0$) are indicated in Figures \ref{fig:Lambda0} and \ref{fig:Lambda12}.
As mentioned at the end of Section~\ref{sec:statesumdef}, at any fixed $\hbar \neq 0$
there exist many other critical points of $\Lambda_{\hbar}$ that can be seen between consecutive pairs of poles and zeroes
in Figures \ref{fig:Lambda0} and \ref{fig:Lambda12};
however, these other critical points become trapped in half-line singularities as $\hbar\rightarrow0$,
and their saddle point approximations are not well-defined.

We now calculate the perturbative invariants $S_n^{({\rm geom})}(u)$ and $S_n^{({\rm conj})}(u)$
by doing a full saddle point approximation of the integral \eqref{Hik41}
on (dummy) contours passing through the two critical points.
We begin by formally expanding $\Upsilon (\hbar,p,u) = \log \Lambda_{\hbar} (p,u)$
as a series in both $\hbar$ and $p$ around some fixed point $p_0$:
\be
\Upsilon(\hbar, p_0+p,u) = \sum_{j=0}^\infty \sum_{k=-1}^\infty\Upsilon_{j,k}(p_0,u)\,p^j\hbar^k\,.
\ee
Our potential $V(p,u)$ is identified with $\Upsilon_{0,-1}(p,u)+2\pi iu$,
and critical points of $V$ are defined by
$\Upsilon_{1,-1}(p,u) = \frac{\partial}{\partial p} V(p,u) = 0$.
Let us also define
\be b(p,u) := -2\Upsilon_{2,-1}(p,u) = -\frac{\partial^2}{\partial p^2}V(p,u)\,. \ee
Then at a critical point $p_0 = p^{(\alpha)}(u)$, the integral \eqref{Hik41} becomes%
\footnote{This expression assumes that $\Upsilon_{j,k}=0$ when $k$ is even,
a fact that shall be explained momentarily.}
\small
\be
Z^{(\alpha)} (M; \hbar, u) = \frac{e^{u+\frac{1}{\hbar}V^{(\alpha)}(u)}}{\sqrt{2 \pi \hbar}}
\int_{C_{\alpha}} dp\, e^{-\frac{b^{(\alpha)}(u)}{2\hbar}p^2} \exp\left[\frac{1}{\hbar}
\sum_{j=3}^\infty \Upsilon_{j,-1}^{(\alpha)}(u)\,p^j
+ \sum_{j=0}^\infty\sum_{k=1}^\infty \Upsilon_{j,k}^{(\alpha)}(u)\,p^j \hbar^k \right],
\label{saddle_integral}
\ee
\normalsize
where $V^{(\alpha)}(u) = V(p^{(\alpha)}(u),u)$,\; $b^{(\alpha)}(u) = b(p^{(\alpha)}(u),u)$,
and $\Upsilon_{j,k}^{(\alpha)}(u)=\Upsilon_{j,k}(p^{(\alpha)}(u),u)$ are implicitly functions of $u$ alone.

We can expand the exponential in \eqref{saddle_integral}, integrate each term using
\be
\int dp\, e^{-\frac{b}{2\hbar}p^2}p^{n} = \left\{
\begin{array}{ll}
 (n-1)!!\left(\frac{\hbar}{b}\right)^{n/2}\sqrt{\frac{2\pi \hbar}{b}} & \;\mbox{$n$ even} \\
 0 & \; \mbox{$n$ odd}
\end{array}
\right.,
\label{gauss_integral}
\ee
and re-exponentiate the answer to get a final result.
The integrals in \eqref{gauss_integral} are accurate up to corrections of order $\CO(e^{-{\rm const}/\hbar})$,
which depend on a specific choice of contour and are ignored.
Following this process, we obtain
\begin{align}
Z^{(\alpha)} (M; \hbar, u)
& = \frac{1}{\sqrt{2 \pi \hbar}} \sqrt{\frac{2\pi \hbar}{b^{(\alpha)}}}
e^{u+\frac{1}{\hbar} V^{(\alpha)}(u)}e^{S_2^{(\alpha)}\hbar+S_3^{(\alpha)}\hbar^2+\ldots} \\
& = \exp\left[\frac{1}{\hbar}V^{(\alpha)}(u)
- \frac{1}{2} \log b^{(\alpha)} + u + S_2^{(\alpha)}\hbar + S_3^{(\alpha)}\hbar^2 + \ldots \right]\,,
\label{final_lambda_int}
\end{align}
where the coefficients $S_n$ can be straightforwardly computed in terms of $b$ and the $\Upsilon$'s.
For example, $S_2^{(\alpha)} = \frac{15}{2(b^{(\alpha)})^3}\Upsilon_{3,-1}^{(\alpha)}+\frac{3}{(b^{(\alpha)})^2}\Upsilon_{4,-1}^{(\alpha)}+\Upsilon_{0,1}^{(\alpha)}$
and $S_3^{(\alpha)} = \frac{3465}{8(b^{(\alpha)})^6}(\Upsilon_{3,-1}^{(\alpha)})^4 +$ (sixteen other terms). In addition, we clearly have
\begin{align} S_0^{(\alpha)}(u) &= V^{(\alpha)}(u)\,, \\
 \delta^{(\alpha)} &= 0\,, \\
 S_1^{(\alpha)}(u) &= -\frac{1}{2}\log\frac{b^{(\alpha)}}{m^2}\,. \label{S1calc}
\end{align}
To actually evaluate the coefficients $\Upsilon_{i,j}(p,u)$,
we refer back to the expansion \eqref{qdl_exp} of the quantum dilogarithm
in Section~\ref{sec:statesumdef}.
We find
\begin{align}
\Upsilon_{j,k}(p,u) &=\frac{B_{k+1}(1/2)\,2^k}{(k+1)! j!}
\left[\text{Li}_{1-j-k} (-e^{p-u})-(-1)^j \text{Li}_{1-j-k} (-e^{-p-u})\right]
\label{jklambda}
\end{align}
when $j\geq 2 \;\mbox{or}\; k \geq 0$, and that all $\Upsilon_{j,2k}$ vanish.

Therefore: to calculate the expansion coefficients $S_n^{(\alpha)} (u)$
around a given critical point, we substitute $p^{(\alpha)}(u)$ from \eqref{HikpGC}
into equations \eqref{Hik41V} and \eqref{jklambda} to obtain $V^{(\alpha)}$,
$b^{(\alpha)}$, and $\Upsilon_{j,k}^{(\alpha)}$;
then we substitute these functions into expressions for the $S_n$ and simplify.
At the geometric critical point $p^{({\rm geom})}(u)$, we obtain
\begin{align}
V^{({\rm geom})}(u) &= \frac{1}{2}\left[\Li(-e^{p^{(\rm geom)}(u)-u})-\Li(-e^{-p^{(\rm geom)}(u)-u})-4p^{(\rm geom)}(u)\, u + 4\pi i u\right]\,,\\
b^{({\rm geom})}(u) &= \frac{im^2}{2} \Delta(m)\,,
\end{align}
and it is easy to check with a little algebra that all the expansion coefficients $S_n^{({\rm geom})}(u)$
reproduce exactly%
\footnote{There appears to be a small ``correction'' of $\frac{1}{2}\log(-1) = \log(\pm i)$ in $S_1^{(\rm geom)}$, comparing \eqref{S1calc} with the value in Table \ref{tab:figeightgeom}. This merely multiplies the partition function by $i$ and can be attributed to the orientation of the stationary-phase contour passing through the geometric critical point. We also allow the usual modulo $2\pi i u$ ambiguity in matching $S_0$.} %
what we found in the previous subsection by quantizing
the moduli space of flat connections. (This has been verified to eight-loop order.)
Similarly, at the conjugate critical point $p^{({\rm conj})}(u)$, we have
\be
V^{({\rm conj})} (u) = -V^{({\rm geom})} (u)\,, \qquad
b^{({\rm conj})} (u) = -\frac{im^2}{2}\Delta(m)\,,
\ee
and more generally $S_n^{(\rm conj)}=(-1)^{n-1}S_n^{(\rm geom)}$. It is not hard to actually prove this relation between the geomtric and conjugate critical points
to all orders by inspecting the symmetries of $e^{\Upsilon(\hbar,p,u)}$.
Thus, we find complete
agreement with the results of Section~\ref{sec:figeightquant}.


\acknowledgments

We would like to thank D.~Auroux, N.~Dunfield, S.~Garoufalidis, K.~Hikami,
T.~Mrowka, W.~Neumann, E.~Witten, and C.~Zickert for useful discussions and correspondence.
Research of SG is supported in part by NSF Grant PHY-0757647
and in part by the Alfred P. Sloan Foundation.
JL acknowledges support from a Marie Curie Intra-European Fellowship.
TD acknowledges support from a National Defense Science and Engineering Graduate Fellowship.
Opinions and conclusions expressed here are those of the authors
and do not necessarily reflect the views of funding agencies.



\begin{thebibliography}{99}

\bibitem{andersen-hansen}
J.E.~Andersen and S.K.~Hansen, \textsl{Asymptotics of the quantum invariants for surgeries on the figure 8 knot},
\newblock J. Knot Theory and Its Ramifications \textbf{ 15} (4), 479--548 (2006).

\bibitem{atiyah-1990}
M.~Atiyah,
\newblock \textsl{ The Geometry and Physics of Knots},
\newblock Cambridge Univ. Press, 1990.

\bibitem{as-1992}
S.~Axelrod and I.~M. Singer, \textsl{ Chern-{S}imons perturbation theory},
\newblock Proceedings of the XXth International Conference on Differential Geometric
Methods in Theoretical Physics, Vol.\ 1, 2 (New York, 1991) (River Edge, NJ),
World Sci. Publishing, 1992, pp.~3--45.

\bibitem{barnatan-1991}
D.~Bar-Natan,
\textsl{ Perturbative Aspects of the Chern-Simons Topological Quantum Field Theory},
\newblock Ph.D. thesis, Princeton Univeristy, June 1991.

\bibitem{barnatan-1991w}
D.~Bar-Natan and E.~Witten,
\textsl{ Perturbative expansion of Chern-Simons theory with noncompact gauge group},
\newblock Commun. Math. Phys. \textbf{ 141}, 423 (1991).

\bibitem{bayen-1978}
F.~Bayen, M.~Flato, C.~Fronsdal, A.~Lichnerowicz and D.~Sternheimer,
\textsl{ Deformation Theory And Quantization. 1. Deformations Of Symplectic Structures},
\newblock Annals Phys.  \textbf{ 111}, 61 (1978).

\bibitem{bloch-2006}
S.~Bloch, H.~Esnault and D.~Kreimer,
\textsl{ On motives associated to graph polynomials},
\newblock Commun. Math. Phys. \textbf{ 267}, 181--225 (2006), {math/0510011}.

\bibitem{cooper-1994}
D.~Cooper, M.~Culler, H.~Gillet, D.~Long and P.~Shalen,
\textsl{ Plane curves associated to character varieties of 3-manifolds},
\newblock Invent. Math. \textbf{ 118}, 47--84 (1994).

\bibitem{cooper-1996}
D.~Cooper and D.~Long, \textsl{ Remarks on the A-polynomial of a knot},
\newblock J. Knot Theory and Its Ramifications \textbf{ 5} (5), 609--628 (1996).

\bibitem{garoufalidis-costin}
O.~Costin and S.~Garoufalidis,
\newblock \textsl{ Resurgence of 1-dimensional sums of Sum-Product type}, \newblock in preparation.

\bibitem{dijkgraaf-2009}
R.~Dijkgraaf and H.~Fuji,
\newblock \textsl{ The Volume Conjecture and Topological Strings},
\newblock {arXiv:0903.2084} [hep-th].

\bibitem{dimofte-2009}
T.~Dimofte and J.~Lenells, \textsl{ Observations on the State Integral Model for $SL(2,\C)$ Chern-Simons Theory}, 
\newblock in preparation.

\bibitem{faddeev-1994}
L.~D. Faddeev,
\newblock Current-Like Variables in Massive and Massless Integrable Models,
\newblock in \textsl{ Quantum groups and their applications in physics (Varenna  1994)},
pages 117--135, Proc. Internat. School Phys. Enrico Fermi, 127, IOS, Amsterdam, 1996,
\newblock {hep-th/9408041}.

\bibitem{faddeev-1999}
L.~Faddeev, \textsl{ Modular Double of Quantum Group},
\newblock (1999), {math/9912078}.

\bibitem{faddeev-2001-219}
L.~D. Faddeev, R.~M. Kashaev and A.~Y. Volkov,
\textsl{ Strongly coupled quantum discrete Liouville theory. I: Algebraic approach and duality},
\newblock Commun. Math. Phys. \textbf{ 219}, 199--219 (2001), {hep-th/0006156}.

\bibitem{freed-2008}
D.~S.~Freed, \textsl{ Remarks on Chern-Simons Theory},
\newblock {arXiv:0808.2507} [math.AT].

\bibitem{garoufalidis-2004}
S.~Garoufalidis,
\newblock \textsl{ On the characteristic and deformation varieties of a knot},
\newblock in Geometry and Topology Monographs, number~7, pages 291--309, 2004.

\bibitem{garoufalidis-2006}
S.~Garoufalidis and J.~Geronimo,
\newblock \textsl{ Asymptotics of $q$-difference equations}, 2006.

\bibitem{garoufalidis-geronimo}
S.~Garoufalidis and J. Geronimo, \newblock \textsl{ A Riemann-Hilbert approach to the asymptotics of 1-dimensional sums of Sum-Product type},
\newblock in preparation.

\bibitem{goncharov-2007}
A.B.~Goncharov,
\newblock The pentagon relation for the quantum dilogarithm and quantized $M_{0,5}$,
\newblock in \textsl{ Geometry and dynamics of groups and spaces},
pages 415--428, Progr. Math. {\bf 265}, Birkh\"auser, Basel, 2008,
\newblock {arXiv:0706.4054} [math.QA].

\bibitem{GruenMor}
D.~Gr\"unberg and P.~Moree, \textsl{Sequences of enumerative geometry: congruences and asymptotics}.
Experim.~Math., to appear.

\bibitem{gukov-2003}
S.~Gukov, \textsl{ Three-Dimensional Quantum Gravity, Chern-Simons Theory, and the A-Polynomial},
\newblock Commun. Math. Phys. \textbf{ 255}, 577--627 (2005), {hep-th/0306165}.

\bibitem{gukov-2006}
S.~Gukov and H.~Murakami,
\textsl{ SL(2,C) Chern-Simons theory and the asymptotic behavior of the colored Jones polynomial},
\newblock (2006), {math/0608324}.

\bibitem{gukov-2008}
S.~Gukov and E.~Witten,
\textsl{ Branes and Quantization},
\newblock {arXiv:0809.0305} [hep-th].

\bibitem{heinonen-1998}
O.~Heinonen, (ed.) \newblock \textsl{ Composite fermions:
A unified view of the quantum Hall regime},
\newblock Singapore: World Scientific, 1998.

\bibitem{hikami-2001-16}
K.~Hikami, \textsl{ Hyperbolic Structure Arising from a Knot Invariant},
\newblock Int. J. Mod. Phys. A \textbf{ 16}, 3309--3333 (2001), {math-ph/0105039}.

\bibitem{hikami-2006}
K.~Hikami,
\newblock Generalized Volume Conjecture and the A-Polynomials -- the Neumann-Zagier
Potential Function as a Classical Limit of Quantum Invariant, 2006.

\bibitem{hitchin-1987}
N.~Hitchin, \textsl{ The Self-Duality Equations On A Riemann Surface,}
\newblock Proc. London Math. Soc.(3) \textbf{ 55}, 59-126 (1987).

\bibitem{kashaev-1997}
R.~M. Kashaev, \textsl{ The hyperbolic volume of knots from the quantum dilogarithm},
\newblock Lett. Math. Phys. \textbf{ 39} 269--275 (1997), {q-alg/9601025}.

\bibitem{kashaev-2000}
R.~M. Kashaev,
\newblock On the spectrum of Dehn twists in quantum Teichmuller theory,
\newblock in \textsl{ Physics and combinatorics (Nagoya, 2000)},
pages 63--81, World Sci. Publ., River Edge, NJ, 2001,
\newblock {math/0008148}.

\bibitem{kontsevich-2001}
M.~Kontsevich and D.~Zagier, \textsl{ Periods},
\newblock in \textsl{ Mathematics unlimited---2001 and beyond},
pages 771--808, Springer, Berlin, 2001.

\bibitem{kontsevich-2006}
M.~Kontsevich, \textsl{ Deformation Quantization Of Algebraic Varieties},
\newblock Lett. Math. Phys. \textbf{ 56} 271-294 (2006), {math/0106006}.

\bibitem{labastida-1999}
J.~M.~F.~Labastida, \textsl{ Chern-Simons gauge theory: Ten years after},
\newblock (1999), {hep-th/9905057}.

\bibitem{LewisZag}
J.~Lewis, D.~Zagier, \textsl{ Period functions for Maass wave forms. I},
\newblock Ann. of Math. \textbf{ 153}, 191--258 (2001).

\bibitem{marino-2004}
M.~Marino, \textsl{ Chern-Simons theory and topological strings},
\newblock Rev. Mod. Phys. \textbf{77} (2005) 675,
\newblock {hep-th/0406005}.

\bibitem{milnor-1982}
J.~Milnor, \textsl{ Hyperbolic geometry: the first 150 years},
\newblock Bull. Amer. Math. Soc. \textbf{ 6}, 9--24 (1982).

\bibitem{mostow-1973}
G.D.~Mostow, \textsl{ Strong rigidity of locally symmetric spaces},
\newblock Ann. Math. Studies \textbf{ 78} (1973).

\bibitem{murakami-2006}
H.~Murakami, \textsl{ A version of the volume conjecture},
\newblock (2006), {math/0603217}.

\bibitem{murthy-2003}
G.~Murthy and R.~Shankar,
\textsl{ Hamiltonian theories of the fractional quantum Hall effect},
\newblock Rev. Mod. Phys.  \textbf{75} (2003) 1101.

\bibitem{neumann-2004}
W.~D. Neumann, \textsl{ Extended Bloch group and the Cheeger-Chern-Simons class}, \newblock Geom. Topol. \textbf{ 8}, 413--474 (2004).

\bibitem{neumann-1997}
W.~D. Neumann and J.~Yang, \textsl{ Bloch invariants of hyperbolic 3-manifolds}, \newblock Duke Math. J. \textbf{ 96}, 29--59 (1999).

\bibitem{nz-1985}
W.~D. Neumann and D.~Zagier, \textsl{ Volumes of hyperbolic three-manifolds},
\newblock Topology \textbf{ 24}, 397--332 (1985).

\bibitem{petronio-2000}
C.~Petronio and J.R.~Weeks,
\textsl{ Partially flat ideal triangulations of cusped hyperbolic 3-manifolds},
\newblock Osaka J. Math.   \textbf{37}, 453--466 (2000).

\bibitem{prasad-1973}
G.~Prasad, \textsl{ Strong rigidity of $Q$-rank 1 lattices},
\newblock Invent. Math. \textbf{ 21}, 255--286 (1973).

\bibitem{sw-1994}
N.~Seiberg and E.~Witten, \textsl{ Electric - magnetic duality,
monopole condensation, and confinement in N=2 supersymmetric Yang-Mills theory},
\newblock Nucl. Phys. B \textbf{ 426} (1994) 19, [Erratum-ibid.  B \textbf{ 430} (1994) 485],
{hep-th/9407087}.

\bibitem{thurston-1980}
W.~P. Thurston,
\newblock The geometry and topology of three-manifolds,
\newblock Lecture notes at Princeton University, Princeton, 1980.

\bibitem{thurston-1982}
W.~P. Thurston, \textsl{ Three Dimensional Manifolds, Kleinian Groups, and Hyperbolic Geometry},
\newblock Bull. AMS \textbf{ 6} (3), 357--381 (1982).

\bibitem{thurston-1986}
W.~P. Thurston, \textsl{ Hyperbolic structures on 3-manifolds. I. Deformations of acylindrical manifolds},
\newblock Ann. Math. \textbf{ 124} (2), 203--246 (1986).

\bibitem{thurston-1999}
D.~Thurston,
\newblock Hyperbolic Volume and the Jones Polynomial,
\newblock in \textsl{ Invariants de Noeuds et de Varietes de Dimension 3 (Greboble 1999)},
Institut Fourier, 1999.

\bibitem{turaev-1992}
V.G.~Turaev and O.Yu.~ Viro, \textsl{ State-sum invariants of 3-manifolds and quantum 6j-symbols},
\newblock Topology \textbf{ 31} 865 (1992).

\bibitem{volkov-2003}
A.~Y. Volkov, \textsl{ Noncommutative Hypergeometry},
\newblock Commun. Math. Phys. \textbf{ 258} (2), 257--273 (2005), {math/0312084}.

\bibitem{witten-1989}
E.~Witten, \textsl{ Quantum field theory and the Jones polynomial},
\newblock Commun. Math. Phys. \textbf{ 121}, 351--399 (1989).

\bibitem{witten-1991}
E.~Witten,
\textsl{ Quantization Of Chern-Simons Gauge Theory With Complex Gauge Group},
\newblock Commun. Math. Phys. \textbf{ 137}, 29 (1991).

\bibitem{woodhouse-1992}
N.~Woodhouse,
\newblock \textsl{ Geometric Quantization},
\newblock New York: Oxford University Press, 1992.

\bibitem{yoshida-1985}
T.~Yoshida, \textsl{ The $\eta$-invariant of hyperbolic 3-manifolds},
\newblock Invent. Math. \textbf{ 81}, 473--514 (1985).


\bibitem{zagVass}
D.~Zagier, \textsl{Vassiliev invariatnts and a strange identity related
to the Dedekind eta-function},
\newblock Topology \textbf{40},  945--960 (2001).

\bibitem{zagDilog}
D.~Zagier, \textsl{The dilogarithm function},
\newblock In {\it Frontiers in Number Theory, Physics and Geometry~II},
P.~Cartier, B.~Julia, P.~Moussa, P.~Vanhove (eds.),
  Springer-Verlag, Berlin-Heidelberg-New York, 3--65 (2006).


\bibitem{zagHans}
D.~Zagier, \textsl{Algebraic and asymptotic properties of quantum invariants of knots},
\newblock in preparation.

\bibitem{zagMod}
D.~Zagier, \textsl{ Quantum modular forms}, \newblock Preprint (2009), 16 pages.

\bibitem{zickert-2008}
C.~Zickert, \textsl{The Chern-Simons invariant of a representation}, \newblock arXiv:0710.2049 [math.GT].


\end{thebibliography}
\end{document}